\documentclass[a4paper,12pt,onecolumn,preprint,doublespace,rsc]{revtex4}
\usepackage[latin1]{inputenc}
\usepackage{xcolor}
\usepackage{array,multirow}
\usepackage{times}
\usepackage{paralist}
\usepackage{amsmath,amssymb}
\usepackage{graphicx}
\usepackage{subfigure}
\usepackage{pdfpages}
\usepackage{anysize}
\usepackage{verbatim}
\usepackage{listings}
\usepackage{tabularx}
\usepackage{float}
\marginsize{1.5cm}{1.5cm}{1.5cm}{1.5cm}
\newcommand{\Eq}[1]{Eq.~(\ref{eq:#1})}

\begin{document}

\title{Structure and water attachment rates of ice in the atmosphere: role of nitrogen}

\author{Pablo Llombart $^{\dag\ddag}$, Ramon M. Bergua$^{\ddag}$, Eva G. Noya
$^\dag$, and Luis G. MacDowell$^{\ddag,*}$}

\affiliation{$^\dag$ Instituto de Qu\'{\i}mica F\'{\i}sica Rocasolano
, CSIC, Calle Serrano 119, 28006 Madrid, Spain}

\affiliation{$\ddag$ Departamento de Qu\'{\i}mica-F\'{\i}sica (Unidad de I+D+i Asociada
   al CSIC), Facultad de
Ciencias Qu\'{\i}micas, Universidad Complutense de Madrid, 28040 Madrid, Spain}

\email[]{lgmac@quim.ucm.es}

\begin{abstract}
In this work we perform  computer simulations  of the ice
surface  in order to elucidate the role of nitrogen in the 
crystal growth rates and crystal habits of snow in the atmosphere. 
In pure water vapor at temperatures typical of ice crystal formation
in cirrus clouds, we find that basal and primary prismatic
facets exhibit a layer of premelted ice, with thickness
in the subnanometer range.   
For partial pressures of 1 bar, well above the expected 
values in the troposphere, we find that only small amounts
of nitrogen are adsorbed. The adsorption takes place onto the 
premelted surface, and hardly any nitrogen dissolves within
the premelting film.  The premelting film thickness does not
change either.  We quantify the resulting change of
the ice/vapor surface tension to be in the hundredth of mN/m  and
find that the structure of the pristine ice surface is not changed
in a significant manner.  
We perform a trajectory analysis of colliding
water molecules, and find that the attachment rates from direct
ballistic collision are very close to unity irrespective of the
nitrogen pressure. Nitrogen is however at sufficient density to
deflect a fraction of trajectories with smaller distance than the mean 
free path.
 Our results show explicitly that the reported differences
   in growth rates measured in pure water vapor and a controlled nitrogen
   atmosphere are not related to a significant disruption of the
   ice surface due to nitrogen adsorption. On the contrary, we show
   clearly from our trajectory analysis that nitrogen slows down
   the crystal growth rates due to collisions between water molecules
   with bulk nitrogen gas. This clarifies the long standing controversy
   of the role of inert gases on crystal growth rates and demonstrates
   their influence is solely related  to  the diffusion limited flow of 
   water vapor across the  gas phase.




\end{abstract}

\maketitle



\section{Introduction}

Modeling of radiation processes in clouds is an essential
requisite for the prediction of climate change.\cite{bartels-rausch13, loyola18} 
In the troposphere, ice is a widespread component, as cirrus clouds
cover about 30 \% of the mid latitudes at any
given time.\cite{baran12} The  ice grains that make
these cirrus clouds, whether in mono-crystalline form or 
as crystal  aggregates play a crucial role on the earth's
climate.\cite{warren08,hesse12,neshyba13,voigtlander18,jarvinen18} 
They  account for a significant amount of the radiation 
budget,\cite{warren08,hesse12,neshyba13,voigtlander18} and concentrate airborne 
chemicals with an important contribution to the
atmosphere's chemistry.\cite{abbat03} 
However,  the underlying microphysics of ice crystallites 
is a major source of uncertainty for climate change models,
while the mechanism for ice growth and surface activity
remains poorly understood.\cite{bartels-rausch13}

In view of this situation, a great number of recent computer simulation 
studies seek to characterize the ice surface
and shed light onto  growth
mechanisms.\cite{conde08,yang13,limmer14,neshyba16,hudait16,kling18,pickering18,mohandesi18,qiu18}
In most such studies,  a bulk ice sample is placed either
in vacuum  or within a saturated water vapor atmosphere,
and only rarely is the interaction of
atmospheric gases considered explicitly.\cite{sandler94,girardet01}  
Of course, acidic gases and polar organic molecules adsorbed on the ice surface
are expected to play a crucial impact on atmospheric
chemistry.\cite{abbat03,girardet01,hudait17,waldner18} 
However, they are usually only found in trace amounts.
As an example, formic acid, which is one of the most abundant strong short 
chain acid in the atmosphere is found in a mixing ratio of 2 parts per billion
in the boundary layer air.\cite{waldner18} Therefore, the study of pristine 
ice surfaces as an initial starting point is warranted.\cite{hudait17}

On the contrary, the two major components of the atmosphere, nitrogen
and oxygen, have far less affinity for ice, but are found at concentrations
two orders of magnitude larger than saturated water vapor. Although
laboratory studies are often performed on a controlled atmosphere
of nitrogen gas to avoid possible
bias,\cite{sazaki12,asakawa15,asakawa16,murata16,mitsui19} the fact is that
their  role on the structure and growth rates remains largely
unknown.\cite{girardet01}

A chemical physicist's first guess clearly
indicates that an inert gas such as nitrogen should have a very small
impact on the properties of the ice surface. Indeed, reported 
thermodynamic data on surface 
tension and phase coexistence data for water essentially ignore the 
role of the surrounding atmosphere.  From Henry's constant data,\cite{sander15}
it is expected that the concentration of nitrogen in water at
0~C and one bar is not more than 10$^{-3}$ mol/L, and likely far smaller
in ice. Furthermore, BET adsorption studies of nitrogen on ice
consistently yield small enthalpies of
adsorption of about $\Delta H_{ads}=8.1$~kJ/mol at 77~K.\cite{adamson67,schmitt87,hoff98} Measurements
using other apolar gases such as methane yield similar results.\cite{hanot99,legagneux02} 
Furthermore, Temperature Programmed Desorption experiments show
that   monolayers of nitrogen on ice have fully desorbed at
temperatures below 50~K, corresponding to estimated adsorption energies
of 9.6~kJ/mol at most.\cite{fayolle15,minissale16,nguyen18}

Intriguingly, laboratory studies of surface structure
and crystal growth rates have recurrently reported significant 
differences between experiments performed on either  a controlled nitrogen 
or pure water vapor
atmosphere.\cite{beckmann82b,beckmann83,kuroda84b,elbaum93,bluhm02,libbrecht17}

In  experiments of ice crystals grown from the vapor, it is well known that
the presence of significant amounts of nitrogen gas 
largely slows down the crystal growth rates.\cite{jayaweera71,kuroda84} However, it has
been usually interpreted that this is not the effect of a significant
perturbation of the ice surface, but rather, the effect of diffusion
limited growth.\cite{gonda70,lamb72} This  occurs when the slow diffusion rate of water vapor
in air is unable to reestablish the depleted ambient vapor density 
close to the growing ice surface. As a result, the interpretation of
growth measurements requires to distinguish the local surface saturation
from the asymptotic bulk vapor saturation. In view of the difficulty
to accurately estimate the full density field at the crystal,
standard experiments aimed
at  unraveling the {\em correct}  surface attachment kinetics have
rather been performed in low pressure chambers with as small nitrogen as 
possible, under the assumption that nitrogen plays no other role than
decreasing the diffusion coefficient of water 
vapor.\cite{lamb72,beckmann82,gonda87,sei89,libbrecht12}

However, some studies on crystal growth, with explicit account of diffusion 
resistance of air reported that the surface saturation corrected
growth rates were strongly influenced
by nitrogen,\cite{beckmann82b,beckmann83,kuroda84b}
 decreasing the surface kinetics selectively on
the prismatic facet by factors as large as 1/100 at T=-11 C, 
and  1/20 at T=-1/20 at -15 C compared to experiments in pure water 
vapor.\cite{beckmann82b,kuroda84b}
In fact, it was claimed that adsorption of nitrogen was essential to
explain the anisotropic growth of ice crystals as
described in the Nakaya diagram.  Particularly, Beckmann and collaborators
postulated a nitrogen poisoning mechanism that is well known
in catalysis.\cite{beckmann83} Unfortunately, this explanation
assumes enthalpies of adsorption ranging from 25 to 40~kJ/mol, 
which  are very much at odds with BET isotherm experiments,
recent Temperature Programmed Desorption
experiments,\cite{fayolle15,minissale16,nguyen18}
and ab-initio
calculations.\cite{adamson67,schmitt87,hoff98,girardet01} On theoretical
grounds, another possible role of adsorbed gases is to proliferate in
between crystal steps and kinks. This can potentially change the
step free energy, and as a result, change significantly the
surface roughness and crystal growth rates.\cite{akutsu01}
However, none of these hypothesis seem to be consistent with
later studies of crystal growth habits at reduced pressure,
 which seem to agree well with results for
the standard Nakaya diagram,\cite{vandenheuvel59,lamb72,sei89}
and exhibit a distinct growth
anisotropy,\cite{lamb71,lamb72,gonda87,sei89,libbrecht12} similar to that
observed in the presence of air.\cite{hallett58,kobayashi67}
Furthermore, similar
slowing down of growth rates has been confirmed with other
inert gases, in support of merely a diffusion limited role on
ice growth.\cite{vandenheuvel59,gonda70} 

Atmospheric gases have also been postulated to play a
significant role in determining the thickness of premelting films
of water adsorbed onto the ice 
surface.\cite{elbaum91,elbaum91b,elbaum93,lied94,dosch95,bluhm02,mitsui19}
 Indeed, conflicting results
for the equilibrium thickness of premelting films have regularly been found, 
with thicknesses differing by several orders of 
magnitude.\cite{dash06,li07b,michaelides17} These discrepancies
can be partly reconciled by taking into account that contamination
can increase the thickness of the premelting
film considerably.\cite{elbaum93,wettlaufer99,bluhm02,mitsui19}
An uncontrolled exposure to the atmosphere can easily promote
such contamination, and this explains why often crystal samples
in air exhibit much larger film thicknesses.\cite{bluhm02} However,
claims that the thickness of premelting films diverges in the
presence of nitrogen,\cite{elbaum93} do not seem warranted in view
of more recent experiments indicating a finite equilibrium thickness
in a controlled nitrogen 
atmosphere.\cite{sazaki12,asakawa15,asakawa16,murata16}

In view of this  account, it seems clear that
the role of adsorbed nitrogen on such significant atmospheric
properties as the equilibrium surface structure and the crystal
growth rate of ice remains largely unknown and controversial.
Certainly, whatever conclusion is made from the
current  evidence is a result of indirect and hypothetical arguments
from macroscopic samples.  Computer simulations could be an invaluable
tool to shed light into this complex problem, as they can elucidate
the microscopic features of the crystal growth mechanism. Indeed,
Libbrecht has urged for such a study, in order to clarify some
conflicting results on the role of nitrogen in recent crystal
growth experiments.\cite{libbrecht12,libbrecht17}

In this work we perform a computer simulation study of the ice
surface under a controlled nitrogen atmosphere. In our study
we first perform computer simulations of  ice 
in presence of  pure water vapor and provide a detailed
structural characterization of the ice/vapor interface
(section \ref{sec:pristine}).  We then perform additional
simulations of ice under nitrogen gas and study the
adsorbed layer of nitrogen (section \ref{sec:nitrogen}), and its
influence on the structural and thermodynamic
properties of the ice interface (section \ref{sec:strucythermo}). 
Finally, we present results on the influence of nitrogen
gas in water attachment kinetics.
(section \ref{sec:attachres}). In section \ref{sec:concl} we
summarize and discuss the significance of our results.
Overall we find that nitrogen gas does not  significantly change
the properties of the ice surface, and conclude
that crystal growth rate
and crystal habit differences observed for ice in nitrogen
are  related only to the complex diffusion limited
flow of water towards the ice surface.

\section{Methods}

\newcolumntype{R}{>{\raggedleft\arraybackslash}X}%
\newcolumntype{R}{>{\raggedleft\arraybackslash}X}%
\newcolumntype{b}{>{\hsize=0.1\hsize}l}
\newcolumntype{o}{>{\hsize=2.0\hsize}l}
\newcolumntype{s}{>{\hsize=0.5\hsize}X}
\newcolumntype{m}{>{\hsize=0.5\hsize}R}

\begin{table}[ht!]
\footnotesize
\begin{tabularx}{0.3\textwidth}{XRR}
\hline
\hline
$Altitude/m$ & $T/K$ & $p\cdot 10^{-5}/$~Pa    \\
\hline
 2750         & 270   &   0.72    \\
 4250         & 260   &   0.60    \\
 5800         & 250   &   0.48    \\
 7310         & 240   &   0.40    \\
 8840         & 230   &   0.32    \\
\hline
\hline
\end{tabularx}
\caption {Average atmospheric temperature and pressure as a function of 
altitude according to the  International Standard Atmosphere (ISO 2533:1975)
\label{alt_temp_press}
}
\end{table}


\subsection{Simulations}

All simulations were performed with the simulation package
GROMACS\cite{gromacs3,gromacs4}.
Simulations were carried out mostly in the 
NVT ensemble. Temperature was controlled using the Bussi-Donadio-Parrinello velocity-rescale
thermostat with a relaxation time of 1~ps\cite{bussi07}. In those simulations carried out
in NpT ensemble, the Berendsen barostat was used to achieve the desired 
pressure. Periodic
boundary conditions were applied on the three directions of space. 
 Lennard-Jones interactions were truncated at a distance of 9~{\AA} and
homogeneous long range corrections applied.\cite{allen17,frenkel02} Coulomb interactions were
evaluated using  Particle Mesh Ewald, with the real space contribution
truncated also at 9~{\AA}. The reciprocal space term is evaluated
over a total of $80\times 64\times 160$  vectors in the 
$x$, $y$, $z$ reciprocal directions, respectively. The
charge structure factors were evaluated with a grid spacing of 0.1~nm
and  a fourth order interpolation scheme. The
Gaussian charge distribution width was set to 0.288146~nm.
Equations of movement were solved using constraint
molecular dynamics, with the GROMACS default LINCS algorithm,
the velocity-Verlet integrator 
and a timestep of 3~fs.

\subsection{Force Field}

Water and nitrogen were described using rigid, non-polarizable 
 point charge
models, TIP4P/Ice~\cite{abascal05} 
for water and TraPPE~\cite{siepmann01} for nitrogen. TIP4P/Ice was fitted to reproduce the melting
temperature of ice Ih, the densities of liquid water and ices II and V, 
and the stability region of ice III\cite{abascal05}, whereas TraPPE was fitted to reproduce the 
vapor-liquid equilibrium of
nitrogen 
 and nitrogen mixtures~\cite{siepmann01}.  It reproduces the full coexistence
curve up to the critical point as well as the vapor pressures.
Crossed Lennard-Jones
   interactions between nitrogen and water are modeled using Lorentz-Berthelot
   rules, with a crossed LJ $\sigma$ parameter determined as the arithmetic 
   average and the crossed LJ $\epsilon$ parameter determined as the geometric 
mean. 

Of course, the use of two widely recognized force fields for the
pure components does not guarantee an accurate description of the crossed
interactions, which is here a crucial issue. For this reason we have performed
a number of calculations to check the validity of the model as regards
H$_2$O-N$_2$ interactions. 

\subsubsection{H$_2$O-N$_2$ dimer interactions}

The binding energy of H$_2$O-N$_2$ dimers has been estimated by a number of
authors using ab-initio, and semi-empirical
methods.\cite{sadlej95,girardet01,tulegenov07}
Using a semi-empirical method, Girardet and Toubin advocate $E_b=4.8$ kJ
/mol, for the optimized binding energy, in agreement
with ab-initio MP2 results by Sadlej et al,\cite{sadlej95} and close to
a more recent MP2 calculation which provide $E_b=5.2$~kJ/mol.\cite{tulegenov07} 

We have performed an energy minimization for the gas phase
H$_2$O-N$_2$ dimer interactions, using GROMACS energy
optimization module. This applies a  a steepest descent algorithm to
find the global energy minimum of the system.  The lowest energy configuration
predicted by our model corresponds to the linear nitrogen molecule aligned
along an OH bond in water, precisely as found in electronic
calculations.\cite{tulegenov07}
The optimized energy yields $E_b=4.97$~kJ/mol, which lies in between
the MP2 results of Sadlej et al.\cite{sadlej95} and Tulegenov et
al.\cite{tulegenov07}, and is less than 5\%  away from either calculation.

\subsubsection{Heat of adsorption of nitrogen on ice}

Adsorption energies of nitrogen on ice at temperatures in the range studied in
this work do not seem to be available. The most significant sources of
information come from the astrophysics \cite{fayolle15,minissale16,nguyen18}
and geophysics community for temperatures below
100~K.\cite{adamson67,schmitt87,hoff98}
In astrophysics, temperature programed desorption experiments (TDP) in
the range between 20 to about 50~K have been
performed on both polycrystalline ice samples (PCI) and amorphous solid
water (ASW).\cite{fayolle15,minissale16,nguyen18} Results
show little differences between one or the other solid phase,\cite{nguyen18} and
provide average adsorption energies
ranging between  9.6 \cite{minissale16} and 9.5 kJ/mols.\cite{fayolle15}
However, it is well
known that at the low temperature of these experiments, nitrogen molecules are
sufficiently mobile to surface hop to the most favorable adsorption sites before
desorption. At higher temperatures, where desorption proceeds much faster,
molecules do not typically have time to reach the most favorable adsorption
sites and the average adsorption energy can be significantly
smaller.\cite{fayolle15,minissale16,nguyen18}
This can be already seen from BET experiments of N$_2$ on snow, that are
performed in the geoscience community as a probe of snow
porosity at a temperature of 77~K.\cite{adamson67,schmitt87,hoff98} Results
 show a very broad range of porosity depending
on sample location and history, but yield fairly consistent average
adsorption energies of ca. 8.1~kJ/mol,\cite{adamson67,schmitt87} already
significantly lower than
results observed for TPD in the range between 20 and 50~K, and somewhat lower
than estimates obtained from the frequency shift of the OH dangling bond stretch
upon adsorption, which provide ca. 8.7 kJ/mol.\cite{devlin95}
These results all fall between the H$_2$O-N$_2$ dimer energy and
the hydration energy of N$_2$ in water at ambient temperature, which,
from the NIST webbook is 10.8~kJ/mol. In view of this discussion,
the ab-initio calculations of Manca and Allouche,\cite{manca01} which suggest
an energy of adsorption of barely 3.1~kJ/mol seem to be somewhat too low.

In section \ref{sec:strucythermo} we calculate the adsorption energies of nitrogen gas
on the ice surface. Our results provide $\Delta H_{ads}=7.4$~kJ/mol for
the basal face and $\Delta H_{ads}=6.5$~kJ/mol for the prismatic face. Bearing
in mind that these results are measured from adsorption data
in the range between 210 and 270~K, i.e.,  150~K above the
BET experiments, and enthalpies of adsorption
are expected to decrease with temperature,\cite{fayolle15,nguyen18}   
our results seem very reasonable.

Table \ref{tab:ads} summarizes available experimental results for H$_2$O-N$_2$ interactions
relevant to our study. The agreement with results from our work
is overall rather satisfactory, in view of the lack of experimental data
available in the range of temperatures considered here.

\begin{table}
\begin{tabular}{ccccc}
 Technique  & Sample & Temperature range / K &  $\Delta
H_{ads}$/kJ$\cdot$mol$^{-1}$ & Reference \\
\hline
\hline
  Semi-empirical &  N$_2$-H$_2$O & 0 & 4.8 &  \cite{girardet01} \\
  ab-initio MP2 &  N$_2$-H$_2$O & 0 & 5.2 &  \cite{tulegenov07} \\
  ab-initio MP2 &  N$_2$-H$_2$O & 0 & 4.8 &  \cite{sadlej95} \\
   MD           & N$_2$-H$_2$O &  0 &  5.0 &  This work \\
\hline
\hline
TPD  & ASW & 20-50 &  9.5 & \cite{fayolle15} \\ 
 TDP  & ASW & 20-50 &  9.6 & \cite{minissale16} \\ 
 BET  & PCI & 77 & 8.1 & \cite{adamson67} \\
 BET  & PCI & 77 & 8.3 & \cite{schmitt87} \\
 IRS  & Ic  & 96 & 8.7 & \cite{devlin95} \\
 HF   & Ih & 0  & 3.3 & \cite{manca01} \\
 MD   & Ih-basal & 230-270  & 7.4  & This work \\ 
 MD   & Ih-pI & 230-270  & 6.5 & This work \\ 
\hline
\end{tabular}
\caption{Adsorption enthalpies of N$_2$ for solid water phases as
estimated with different techniques. TDP: Temperature Programmed Desorption;
BET: Brunauer-Emeret-Teller isotherms; IRS: Infrared spectroscopy; HF-MP2:
Hartree Fock plust M\"oller Plesset perturbation theory.
Samples are: ASW, Amorphous solid water; PCI: polycrystalline ice; Ic: Cubic ice.
Ih: hexagonal ice.
}
\label{tab:ads}
\end{table}


\subsection{Determination of the equation of state of nitrogen gas}

Prior to the simulations of the ice-vapor interface, we carried out a series of
simulations aimed at determining the equation of state for nitrogen. This
equation is
later used to relate the number of adsorbed nitrogen molecules with the
corresponding bulk pressure in the ice-vapor system.
The simulations covered densities within $\rho=$0.5-2 $kg\cdot m^{-3}$
and temperatures within $T=$230-270~K (at 5~K intervals),
 corresponding
to  pressures between about 0.30 to 1.1 bars that bracket
the expected pressure of the International Standard Atmosphere
(c.f. Tab.\ref{alt_temp_press}) . 
All simulations were carried
out in a cubic box of length 21.5~nm, containing between 120 and 420 nitrogen molecules
depending on density. Each of these thermodynamic states was simulated
during 3~ns averaging every 75 ps for the calculation of thermal averages. 
Pressure was estimated as an ensemble average using the virial route\cite{allen17}.
For each isotherm, the pressure as a function of density was fitted to:
\begin{equation}
P(T)= \rho k_{B}T(1+B_{2}(T)\rho)
\label{eq:eosN2}
\end{equation}
where $k_B$ is the Boltzmann constant and $B_2(T)$ is the second virial coefficient.
The values of $B_2(T)$ obtained at each temperature were then fitted to the function
$B_2(T)=A_0-A_1 T^{-1}$. Eq.~\ref{eq:eosN2} together with this fit can be used
to predict the pressure of nitrogen for any particular values of density and pressure
within the range of simulated densities and pressures.

\begin{figure}[H]
\centering
\resizebox*{9cm}{!}{\includegraphics{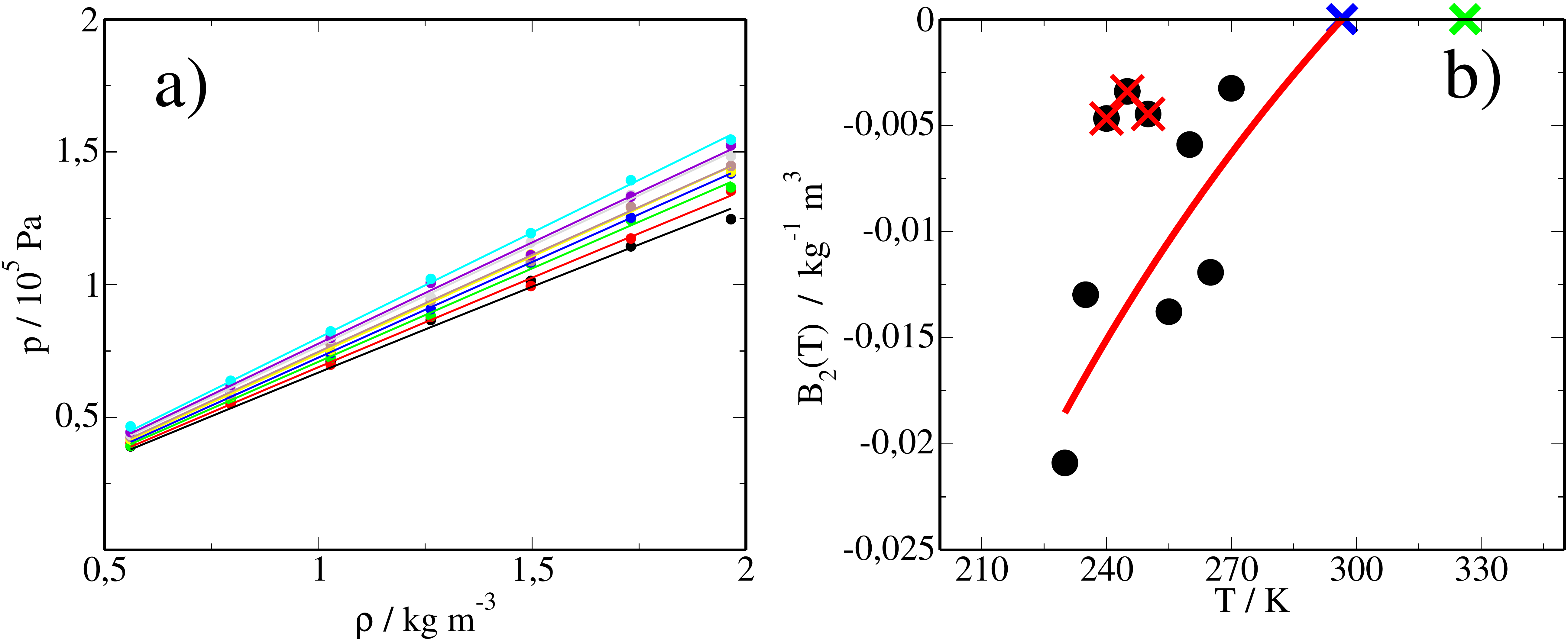}}
\caption{
   (a) Pressure as a function of nitrogen density, with simulation results shown
   as circles. In order of
   decreasing slope, studied temperatures are $230$ (black),
	  $235$ (red), $240$ (green), $245$ (blue), $250$ (yellow), $255$ (gray),
	  $260$ (brown), $265$ (violet) and $270$ (cyan) respectively.
         Fits to Eq. \ref{eq:eosN2} for each isotherm are shown as full 
	   lines with the same color code.
         (b) Second Virial Coefficient, $B_{2}(T)$, as a function of temperature. Black circles 
	   correspond to the slopes obtained from the fit to the isotherms, while
          the red line represents the fit to the function
          $B2(T) = 0.06402 (kg^{-1}m^{3}) - 18.990 (kg^{-1}m^{3} K) T^{-1}(K^{-1}) = A_0-A_1 T^{-1}$.  
	    Symbols with red crosses were  not taken into account for the  fit.
	    The blue cross is the extrapolated Boyle temperature obtained from the
	    simulated results.  The green cross corresponds to the experimental Boyle 
	    temperature.\cite{estrada07} 
}
 \label{fig:eosN2}
\end{figure}

The equation of state data and the corresponding
virial coefficients are collected in detailed form in the supplementary
material.  The pressure isotherms
and  fits to the virial coefficients are displayed in Figure \ref{fig:eosN2}. 
Whereas the empirical model employed for nitrogen was fitted at far lower 
temperatures to match coexistence liquid densities, we find a reasonable
agreement for the Boyle temperature, which, by extrapolation from our results is
$T_B=296$~K, compared to the reported experimental value of
$T_B=326$~K.\cite{estrada07}

\subsection{Simulation of the ice-vapor interface in the presence of nitrogen}

\label{sec:simmet}
The ice-vapor interface was built using the following procedure. We start
by generating a slab of ice containing $8\times8\times5$
 pseudo-orthorhombic unit cells with 16 molecules each, whence,
a total of 5120 water molecules.
At each temperature, the bulk solid is simulated during 6~ns in
the NpT ensemble and $p=1$~bar. The simulation cell is then rescaled to its
equilibrium value, and placed inside a larger elongated cell along the direction
$z$, perpendicular to the interface.  A number of nitrogen molecules are then randomly 
inserted in the vacuum region of the box.
Simulation boxes typically had a lateral size $L_x=7$~nm and $L_y=6$~nm. The
perpendicular size was set to $L_z=15$~nm for pure ice under water vapor and 
$L_z=39$~nm for ice in a nitrogen atmosphere.  For the latter simulations, the gas phase was 
formed with 15 to 50 nitrogen molecules.  Detailed information about system size
and dimensions are provided in the supplemental material.

The system is evolved in the NVT ensemble during 15~ns to 
allow the premelting layer to equilibrate.
Averages are taken over the next 35~ns, saving configurations every 75~ps.
The density profile of the vapor phase established mainly by
nitrogen molecules is measured, and these data are used to infer
the ambient pressure from \Eq{eosN2}. Amounts of nitrogen
are selected such that, for each temperature, 
the pressure ranges from the expected value for the
International Standard Atmosphere 
(c.f. Tab.\ref{alt_temp_press}) to a pressure somewhat above 1 bar.

The amount of nitrogen adsorbed on the ice surface at each thermodynamic state was
quantified by measuring the surface excess or Gibbs adsorption:
\begin{equation}\label{eq:nads}
\Gamma_{\text{N}_\text{2}}= (n_{\text{N}_\text{2}}-\rho_{\text{N}_\text{2}}V_{\text{N}_\text{2}})/A
\end{equation}
where $n_{\text{N}_\text{2}}$ is the total number of nitrogen molecules 
within the simulation box,
$\rho_{N2}$ is the bulk  density of nitrogen, $V_{\text{N}_\text{2}}$ the volume of the vapor phase,
and $A$ the area of the system in a plane parallel to the interface.
Relative to the ice/vapor interface, the volume of the vapor phase is
determined for an equimolar dividing surface as
$V_{\text{N}_\text{2}}=A(L_z-H)$,
where H is the thickness of the ice slab, $H=N_w/(A\rho_{ice})$ and $N_w$ is the
total number of water molecules in the condensed phase.
The change of the surface tension, $\gamma$, as a consequence of a change of the pressure induced
by nitrogen is accounted for by the Gibbs equation:\cite{rowlinson82b}
\begin{equation}
d\gamma=-\Gamma_{\text{N}_\text{2}} RT \frac{dp}{p}
\label{eq:gibbs}
\end{equation}
where $R$ is the gas constant, $T$ is temperature and $p$ is the partial
pressure of nitrogen. At the low pressures considered in this study,
the nitrogen surface excess increases linearly with pressure 
$\Gamma_{\text{N}_\text{2}}=a p$,  and  the slope $a$ can be adjusted
to the simulation data. Substitution of this expression into Eq.~\ref{eq:gamma},
yields:
\begin{equation}
\Delta\gamma = - \Gamma_{\text{N}_\text{2}} RT = -a RT p   
\label{eq:gamma}
\end{equation}

\subsection{Intrinsic surfaces and order parameter}

In the simulations, the ice slabs in vacuum develop a thin layer of disordered
quasi-liquid like molecules. The properties of this layer are best characterized
by defining ice-liquid and liquid-vapor surfaces that sandwich the premelting
layer.\cite{benet16,benet19} 
The thickness of the liquid layer formed on the surface of
ice can be determined from the average distance between the surfaces.

\subsubsection{Order parameter}

In order to locate the surfaces, we need first of all to establish 
a criterion to distinguish between solid and liquid water molecules. 

\begin{figure}[H]
\centering
\resizebox*{5cm}{!}{\includegraphics{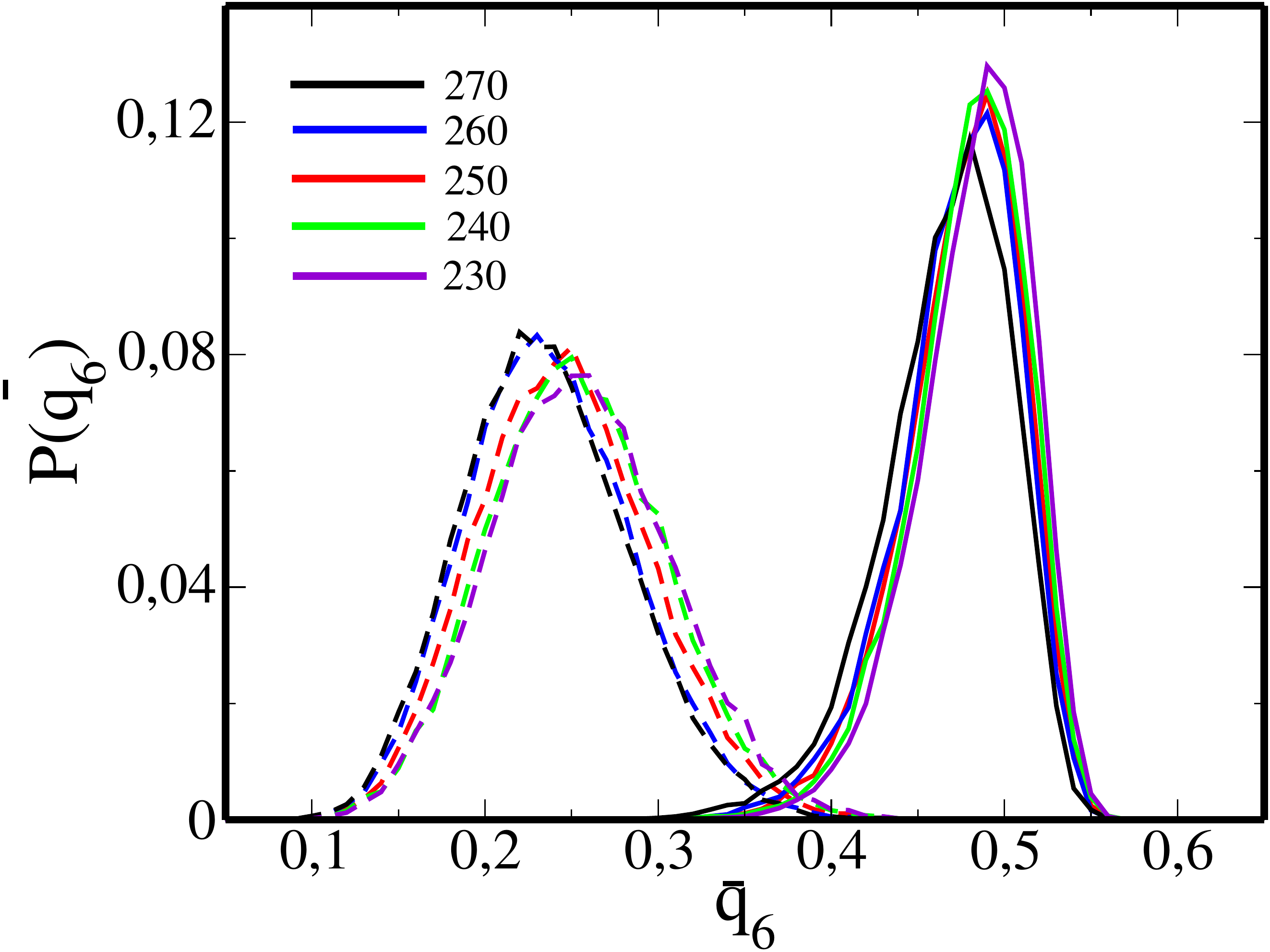}}
\resizebox*{5cm}{!}{\includegraphics{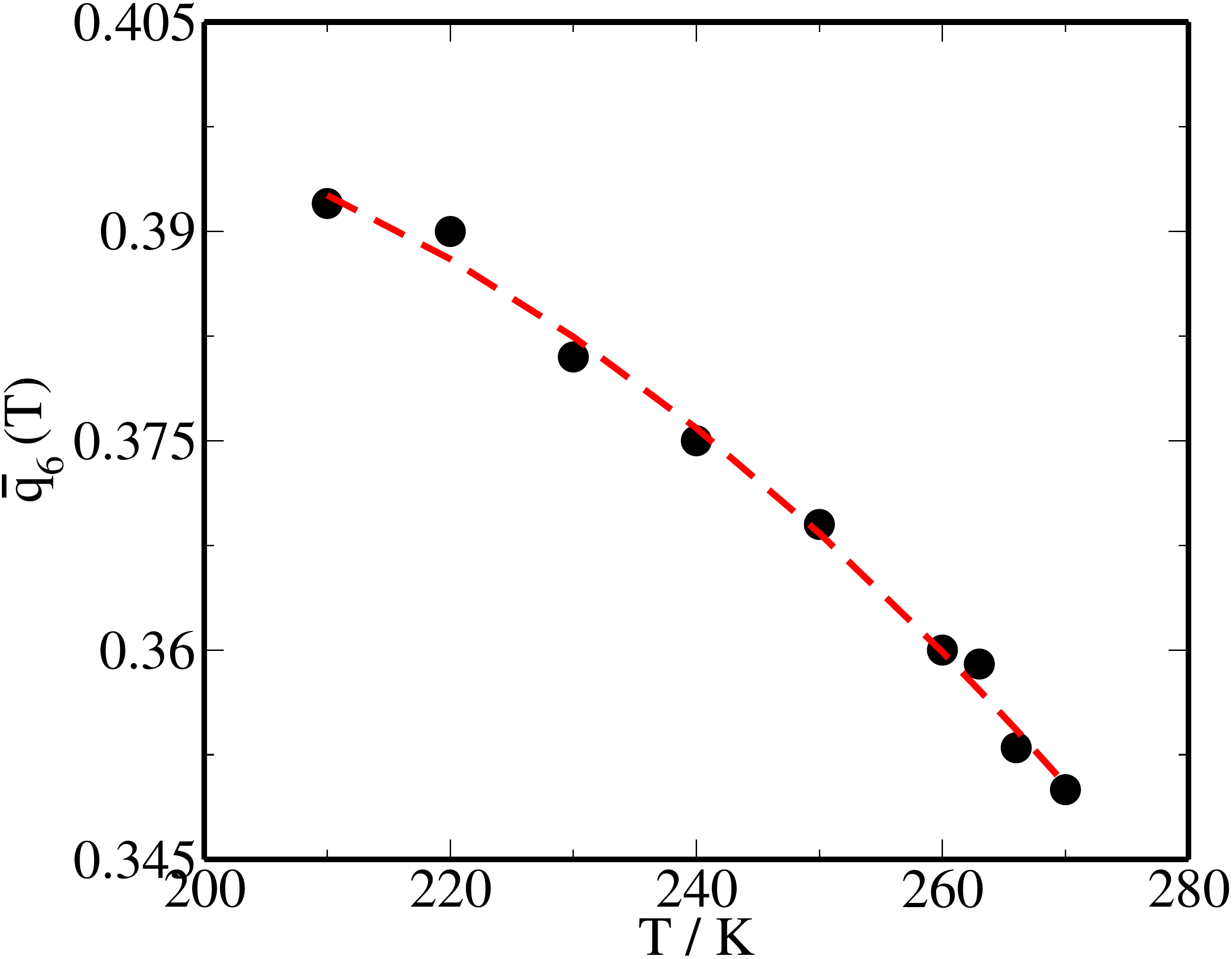}}
\resizebox*{4cm}{!}{\includegraphics{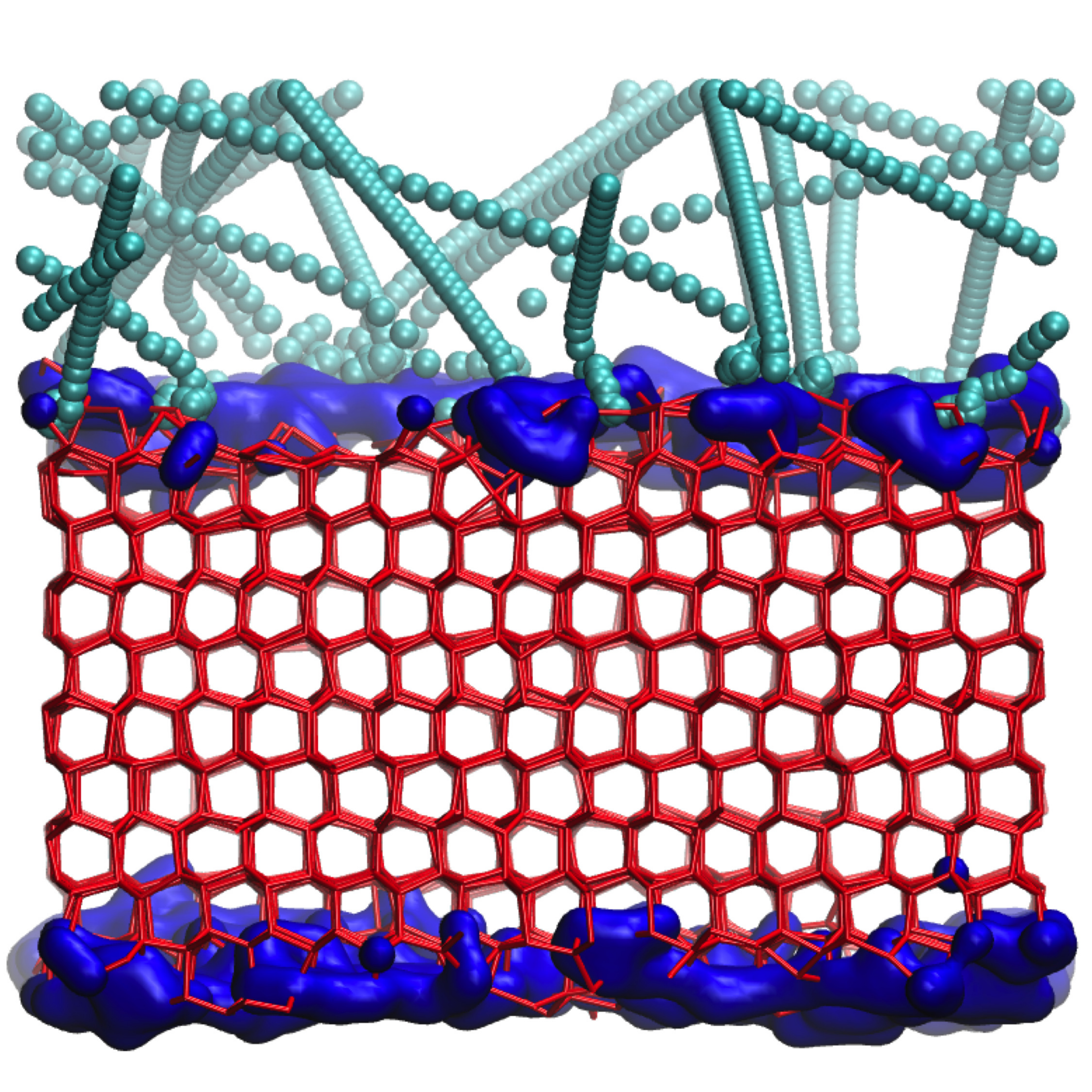}}
\caption{
Left: Probability distribution of the $\bar q_{6}$ on the solid (full lines) and liquid
(dashed lines) phases for temperatures between 230~K and 270~K as indicated in
the color code.  
 Middle: Limiting values of $\bar q_{6}$ used to discriminate liquid from
          solid particles as a function of temperature. These values were obtained 
          using the mislabeling criterion described by Espinosa \emph{et al.}~\cite{espinosa16}.
	  These limit values can be accurately fitted to a quadratic polynomial function:
          $\bar q_{6}(T) = 0.26266 + 0.0016474 T - 4.8982\cdot 10^{-6} T^{2}$.
          Right: Atomistic view of the ice-vapor interface at T=230~K and of the
          trajectory of 20 water molecules that are shot to the ice surface. 
          Water molecules identified with $\bar q_{6}$ as solid are shown as red sticks,
          and liquid molecules as dark blue spheres. The trajectories followed by the  molecules
          are depicted as cyan spheres. 
}
 \label{fig:q6_temp_sample}
\end{figure}

For that purpose 
we used the local averaged order parameter, $\bar q_{6}$, proposed
by Lechner and Dellago~\cite{lechner08}, which averages the Steinhardt's local
parameter, $q_6$,\cite{steinhardt83} over first neighbor shells. 
This order parameter adopts different values 
depending on the local environment of the water molecule. Solid molecules adopt large
values of $\bar q_{6}$, whereas molecules in liquid environments exhibit low
values of $\bar q_{6}$. 
Figure \ref{fig:q6_temp_sample} shows distributions of the
probability $P(\bar q_{6})$ to observe a given value of $\bar q_{6}$ in either
the solid or liquid phase for the temperatures studied in this work.  As
found in our previous work,\cite{benet14,benet14c} the distribution is close
to Gaussian for both the solid and liquid phases, with peaks that
are well separated and show only a small 
region of overlap. 
 A  threshold value of $\bar q_{6}$ used
to discriminate between solid and liquid particles is obtained by plotting the distribution
of $\bar q_{6}$ for molecules in bulk solid and bulk liquid environments, and finding
the limiting value of $\bar q_{6}$ that leads to minimum mislabeling, as 
described in Ref.~\onlinecite{benet14,benet14c,espinosa16}. The threshold
values obtained at each temperature are provided in Fig.~\ref{fig:q6_temp_sample}.
A sample configuration in which $\bar q_{6}$ was used to label the particles as solid
or liquid is also shown in Fig.~\ref{fig:q6_temp_sample}.

Notice that the identification of individual molecules as 'solid' or 'liquid'
like is not without a degree of arbitrariness, since bulk phases can only be
properly defined from ensemble averages of appropriate order parameters over
long times. This
difficulty is particularly significant for water, which can exhibit
a large number of allotropes, and could thus have a number of different
order parameters consistent with the solid phase. Particularly at the
solid/liquid interface, a number of different local environments that
resemble bulk ice phases have been identified  based on the number of staggered
and eclipsed bonds, such as hexagonal ice,
cubic ice, or ice clathrate forms, as well as purely interfacial varieties such
as interfacial hexagonal ice.\cite{nguyen15}  The total number of
liquid-like molecules will therefore depend on the decision as to
how to attribute allotropic forms different from ice Ih into 'solid-like'
or 'liquid-like' environments. By using the $\bar q_{6}$ parameter,
we are able to identify local environments consistent with hexagonal ice,
but we cannot distinguish disordered liquid-like environments from other
more ordered environments such as cubic ice, ice clathrates or interfacial
hexagonal ice. It must therefore be understood that our choice of
ice-liquid surface separates essentially bulk ice Ih from a loosely
defined premelting layer which includes liquid-like environments as well
as relatively ordered allotropic forms. Alternatively, using
the CHILL+ algorithm, one can distinguish all such  forms.\cite{nguyen15} In the conventional 
application of the
CHILL+ algorithm, ice clathrate and ice Ic, as well as interfacial
ice Ih environments, are attributed
to the solid phase, and the quasi-liquid layer is made only of the disordered
molecules. Therefore, 
the premelting layer thickness somewhat differs, in a similar
way as it does by using different experimental
probes.\cite{elbaum93,lied94,dosch95,bluhm02,mitsui19,dash06,li07b,michaelides17}
Coincidentally,
we have checked that the number of molecules in ice clathrate, ice Ic and interfacial ice Ih
environments remains almost constant within the temperature interval
studied in our work, so that the premelting layer thickness defined
in either way only differs by an almost constant offset (supplemental material).
For consistency with our previous 
work,\cite{benet14,benet14c,benet16,benet19} we will therefore use
the $\bar q_{6}$ parameter as a criterion for discriminating 
liquid-like from solid-like environments.

\subsubsection{Intrinsic surface}

Once that water molecules
have been identified as solid or liquid, the largest 
cluster is searched using
a cluster analysis algorithm, in which two water molecules are considered to be 
nearest neighbors if the distance between their oxygen atoms is smaller than
3.5~{\AA}. 

The outermost layer of oxygen atoms on this cluster corresponds to water-like
molecules. To locate the liquid-vapor surface at a point $(x,y)$ on the 
plane of the interface, $h_{wv}(x,y)$ we search for all liquid like atoms within a rectangular
area of $3\sigma\times 3\sigma$, and define the surface height as the
average value of the four outermost atoms. Having found the liquid-vapor
surface at that point, we then locate the solid-like atoms within a 
rectangular unit-cell plane about $(x,y)$. The four outermost solid-like
atoms within that area determine the solid-liquid surface location,
$h_{iw}(x,y)$.

A local liquid-layer thickness at point $(x,y)$ can be calculated as
\begin{equation}
\delta h(x,y) = | h_{iw}(x,y) - h_{wv}(x,y)|
\end{equation}
The average thickness is determined by averaging the
local heights over a mesh with twice as many points as unit cells
on the surface.

	Further information about the structure of the ice-liquid and liquid-vapor
interfaces can be obtained from the density profiles measured along the direction perpendicular
to the interface (i.e. along $z$):
\begin{equation}
   \rho_{\alpha} (z) = \frac{N_{\alpha} (z)}{L_x L_y \Delta z}
\end{equation}
where $\alpha$ denotes either solid-like or liquid-like molecules, and $N_{\alpha}(z)$ is the number of 
molecules of that type in a slice of simulation box centered at $z$ and of width $\Delta z$.

\begin{figure}[H]
\centering
\resizebox*{9cm}{!}{\includegraphics{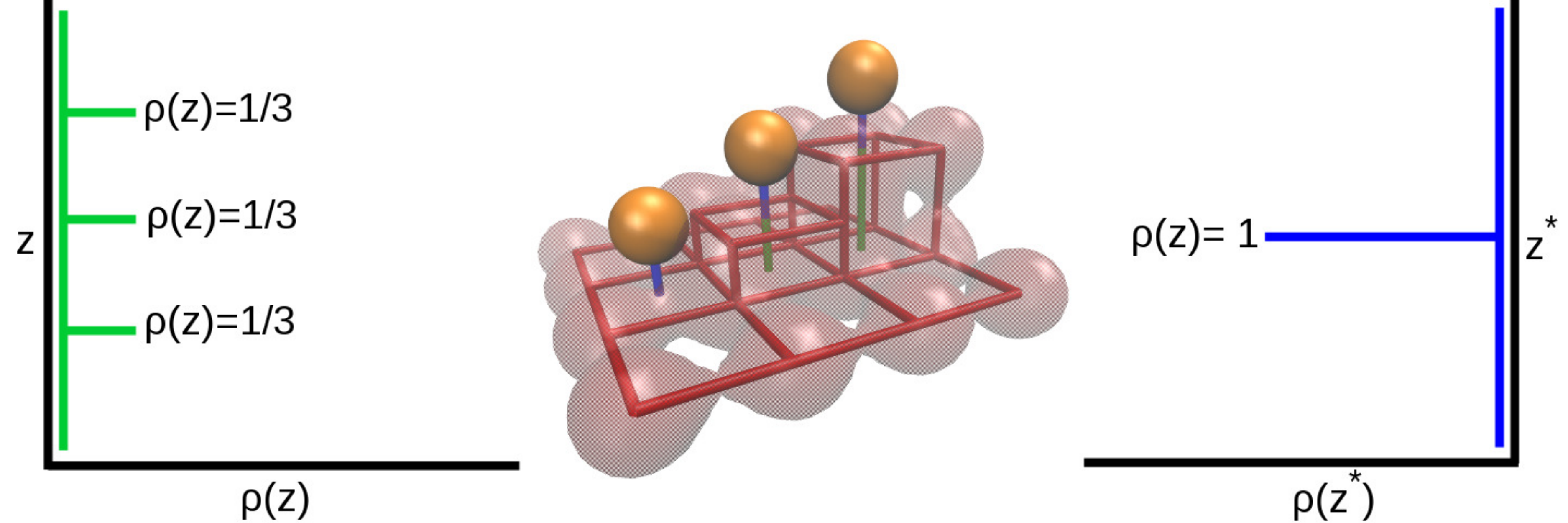}}
\caption{Schematic illustration of the difference in the density profiles as a function
         of absolute $z$ coordinates (measured with respect to the simulation box reference 
         frame) or as a function of the $z^*$ coordinate measured with respect to the
         local height of the ice-liquid interface. Pink faded spheres represent solid 
         molecules, and orange spheres liquid molecules. The red grid shows the local
         ice-liquid interface. The three orange molecules
         are assigned to three different bins when measured in absolute values
         of $z$ (left scale), whereas to only one single bin when measured with 
         respect to the local solid height (right scale). 
        }
 \label{fig:explica_intrinseco}
\end{figure}

Besides analyzing the density profile as a function of the absolute $z$-coordinate measured
with respect to a fixed reference system, we also evaluated the density 
profiles as a function of the local distance to either the ice-liquid or
liquid-vapor surfaces:
\begin{equation}
z_{i,\alpha}^{\ast} = z_i - h_{\alpha} (x,y)
\end{equation}
where $\alpha={iw,wv}$ stands here for either ice-water or water-vapor surfaces.
Note that adopting a local definition of the distance
to the interface, the slice over which the number of water molecules is counted is no
longer planar, as in the usual calculation of the density profile. This is
illustrated for positions measured relative to 
the ice-water surface in Figure~\ref{fig:explica_intrinseco}.

\subsection{Estimation of water attachment rates}

\label{sec:attach}

Finally, we performed a last set of simulations aimed at measuring the attachment rate of water vapor 
to the ice surface. For this purpose, we performed the following experiment. A total of 400 water molecules
were shot against 20 different thermal configurations of ice. Each water molecule shot was
initially placed at a distance of 2~nm from the ice surface and assigned a velocity randomly chosen
from the corresponding Maxwell-Boltzmann distribution, with 
the initial $z$ component directed to the interface. The
simulations were carried out in the microcanonical ensemble to avoid kinetic energy dissipation before
and after the collision. We used a time step of 0.6~fs in these
simulations to ensure energy conservation.  Using the initial velocity we
estimate the time required for the water molecule to reach the surface in a
ballistic trajectory, and allow the simulation to proceed for twice as much
time. At the end of the simulation we perform a cluster analysis over
all molecules and determine
whether the colliding molecule is attached to the condensed phase or not.
From the ensemble of trajectories, we can estimate
the attachment rate ($\alpha$), which is defined as the ratio between the water
molecules that are incorporated into the ice surface 
after the impact ($N_{ads}$) and the total number of water molecules that 
are shot against the ice surface ($N_{total}$):
$\alpha=N_{ads}/N_{total}$. Notice that in this way we are measuring a direct attachment rate,
not including the net rate due to evaporation events.
These measurements were carried out at three different temperatures,
$T=$ 230, 260 and 270~K and at two nitrogen partial pressures, namely,  
zero and a 1~bar.

\section{Results}

In this section we present results for the structure and attachment kinetics of
ice in the presence of nitrogen. Our study covers a temperature range between
230~K to 270~K, which spans most of the relevant temperature of ice crystal
growth in the atmosphere as described in the Nakaya diagram. For the pressure,
we have chosen conditions pertaining to the International Standard Atmosphere
at those temperatures (c.f. \ref{alt_temp_press}). This corresponds to pressures
well above typical values in cirrus clouds, and therefore provide
an upper limit for the influence of nitrogen on the ice surface. The
detailed conditions studied in our simulations as well as the collected results
may be found in Table \ref{tab:results1}.

\begin{table*}[ht]
\footnotesize
\begin{tabularx}{0.9\textwidth}{l| R R R R R |R R R R R}
\multicolumn{1}{c}{} & \multicolumn{5}{c}{Basal}  &  \multicolumn{5}{c}{Prismatic I}   \\
\hline
\hline
  T  & p             & $\delta h$ &  $\Gamma$        & $K$                &
$\Delta\gamma$ & p             & $\delta h$ &  $\Gamma$        &  $a$               & $\Delta\gamma$ \\
 $K$ & $10^{-5} Pa$  & \AA        &  $nm^{-2}$       & $10^{-11}m/J$    & $mN/m$         & $10^{-5}Pa$   & \AA        &  $nm^{-2}$       & $10^{-11}m/J$    & $mN/m$ \\
\hline
 \textcolor{blue}{ 230 }   &      \textcolor{blue}{0.34}   &    4.1 & 0.008  & 2.52 & -0.03  &  \textcolor{blue}{0.37}  & 3.8 & 0.008 & 2.39  & -0.03   \\
 \textcolor{blue}{ 230 }   &      \textcolor{cyan}{0.67}   &    4.4 & 0.016  & 2.52 & -0.05  &  \textcolor{cyan}{0.72}  & 3.8 & 0.017 & 2.39  & -0.06  \\
 \textcolor{blue}{ 230 }   &      \textcolor{green}{1.01}  &    4.3 & 0.026  & 2.52 & -0.08  &  \textcolor{green}{1.10} & 4.0 & 0.027 & 2.39  & -0.08 \\
 \textcolor{cyan}{ 240 }   &      \textcolor{blue}{0.44}   &    4.8 & 0.010  & 2.18 & -0.03  &  \textcolor{blue}{0.49}  & 4.1 & 0.010 & 2.00  & -0.03  \\
 \textcolor{cyan}{ 240 }   &      \textcolor{cyan}{0.78}   &    5.2 & 0.016  & 2.18 & -0.06  &  \textcolor{cyan}{0.87}  & 4.1 & 0.017 & 2.00  & -0.06  \\
 \textcolor{cyan}{ 240 }   &      \textcolor{green}{0.94}  &    5.1 & 0.019  & 2.18 & -0.07  &  \textcolor{green}{1.05} & 4.1 & 0.021 & 2.00  & -0.07  \\
 \textcolor{cyan}{ 240 }   &      \textcolor{orange}{1.00} &    5.1 & 0.023  & 2.18 & -0.07  &  \textcolor{orange}{1.12}& 3.9 & 0.023 & 2.00  & -0.07  \\
 \textcolor{green}{ 250 }  &      \textcolor{blue}{0.54}   &    5.6 & 0.011  & 1.92 & -0.04  &  \textcolor{blue}{0.56}  & 4.9 & 0.012 & 1.89  & -0.04  \\
 \textcolor{green}{ 250 }  &      \textcolor{cyan}{0.66}   &    5.6 & 0.012  & 1.92 & -0.04  &  \textcolor{cyan}{0.71}  & 5.0 & 0.013 & 1.89  & -0.05  \\
 \textcolor{green}{ 250 }  &      \textcolor{green}{1.03}  &    5.6 & 0.018  & 1.92 & -0.07  &  \textcolor{green}{1.05} & 5.0 & 0.020 & 1.89  & -0.07  \\
 \textcolor{green}{ 250 }  &      \textcolor{orange}{1.10} &    5.5 & 0.021  & 1.92 & -0.07  &  \textcolor{orange}{1.13}& 5.0 & 0.021 & 1.89  & -0.07  \\
 \textcolor{green}{ 250 }  &      \textcolor{red}{1.16}    &    5.5 & 0.024  & 1.92 & -0.08  &  \textcolor{red}{1.17}   & 5.0 & 0.022 & 1.89  & -0.08  \\
 \textcolor{orange}{ 260 } &      \textcolor{blue}{0.66}   &    6.3 & 0.011  & 1.60 & -0.04  &  \textcolor{blue}{0.71}  & 5.8 & 0.011 & 1.52  & -0.04  \\
 \textcolor{orange}{ 260 } &      \textcolor{cyan}{0.82}   &    6.1 & 0.015  & 1.60 & -0.05  &  \textcolor{cyan}{0.88}  & 5.6 & 0.013 & 1.52  & -0.05  \\ 
 \textcolor{orange}{ 260 } &      \textcolor{green}{0.98}  &    6.2 & 0.016  & 1.60 & -0.06  &  \textcolor{green}{1.06} & 5.6 & 0.016 & 1.52  & -0.06  \\
 \textcolor{orange}{ 260 } &      \textcolor{orange}{1.05} &    6.2 & 0.017  & 1.60 & -0.06  &  \textcolor{orange}{1.13}& 5.6 & 0.017 & 1.52  & -0.06  \\
 \textcolor{orange}{ 260 } &      \textcolor{red}{1.10}    &    6.2 & 0.016  & 1.60 & -0.06  &  \textcolor{red}{1.20}   & 5.8 & 0.018 & 1.52  & -0.07  \\
 \textcolor{red}{ 270 }    &      \textcolor{blue}{0.76}   &    8.9 & 0.012  & 1.44 & -0.04  &  \textcolor{blue}{0.81}  & 8.3 & 0.012 & 1.45  & -0.04  \\
 \textcolor{red}{ 270 }    &      \textcolor{cyan}{0.88}   &    9.1 & 0.013  & 1.44 & -0.05  &  \textcolor{cyan}{0.94}  & 8.1 & 0.014 & 1.45  & -0.05  \\
 \textcolor{red}{ 270 }    &      \textcolor{green}{1.00}  &    9.3 & 0.015  & 1.44 & -0.05  &  \textcolor{green}{1.07} & 8.4 & 0.015 & 1.45  & -0.06  \\
 \textcolor{red}{ 270 }    &      \textcolor{orange}{1.05} &    9.2 & 0.013  & 1.44 & -0.06  &  \textcolor{orange}{1.12}& 8.5 & 0.016 & 1.45  & -0.06  \\
 \textcolor{red}{ 270 }    &      \textcolor{red}{1.10}    &    9.1 & 0.015  & 1.44 & -0.06  &  \textcolor{red}{1.17}   & 8.2 & 0.018 & 1.45  & -0.06  \\ 
\hline
\hline
\end{tabularx}
\caption{
   Summary of systems studied and
collected results for premelting thickness, $\delta h$, nitrogen adsorption,
$\Gamma$, Henry adsorption constant, $a$ and change in surface tension, 
$\Delta \gamma_{iv}$,
as a function of temperature and pressure for basal (left) and
primary prismatic  (right) facets. The color code employed in
the table labels defines the color code for the forthcoming figures.
  \label{tab:results1}
  }
\end{table*}

\begin{figure*}[h]
\centering
\subfigure[ basal]{\label{fig:rhosolid_basal}
\resizebox*{4cm}{!}{\includegraphics{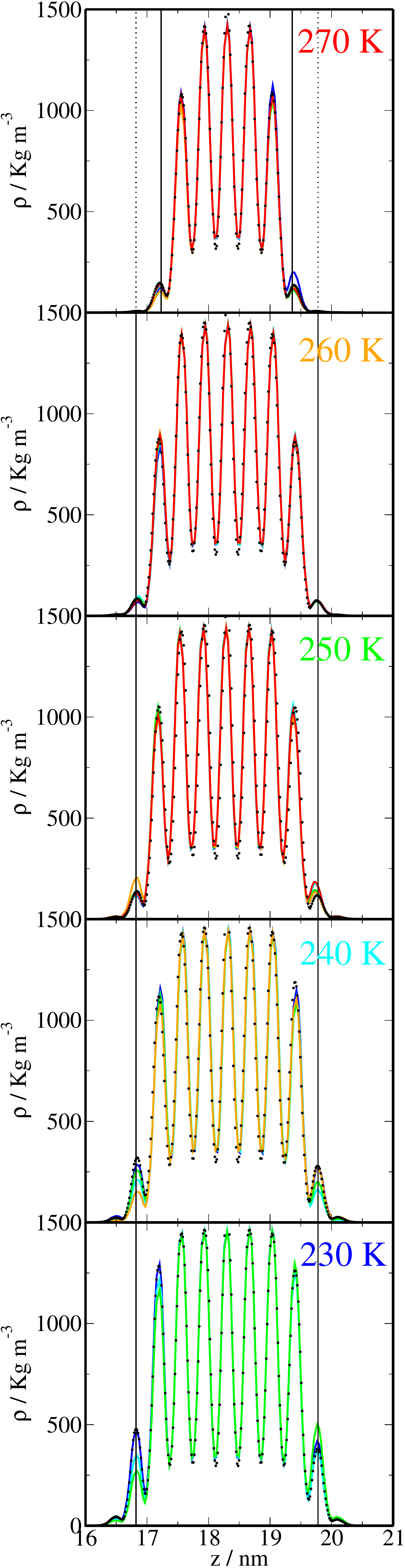}}}   
\subfigure[ basal]{\label{fig:rholiquid_basal}
\resizebox*{4cm}{!}{\includegraphics{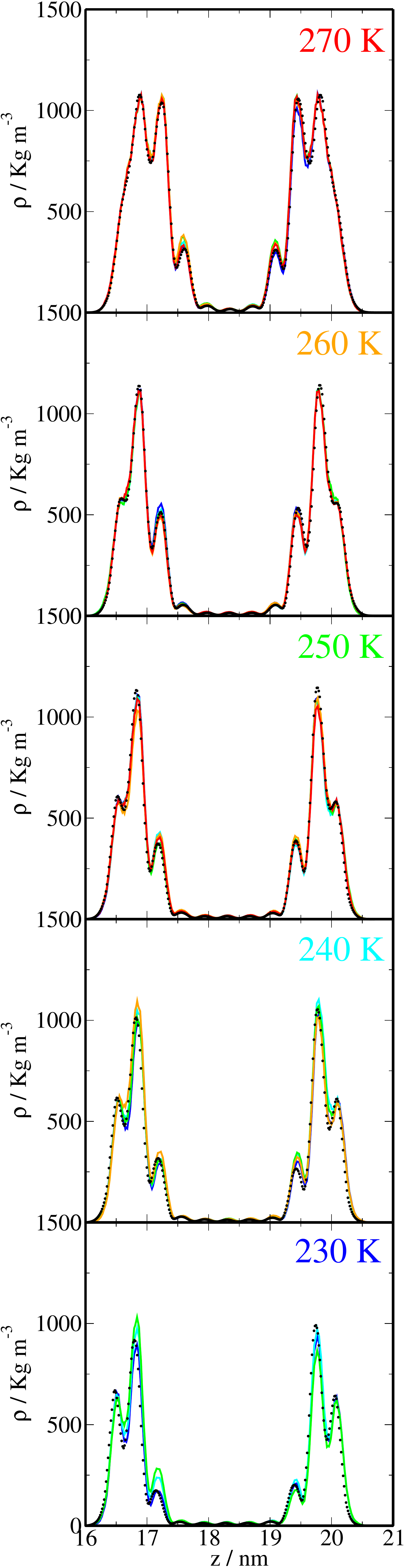}}}  
\subfigure[ pI]{\label{fig:rhosolid_pi}
\resizebox*{4cm}{!}{\includegraphics{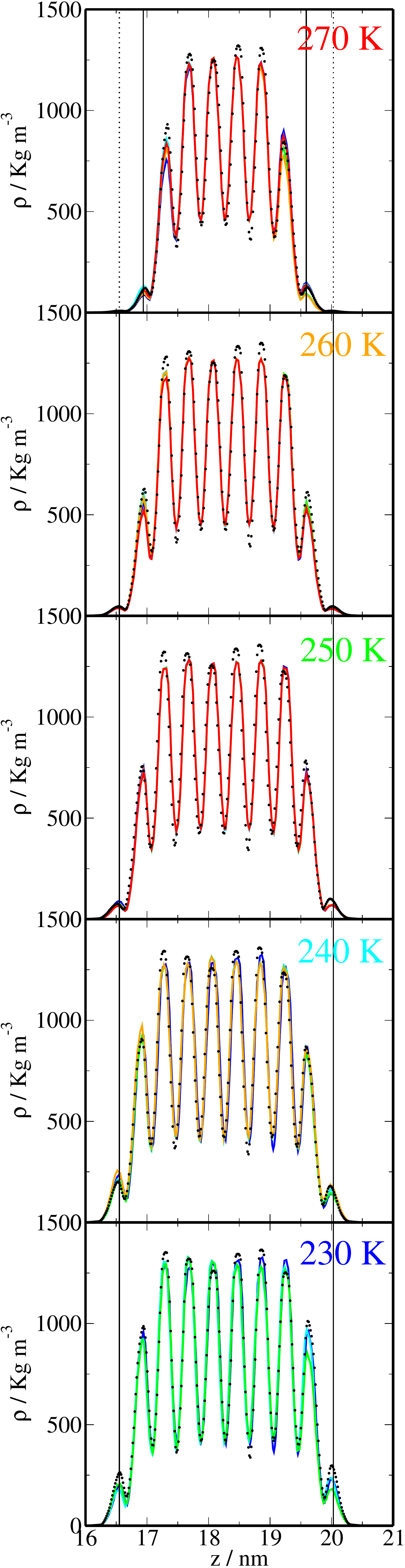}}}  
\subfigure[ pI]{\label{fig:rholiquid_pi}
\resizebox*{4cm}{!}{\includegraphics{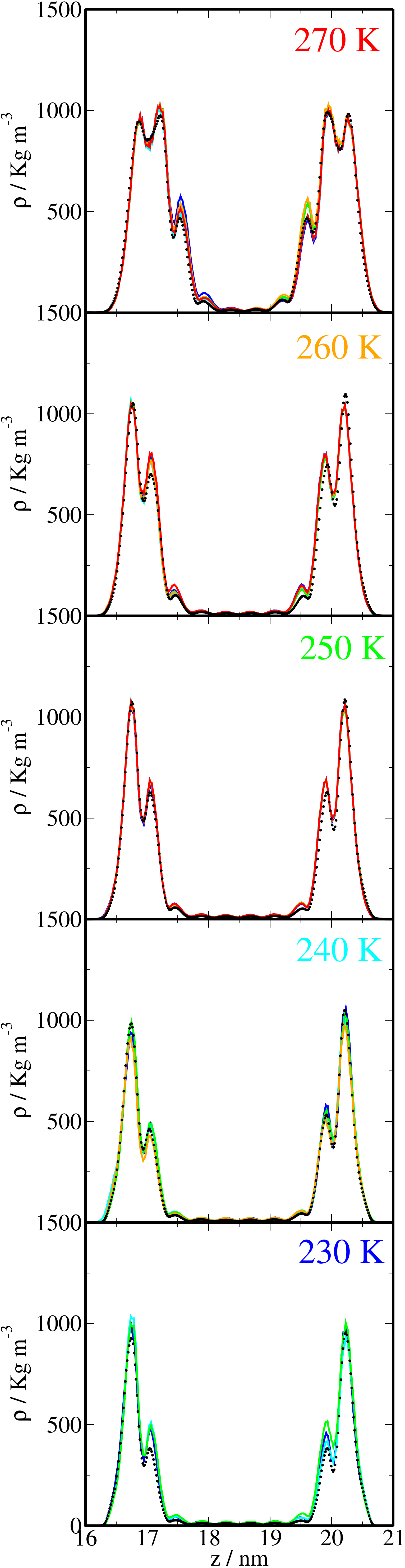}}}  
\caption{
Density profiles of water as a function of  temperature and nitrogen
pressure for basal (left) and primary prismatic (right) planes. 
Columns (a,c) correspond to density profiles of solid like  molecules,
while columns (b,d) correspond to density profiles of liquid like molecules.
Black circles correspond to 
density profiles in pure water-vapor (zero nitrogen pressure), while
colored lines stand for density profiles with  cold to warm colors in 
order of increasing nitrogen pressure, with specific values of pressure
as described in table \ref{tab:results1}. For the density profile of solid
like molecules, the full vertical line indicates the outermost interfacial
bilayer. Notice how the location of this bilayer moves one lattice spacing
towards the bulk phase at T=270~K.
}
 \label{fig:perfil_agua_solido_liquido}
\end{figure*}

\subsection{Structure of the pristine ice/vapor interface}

\label{sec:pristine}

A priori, the study of the ice/vapor interface is complicated by the presence of
a thin layer of disordered ice, which grows as the temperature rises along the
ice/vapor sublimation line.\cite{dash06,michaelides17} Depending on the
community, this layer is known as a premelting film or quasi-liquid layer. Here
we will stick to the standard terminology of wetting and adsorption physics,
and denote the growing adsorbed layer as a premelting film.

Because of the complex structure of this interface, it is convenient to
label surface molecules in terms of a suitable order parameter that
can distinguish between water-like or disordered molecules and solid-like
or ordered molecules.\cite{lechner08} This allows us to give a detailed account of the
complex ice/vapor interface with reference to two simpler ice/water and
water/vapor surfaces, and to plot separate solid-like and liquid-like
density profiles of water molecules.\cite{benet16,benet19}


Figure \ref{fig:perfil_agua_solido_liquido} shows the density of
solid-like and liquid-like molecules as a function of the perpendicular
distance to the interface for both basal and pI facets in a temperature
interval between 230 to 270~K. At low temperature (230~K), the solid-like density
profile (\ref{fig:rhosolid_basal} and \ref{fig:rhosolid_pi}) 
reveals an ordered crystalline solid with high density
peaks that correspond to complete bilayers
within the bulk of the crystal (five for the basal facet and six for the
pI facet in our simulation setup). Each such bilayer contains the molecules
of full stacked hexagonal rings of ordered
ice.\cite{sazaki12,asakawa15,asakawa16,murata16,michaelides17}
As we move away from the bulk solid
towards the vapor phase, we find a decay of the solid-like density profile.
The interface consists of  two partially formed
bilayers of intermediate and small density. The outermost bilayer is
indicated in the plot with a black vertical line. 

As temperature increases
towards the melting point, the height of the interfacial bilayers
decreases smoothly up to 260~K. However, in the range between 260~K and
270~K, one full bulk like solid bilayer per interface melts into the
premelting film. A similar bilayer melting has been observed previously
for the TIP4P/ice model that is consistent with Sum Frequency Generation
experiments, albeit  at a somewhat lower temperature of about
250~K.\cite{sanchez17,michaelides17} The difference with  previous simulations 
lies in the improved density representation, which reveals
the bilayer melting process more clearly. 

Interestingly, the structure
of the solid like density profile at the interface remains qualitatively
similar along the full temperature range studied, even after the
first full bilayer melting (i.e., remains composed of two partial bilayers
with intermediate and small densities). This process of melting can
be observed also with the complementary density profiles of liquid-like
molecules depicted in \ref{fig:rholiquid_basal} and \ref{fig:rholiquid_pi}.
In the range between 230~K and
260~K the profiles reveal an adsorbed film of water molecules which
is made from three partial bilayers for the basal face and only
two partial bilayers for the pI plane. As temperature rises in this
range, the density of the  profile increases slightly, but in
the range between 260 to 270~K, two of the partial bilayers
attain bulk like densities, as found recently for the related
TIP4P/2005 model.\cite{benet16}
The  premelting film is never thick enough
for the density of liquid molecules to attain a plateau value,
but the maxima of the density peaks clearly attains order of
magnitude values expected for bulk liquid water

   The density profiles shown here describe the laterally averaged structure
   of the ice/vapor interface. A hint on the surface structure in the direction
   parallel to the interface is given by the partially filled  density
   peaks found in the outermost layer. Their location remains congruent with the
   lattice spacing, but the height is significantly smaller than that observed
   in bulk. However, a difference is observed between basal and pI planes. In
   the former, the small peak is separated from a neighboring peak by a region
   of depleted density. In the latter, on the contrary, the peak rather appears
   as a shoulder and hardly exhibits a density minimum separating it from
   the next peak. This suggests that the outermost region of the ice surface,
   at least for the basal plane, exhibits
   interrupted regions of fully formed layers, leading to a stepped
   surface.\cite{benet19}

   \newpage
   \clearpage

\begin{figure}[H]
\centering
\resizebox*{9cm}{!}{\includegraphics{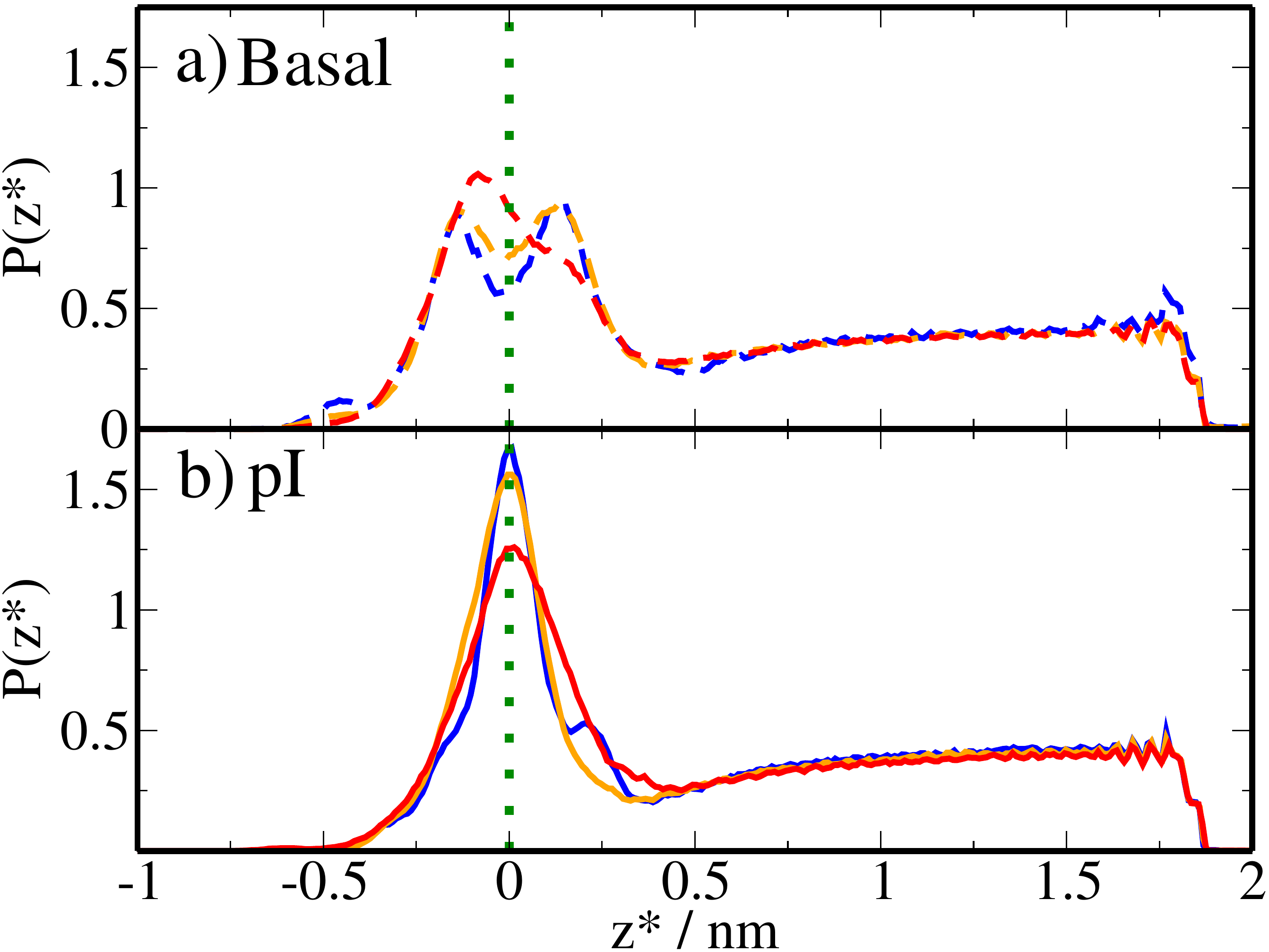}}
\resizebox*{9cm}{!}{\includegraphics{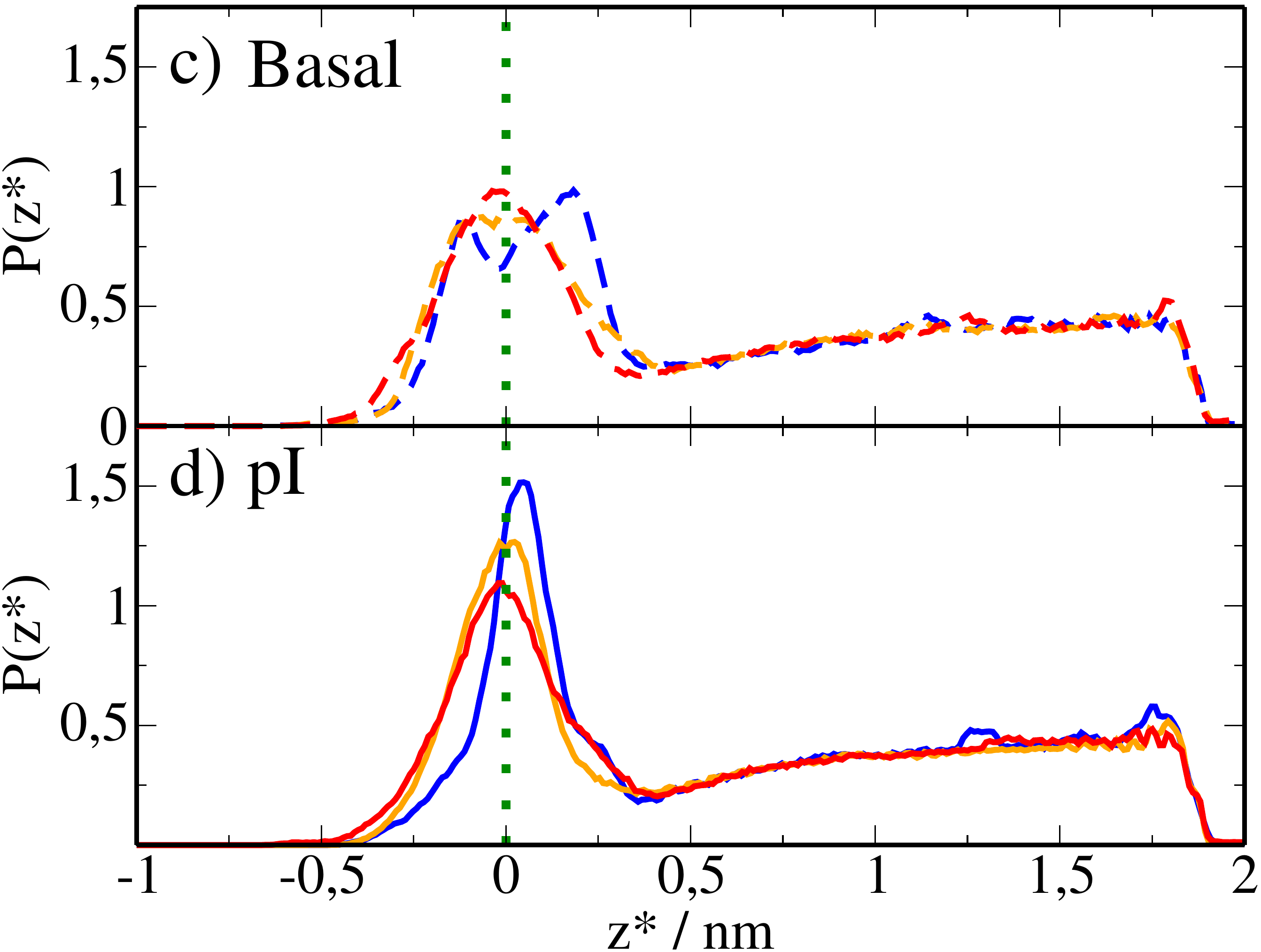}}
\caption{
         Probability $P(z^{\ast})$ of finding impinging water molecules at a distance $z$ for T=230~K (blue),
         260 (orange) and 270 (red) at pressures of 0~bar (a,b) and 1~bar (c,d).
}
 \label{fig:z_time_0bar}
\end{figure}

   In a forthcoming section, we study the attachment coefficient of water
   molecules on the ice surface. These are calculated by shooting water
   molecules against the ice substrate. Here, we exploit these data in order
   to obtain a scattering profile of the substrate. Merely by calculating the
   probability $P(z^{\ast})$ to observe an impinging water molecule at a distance $z$
   away from the average ice surface location, we find a revealing clue as to
   significant differences between basal and pI planes. Figure
   \ref{fig:z_time_0bar}.a shows plots of $P(z^{\ast})$ obtained from trajectories
   aimed at an ice substrate equilibrated under pure water vapor. The results
   show a region  of slowly decaying probability at large $z$ that corresponds to
   close to ballistic trajectories of the water molecules, and then exhibits
   a high probability region corresponding to the sticking of the water
   molecules onto the ice substrate. Interestingly, for the basal face we
   find at the lowest temperature T=230~K a bimodal distribution, indicating
   that the water vapor molecules shot against the substrate have a choice
   of two preferred locations. The distance between the two maxima is
   about 0.25~nm, which corresponds to the distance between adjacent planes
   in an ice bilayer (or alternatively, the perpendicular distance between
   the oxygen molecules of the stacked hexagonal rings). This bimodal structure 
   smoothes at the temperature T=260~K and then eventually disappears at T=270~K. For
   the pI plane, on the other hand, the distribution is unimodal for
   all temperatures studied, but clearly broadens as the temperature raises.

   These results are clearly supportive of a stepped surface for the
   instantaneous configurations of basal facets, with stacks of half filled
   planes of half a bilayer and provide further evidence of
   the horizontally inhomogeneous distribution on the basal ice
   surface, 
   as recently noted.\cite{hudait17,pickering18,qiu18}
   The structure of the pI facet on the contrary,
   does not appear to have this form of surface disorder, 
   at least on the outermost liquid-vapor surface.

\subsection{Structure of the adsorbed nitrogen layer}

\label{sec:nitrogen}

\begin{figure}[H]
\centering
\subfigure[ basal]{\label{rhonitro_basal}
\resizebox*{4cm}{!}{\includegraphics{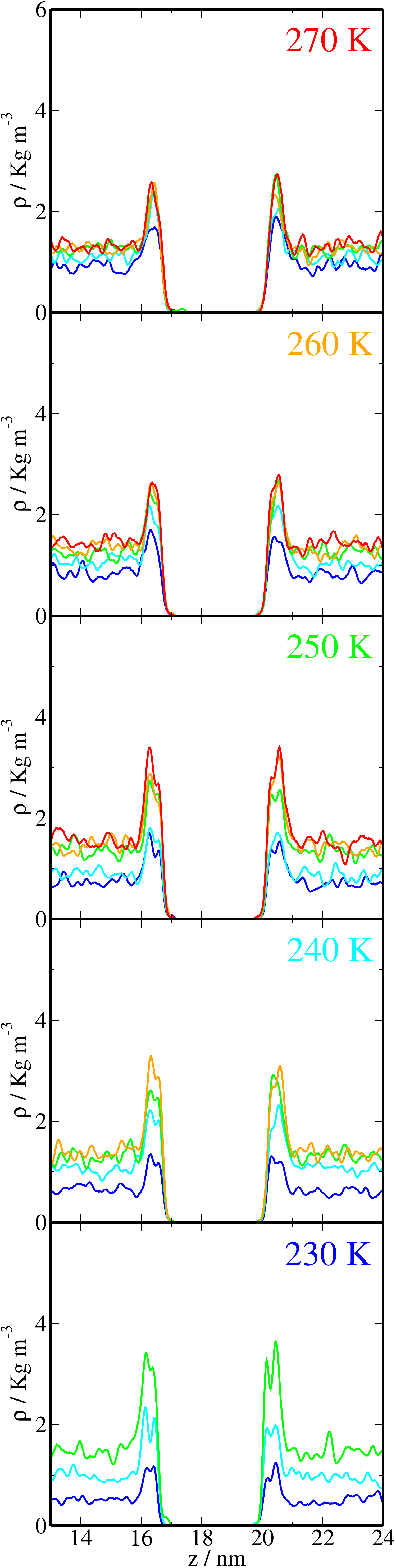}}}   
\subfigure[ pI]{\label{rhonitro_pi}
\resizebox*{4cm}{!}{\includegraphics{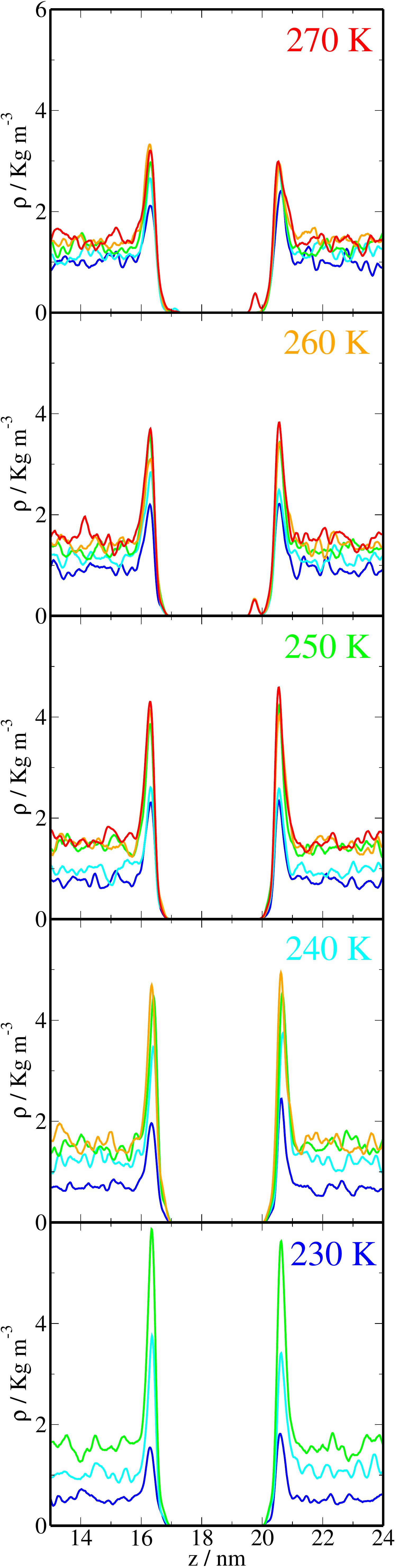}}}  
\caption{
   Density profiles of nitrogen as a function of perpendicular distance to the
   interface $z$. For each temperature, density profiles correspond to pressures
   as indicated in Table \ref{tab:results1}, with pressure increasing in the
   range from ca. 0.3 to ca. 1~bar from cold  to warm colors.
}
 \label{fig:perfil_nitro_abs}
\end{figure}

Having studied the structure of the pristine ice/vapor interface,
we now consider the structure of the adsorbed nitrogen layer on the
ice surface. Figure \ref{fig:perfil_nitro_abs} shows the density
profile of nitrogen molecules as a function of the perpendicular
distance to the interface for both basal and pI planes. Results
are given for temperature and nitrogen pressure made to match
conditions of the International Standard Atmosphere for 
altitude in the range between 2500 and 9000 m, which
correspond to temperatures in the range 230 to 270~K studied
above (c.f table \ref{alt_temp_press}).

In all cases we find a plateau region of constant nitrogen density
corresponding to the bulk gas phase, a small but significant
adsorption layer at the interface, and a completely depleted
central region corresponding to the highly nitrogen 
insoluble ice phase. Interestingly, the density profiles
depicted here suggest a significant difference in the structure
of adsorbed nitrogen at the basal and pI planes, with
local densities that are roughly between
twice (for the basal plane) and four times larger (for the pI
plane) than the bulk vapor value. Also, the adsorption
profile for the basal plane appears to be about twice as broad
than the corresponding profile for the pI plane.

Unfortunately, these 'absolute' density profiles do not quite
reveal the position of the adsorption peak relative to the
complex ice/vapor interface.  Particularly, it does not reveal
whether the nitrogen is adsorbed on the 
premelting layer, or rather, is solubilized within. To study
this issue further, we exploit the method of 'intrinsic' surfaces,
which allows us to calculate local ice/film,
$h_{iw}({\bf x})$ and  film/vapor, $h_{wv}({\bf x})$ 
surfaces for instantaneous configurations during the simulation 
(c.f. section \ref{sec:simmet}). From knowledge of the instantaneous
local surfaces, we calculate so called 'intrinsic' density
profiles, which are given as (c.f.
\ref{fig:explica_intrinseco}):\cite{jorge10,sega15}
\begin{equation}
\rho_{\alpha}(z^{\ast}) = \langle \rho(z_{i}-h_{\alpha}(x,y)) \rangle
\end{equation} 
where $\alpha$ is the surface label corresponding to either $wv$ or $iw$
surfaces.

\begin{figure}[H]
\centering
\subfigure[ basal]{\label{fig:perfil_nitro_basal_int_sol}
\resizebox*{4cm}{!}{\includegraphics{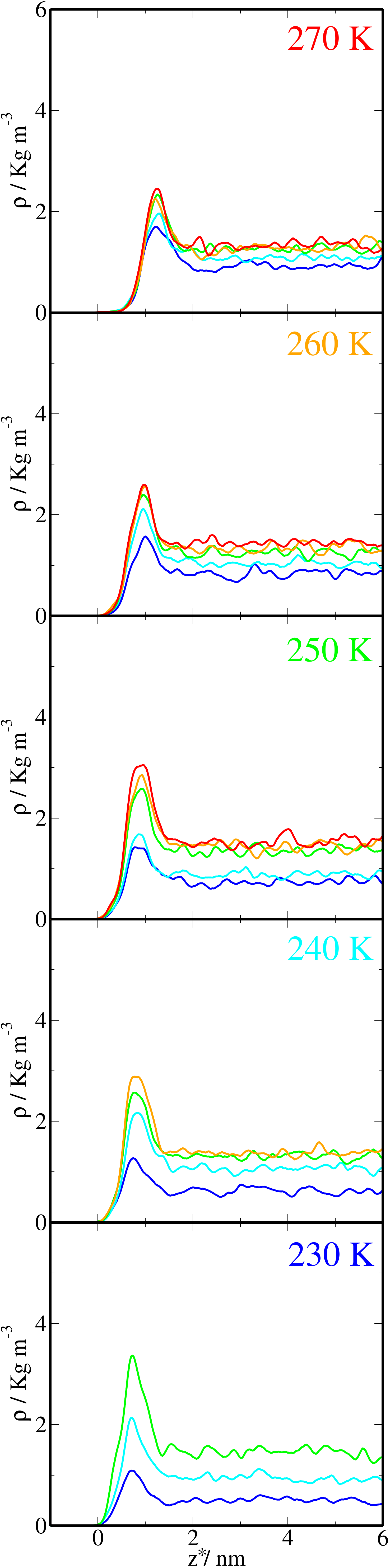}}}  
\subfigure[ pI]{\label{fig:perfil_nitro_pi_int_sol}
\resizebox*{4cm}{!}{\includegraphics{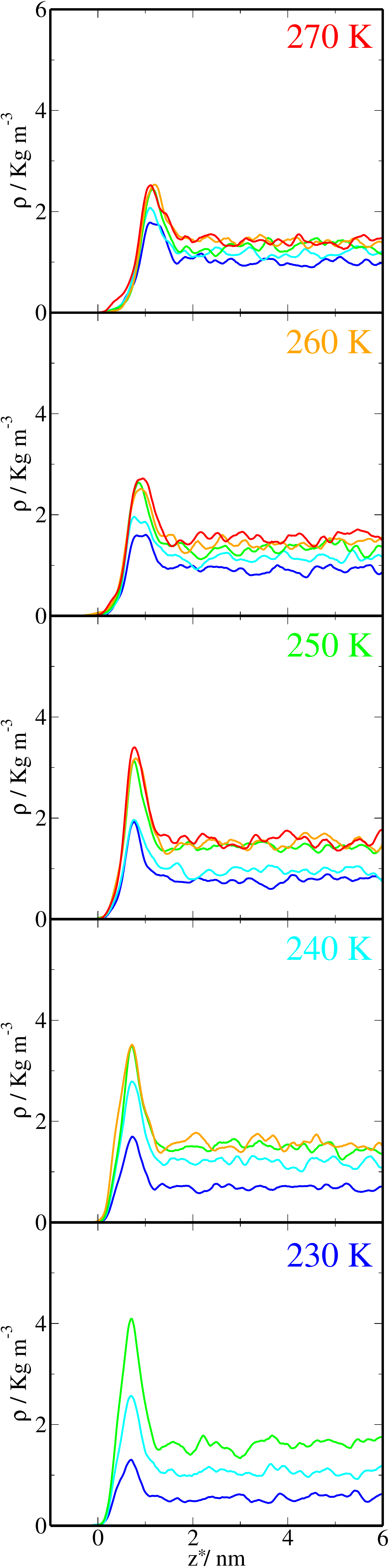}}}  
\caption{
         Density profiles of nitrogen as a function of perpendicular distance to the
         ice-liquid surface $z-h_{iw}(x,y)$. For each temperature, density profiles correspond to pressures
         as indicated in Table \ref{tab:results1}, with pressure increasing in the
         range from ca. 0.3 to ca. 1~bar from cold colors to warm colors.
}
 \label{fig:perfil_nitro_int_sol}
\end{figure}

Figure \ref{fig:perfil_nitro_int_sol} shows the intrinsic density
profile of nitrogen molecules as  measured relative to the solid/liquid
surface. In all cases, for both the basal and pI planes,
the nitrogen adsorption peak is located
at positive values of $z$, and reveals essentially zero nitrogen
density within the bulk solid phase. Furthermore, the relative position
of the adsorption peak moves away from zero as the temperature is
raised. This is indicative of the growing premelting layer, which
likely carries the adsorbed nitrogen away from the solid phase. 

\begin{figure}[H]
\centering
\subfigure[ basal]{\label{fig:perfil_nitro_basal_int_liq}
\resizebox*{4cm}{!}{\includegraphics{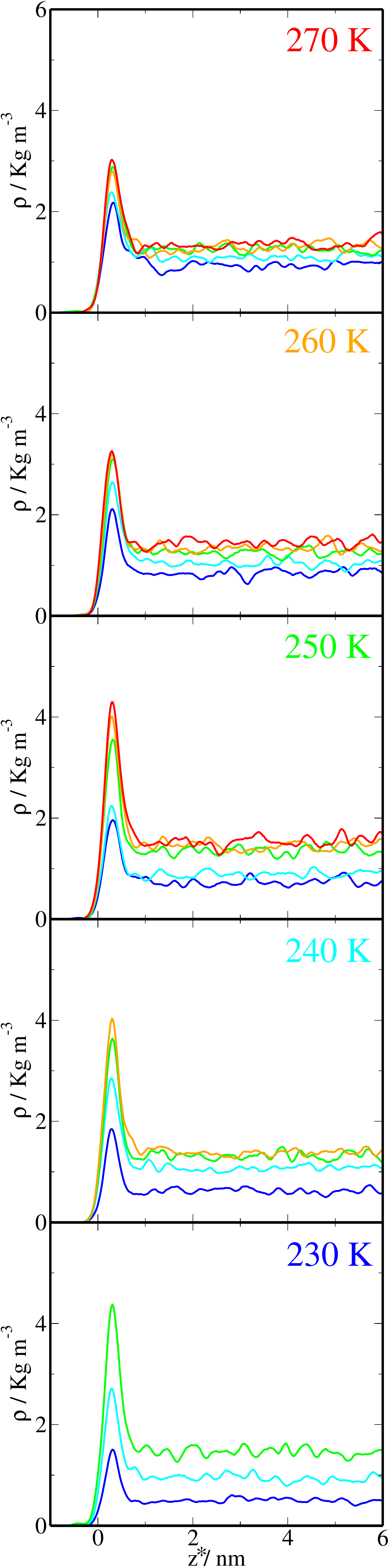}}}  
\subfigure[ pI]{\label{fig:perfil_nitro_pi_int_liq}
\resizebox*{4cm}{!}{\includegraphics{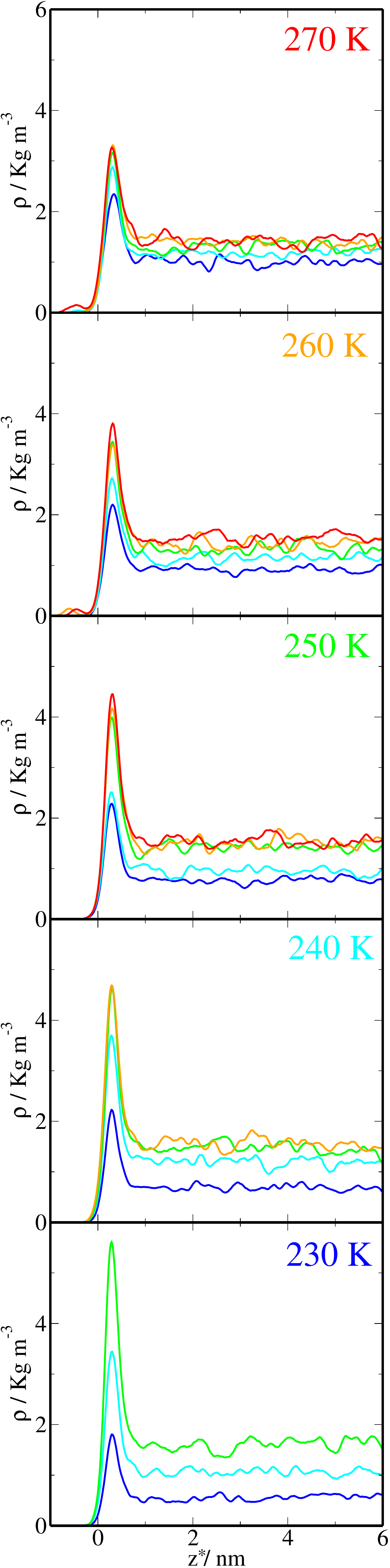}}}  
\caption{
         Density profiles of nitrogen as a function of perpendicular distance to the
         liquid-vapor surface $z-h_{wv}(x,y)$. For each temperature, density profiles correspond to pressures
         as indicated in Table \ref{tab:results1}, with pressure increasing in the
         range from ca. 0.3 to ca. 1~bar from cold colors to warm colors.
}
 \label{fig:perfil_nitro_int_liq}
\end{figure}

We can check this by plotting the intrinsic density profile of
nitrogen molecules measured relative to the water/vapor surface,
as shown in Figure \ref{fig:perfil_nitro_basal_int_liq}. We find 
again that the adsorption peak lies always at 
positive $z$, outside the premelting layer. The difference 
is that now its location with respect to the film/vapor surface
does not change with temperature. This confirms that the adsorbed
nitrogen sticks on the film/vapor surface, and is carried away from
the bulk solid as the premelting film grows. 
 This behavior is similar to that observed for
hydrocarbons and even far more soluble small organic molecules
such as glyoxal, which
have been observed to remains on the ice surface rather than
sink into the premelting layer as ions  do.\cite{hudait17,waldner18}

Also notice that only occasionally do we find finite nitrogen densities 
at negative $z$, whence, only a small fraction
of all adsorbed nitrogen is dissolved into the premelting layer.
A finite but very small amount of nitrogen may be seen  to
penetrate the premelting film for the pI plane at the two highest temperatures, 
but is negligible in most other cases.


A striking feature of the intrinsic density profiles is that
the adsorption peaks of both basal and pI planes are now
very similar,  exhibiting almost equal height and breadth.
This means that at a given fixed point on either the
water/vapor or ice/water surfaces, the nitrogen density
profiles look just the same for both the basal and pI 
planes. The apparent differences in adsorption suggested
by the absolute density profiles of Figure \ref{fig:perfil_nitro_abs}
is not really given by a distinct interaction of nitrogen
with either plane, but rather, by a difference in the
structure of the pristine crystal planes. The broader peak
observed in the absolute density profiles for the basal
plane is rather related to a surface structure with crystal
steps and pockets, which can accommodate the nitrogen molecules
at different absolute heights, as discussed previously. 
 This reaffirms the observations of the inhomogeneity
   of the ice surface observed recently for the mW
   model.\cite{hudait17,pickering18,qiu18}

Interestingly, the picture that emerges for the ice surface in contact with low
pressure nitrogen is similar to that observed for ice in contact with
methane at 100 bar.\cite{shepherd12} Despite this much higher pressure,
the structure of the premelting layer remains  unaffected by the influence
of  methane relative to the structure found in pure water vapor, provided
results are compared for the same degree of under-cooling. In connection
with the study of Shepherd et al., we expect that as the premelting
layer becomes sufficiently thick, the solubility of nitrogen 
within the premelting film should become similar to that found in bulk
water. However, here the partial pressures are very small, whence,
the amount of adsorbed nitrogen can hardly be measured in our simulations.

\subsection{Structural and thermodynamic implications of nitrogen adsorption}

\label{sec:strucythermo}

The study of nitrogen densities seems to indicate a rather passive role of
the adsorbed nitrogen, which does hardly dissolve within  the premelting film,
but rather, remains adsorbed onto the disordered water layer. Furthermore,
the surface enrichment is quite moderate, with a local density increase
at the ice surface that is only a factor of two greater than the very small
bulk gas density at atmospheric pressure. 

To shed further light into this issue, we now look at the density of solid-like
and liquid-like water molecules in the presence of nitrogen at different
pressures (Figure \ref{fig:perfil_agua_solido_liquido}). The results confirm
that nitrogen gas does not result in any significant structural change of
the ice/vapor interface. Indeed, neither the density of solid-like, nor
the density of liquid-like water molecules seems to change with nitrogen
pressure in any significant manner, for nitrogen pressures up to one bar.
In fact, the density profiles corresponding to different nitrogen pressures
can hardly be distinguished from the result of ice in pure water vapor.

Similarly, the plots of $P(z^{\ast})$, obtained from the distribution of 
$\rm{H_{2}O}$(g) colliding against the substrate, remain also very similar
to those observed in the absence of nitrogen, as shown in
Figure \ref{fig:z_time_0bar}.b. In this case, the distributions do
not appear to be identical, but we attribute the small differences mainly
to the small amount of equilibrated ice configurations employed in this
study, which was just 20, compared to the statistics gathered for the
density profiles, which is taken over more than 400 configurations.

\begin{figure}[H]
\centering
\resizebox*{9cm}{!}{\includegraphics{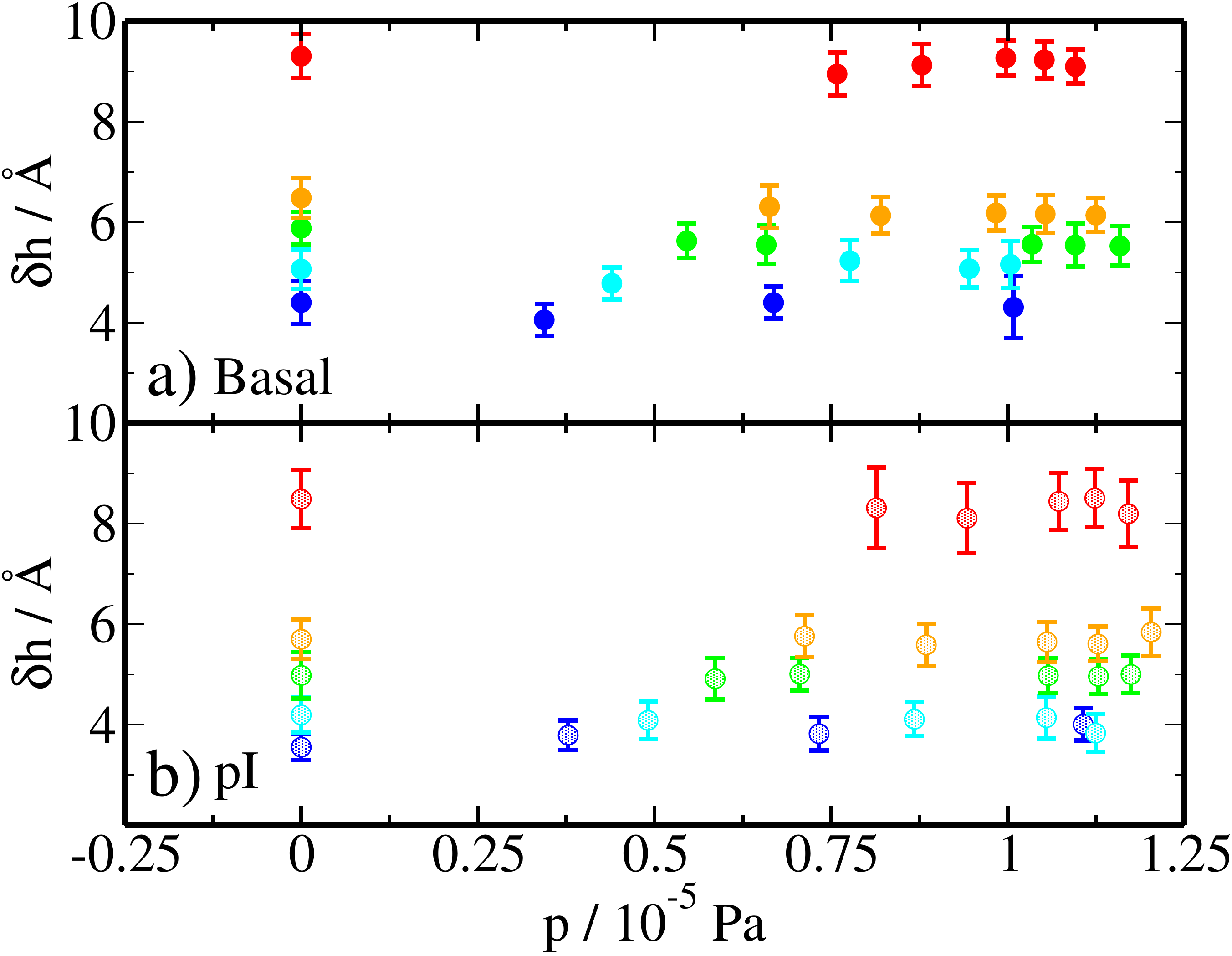}}
\caption{
Premelting layer thickness, $\delta h$, for basal and primary prismatic
planes as a function of nitrogen pressure for T=230~K (blue), 240~K (cyan), 250~K
(green), 260~K (orange) and 270~K (red).  
}
 \label{fig:espesores}
\end{figure}

As a final illustration of the small role of nitrogen gas on the structure of
the ice/vapor interface, we now concentrate on the influence of nitrogen gas
on the thickness of the premelting layer, $\delta h$, calculated
as described in section \ref{sec:simmet}. Table \ref{tab:results1}
collects the results for all systems studied, while Figure \ref{fig:espesores}
shows the premelting layer thickness as a function of nitrogen pressure
for a set of temperatures in the range between 230~K and 270~K.
The results indicate that the premelting layer thickness does not show
any significant change with nitrogen pressure, for pressures up to
1 bar, in all the temperature range studied, including the highest temperature
of 270~K. Notice that this corresponds to about 2~K less than the melting
point of the TIP4P/ice model used here. The results are consistent with
recent experiments, which indicate that the premelting film in presence
of nitrogen remains in the subnanometer range up to 1~K below
melting (similar to results found in pure water vapor, 
c.f.\cite{bluhm02}), but increases very much in the scale of tenths
of Kelvin away from the triple point.\cite{mitsui19}
This seems reasonable, since at
such small distances away from the triple point the premelting film
could thicken considerably and incorporate nitrogen and other impurities
by dissolution. These small perturbations slightly change the
intermolecular forces between the film and ice, but very close to
the triple point such small changes in the intensity of the intermolecular
forces can  can significantly shift the equilibrium
thickness.\cite{wettlaufer99}

As a summary, the structural analysis suggests that i) the ice/vapor surface
essentially behaves as a high-energy inert substrate as regards nitrogen adsorption,
with no significant disruption of the pristine interface 
and ii) there does not appear to be any apparent difference
between the adsorption of nitrogen on basal or pI planes, other than that
related with the differences between pristine interfaces under pure
water vapor.

In order to quantify these qualitative statements, we now consider the
thermodynamic implications of nitrogen adsorption on the ice/vapor interface.
Firstly, we calculate the surface adsorption of nitrogen, $\Gamma_{\rm{N_{2}}}$, as given
in terms of adsorbed molecules per unit surface (c.f.\Eq{nads}). 

\begin{figure}[H]
\centering
\resizebox*{9cm}{!}{\includegraphics{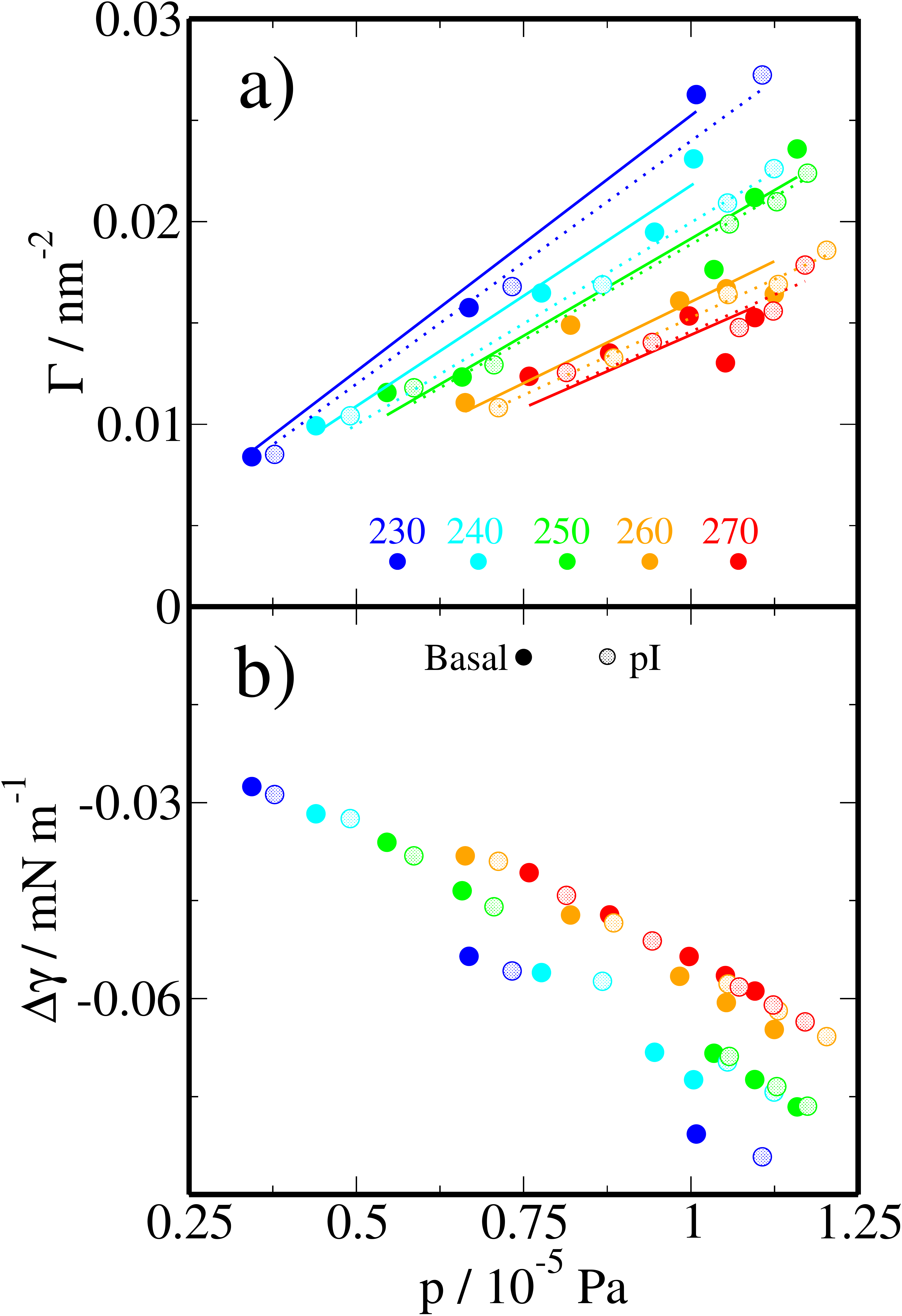}} \\ 
\resizebox*{8cm}{!}{\includegraphics{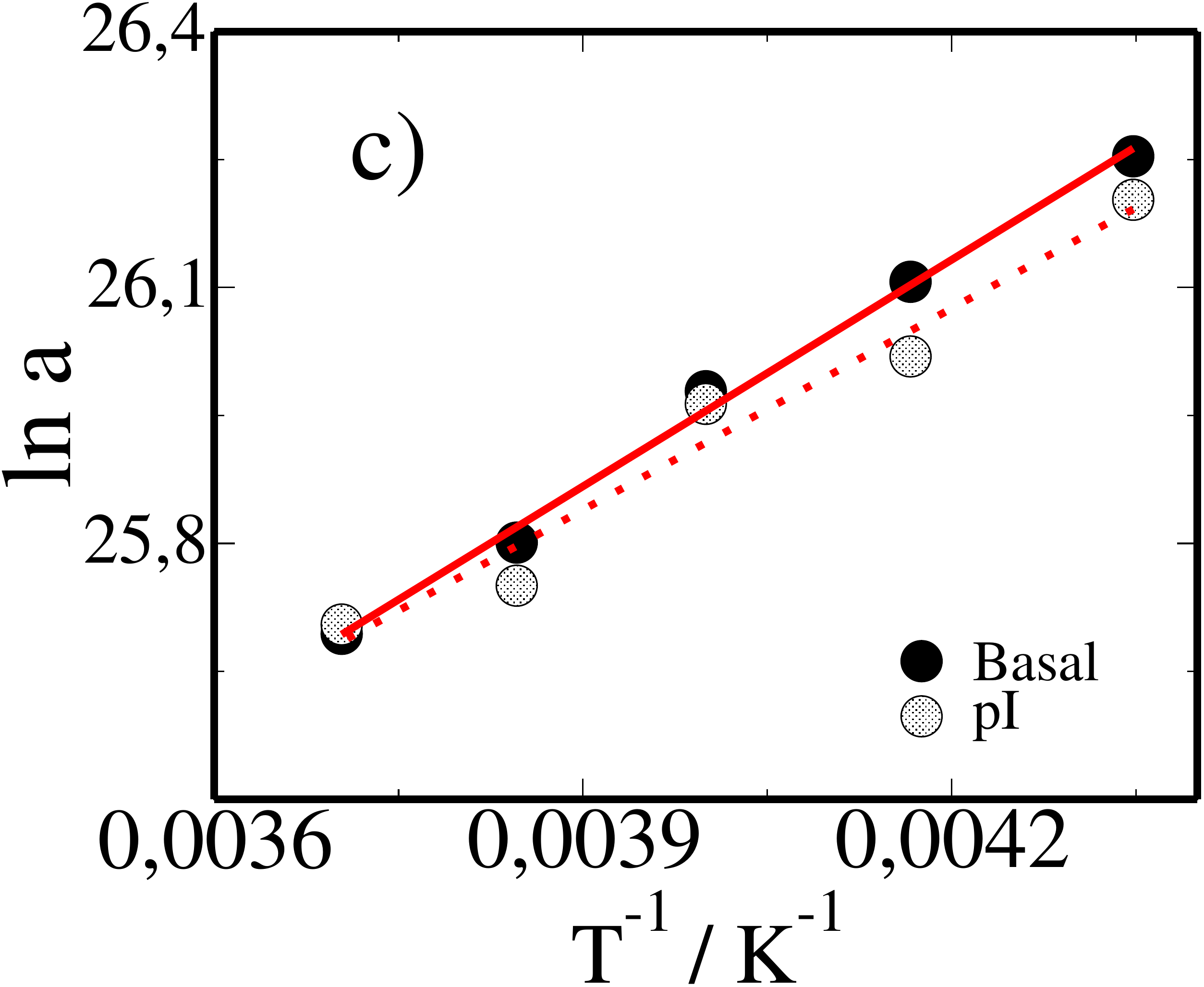}}
\caption{ Energetics of nitrogen adsorption. 
         (a) Adsorption isotherms as a function of pressure, $\Gamma_{\rm{N_{2}}}$.
	   (b) Change of the ice/vapor surface tension due to nitrogen adsorption,
	   $\Delta\gamma$, as a function of pressure.
	   (c) Arrhenius plots of Henry adsorption constants, $a$.
   Full symbols correspond to data for the basal plane, and empty symbols for the pI plane.
   Temperature color code as indicated in the legends on the plot.
 \label{fig:gamma}
}
\end{figure}

Fig.\ref{fig:gamma}.a shows $\Gamma_{\rm{N_{2}}}$ as a function of nitrogen pressure for
temperatures in the range 230 to 270~K, for both basal and pI planes.
Results obtained in a  range of conditions relevant for most practical atmospheric conditions
yield adsorptions that are essentially linear in the pressure, and
correspond to a simple Henry adsorption law of the form $\Gamma_{\rm{N_{2}}} = a p$,
with $a$, the Henry adsorption constant of nitrogen on ice.  As regards the
temperature dependence, the adsorption  increases as temperature is decreased,
as reflected by the slope of the adsorption isotherms. Overall, in all the
regime of atmospheric conditions explored in this study, we find very small
adsorption, and relatively small temperature and pressure dependences. Indeed,
reported results for $\Gamma_{\rm{N_{2}}}$ hardly change by a factor of two from the highest
to the lowest temperature, while the absolute values lie all in the order of
$10^{-2}$ molecules per nanometer squared, whence, corresponding roughly
to a surface concentration of one molecule  per lateral area of $30\times 30$ 
molecular diameters. 
This adsorption is however not small if we take into
account the small solubility of nitrogen gas in water. Indeed, at the
triple point the bulk solubility of nitrogen gas in water in terms of
mole fractions is only $x_{\rm{N_{2}}}\approx 2\cdot10^{-5}$.\cite{battino84,sander15} 
This can be compared with an effective solubility of nitrogen on the premelting
film, which we can estimate as $\chi_{\rm{N_{2}}}=\Gamma_{\rm{N_{2}}} A/N_{\rm{H_{2}O}}$,
and provides $\chi_{\rm{N_{2}}}=1.133\cdot10^{-4}$ and $1.125\cdot10^{-4}$ for
basal and pI facets. Whence, the premelting film is enriched in nitrogen
by an order of magnitude, although, as observed earlier, the nitrogen does
not actually dissolve within, but rather, remains adsorbed atop  the
water/vapor surface. On the other hand, the maxima of the density profiles
at the ice/vapor interface is between two and three times larger than the
bulk gas density (c.f. Fig.\ref{fig:perfil_nitro_abs} ), but exhibits essentially no penetration
into the bulk ice phase. Overall, the competition between partitioning and
adsorption at the air/ice interphase is completely dominated by
adsorption atop the premelting film.

In order to understand the thermodynamic significance of these results,
we now use the adsorption constants  to quantify the
change in the ice/vapor surface tension resulting
from the nitrogen adsorption at the interface. This can be estimated
readily from the adsorption isotherms and the Gibbs adsorption
equation (c.f. \Eq{gibbs}-\Eq{gamma}). 
The results are shown in Figure \ref{fig:gamma}.c, and clearly reveal
an extremely small surface activity of nitrogen gas. The change
in the ice/vapor surface tension resulting from the nitrogen adsorption
is of order $-10^{-2}$mN/m, which is to be compared with the estimated
ice/vapor surface tension,  on the order of 100
mN/m.\cite{fletcher70,pruppacher10} Whence,
nitrogen at conditions relevant to atmospheric sciences decreases the
ice/vapor surface tension to a negligible extent. The results also
reveal an extremely small surface anisotropy of the adsorption process,
with only a slightly larger surface activity on the basal plane than
on the pI plane.

To further characterize the energetics of nitrogen adsorption on the
ice surface, we fit $\Gamma_{\rm{N_{2}}}$ to  the linear law,
$\Gamma_{\rm{N_{2}}} = a p$, and obtain adsorption constants from the the
slope. These Henry's adsorption constants are expected to follow an Arrhenius
behavior. This is tested and confirmed in Figure \ref{fig:gamma}.c. From
the slope of the Arrhenius plots, we find adsorption enthalpies of 
7.4 and 6.5 kJ/mol for the basal and pI surfaces, respectively
(c.f. Table \ref{tab:ads}). On the other
hand, experimental measurements of nitrogen adsorbed on ice from BET isotherms
yield an excess enthalpy of adsorption of about 2.5~kJ/mol at
77~K,\cite{adamson67,schmitt87,hoff98} as measured relative to nitrogen
enthalpy of condensation at 77~K. 
 Notice however that BET experiments
   report binding energies relative to the enthalpy of vaporization
   of nitrogen at 77~K. From standard phase-change thermodynamic
   data available at the National Institute of Standards and
   Technology web page, we find an enthalpy of vaporization of 
   5.6~kJ/mol, which added to the 2.5~kJ/mol excess enthalpy of adsorption
yields $\Delta H_{ads}=5.6+2.5 = 8.1$~kJ/mol. 
Estimates 
 obtained from the frequency shift of the OH dangling bond stretch
upon adsorption at 96~K,  provide ca. 8.7 kJ/mol.\cite{devlin95},
while TPD experiments
provide ca. 9.6~kJ/mol.\cite{fayolle15,minissale16,nguyen18} However, as acknowledged, 
TPD probes only  the high-value tail of the distribution
of binding energies, so that the latter values are expected to be an upper bound for 
desorption energies at higher temperatures.\cite{fayolle15,minissale16,nguyen18}
Since most of the conclusions of this work follow essentially from the low adsorption
energies of nitrogen on ice, it follows that improvements on the force field
employed will hardly change them in any significant way, 
since we provide adsorption energies that are barely 2.5~kJ/mol below
the upper conceivable bound of 9.6~kJ/mol.

Our estimate from the analysis of adsorption constants thus
looks rather reasonable, and at any rate provides adsorption
energies far smaller than
the calculated binding energies of water ad-atoms on ice, which
range from 30 to 70~kJ/mol.\cite{watkins11}

\section{Surface attachment kinetics}

\label{sec:attachres}

In the previous sections, we have undertaken a study of the structure and
thermodynamics of the ice surface in presence of nitrogen. The results reveal
an almost negligible   surface activity  of nitrogen, and only a very small
surface anisotropy as regards nitrogen adsorption. Accordingly, the reported
influence of nitrogen gas on crystal growth rates must be  related to
kinetic effects.

Of course, significant amounts of crystal growth cannot be observed within
the time scale of computer simulations. The success of the kinetic theory
of gases and crystal growth theory is to write down the macroscopic
net crystal growth rate in terms of microscopic  properties that can be
quantified without the need to monitor the process over macroscopic
length and timescales.

In standard applications, the crystal growth rates are reported in terms of
the maximum flux of water vapor molecules impinging onto the ice surface,
which, from elementary kinetic theory is given as:
\begin{equation}\label{eq:jmax}
J_{max} = \frac{1}{4} \langle v \rangle \rho_{H_2O}
\end{equation} 
where $\langle v \rangle$ is the thermal average speed of water molecules and
$\rho_{H_2O}$ is the water vapor density in the neighborhood of the surface.

In practice, the crystal growth rate depends on the flux of vapor towards
the surface, but also on the sublimation rate. Accordingly, the overall
growth rate can be expressed as:
\begin{equation}
   R = \alpha J_{max} - J_{ev}
\end{equation} 
where $\alpha$ is the (microscopic) attachment coefficient, which dictates
the probability that a waver molecule in a ballistic trajectory will stick
to the ice surface, and $J_{ev}$ is a net flux of sublimation.
In view of the difficulties to measure $\alpha$ and $J_{ev}$ separately,
the crystal growth rate is usually given experimentally as:
\begin{equation}
  R = \alpha_{\rm eff} J_{max}
\end{equation} 
where $\alpha_{\rm eff}$ is now  the (macroscopic) accommodation
coefficient. This lumps into one single empirical parameter all the complicated surface
kinetics, including the adsorption probability of impinging water molecules,
the thermal equilibration and surface diffusion into the premelting film, 
the accommodation of a water-like molecule into the crystal lattice, and the
decrease of the crystal growth rate due to sublimation
events.\cite{haynes92,neshyba09,mohandesi18}

A great number of experiments aimed at calculating  $\alpha_{\rm eff}$ in 
atmospheric conditions have been performed, but 
unfortunately, results  seem to depend considerably on experimental details and data
analysis.\cite{haynes92,delval04,libbrecht12,skrotzki13,kong14} Most reported
results vary between $\alpha_{exp}=0.1$ to $\alpha_{exp}=1$, but
occasionally provide coefficients well smaller than 0.1, leaving
a great room for improvement.\cite{iupac09}

One major source of discrepancy is related to the proper measurement of
the net flux of impinging molecules. The reason is that this result is
reliable only when the vapor density $\rho_{H_2O}$ is that of water vapor
molecules at a distance from the surface in the order of the mean free path. Because most
experiments are performed at ambient pressure, and necessitate an inert
carrier gas to ensure thermal equilibrium, the macroscopic flow of water
vapor is not related to ballistic trajectories implied in \Eq{jmax}, but
rather, is a slow random walk process. When the rate of attachment of
water molecules into the surface is of the order of the rate of diffusion
or larger, the density becomes considerably depleted at the neighborhood
of the ice surface, and the calculation of $J_{max}$ is then a complicated
diffusion limited process. Disentangling this diffusion process from the
attachment kinetics on the ice surface has revealed to be rather  complicated 
and ambiguous.

\begin{figure}[H]
   \centering
   \resizebox*{9cm}{!}{\includegraphics{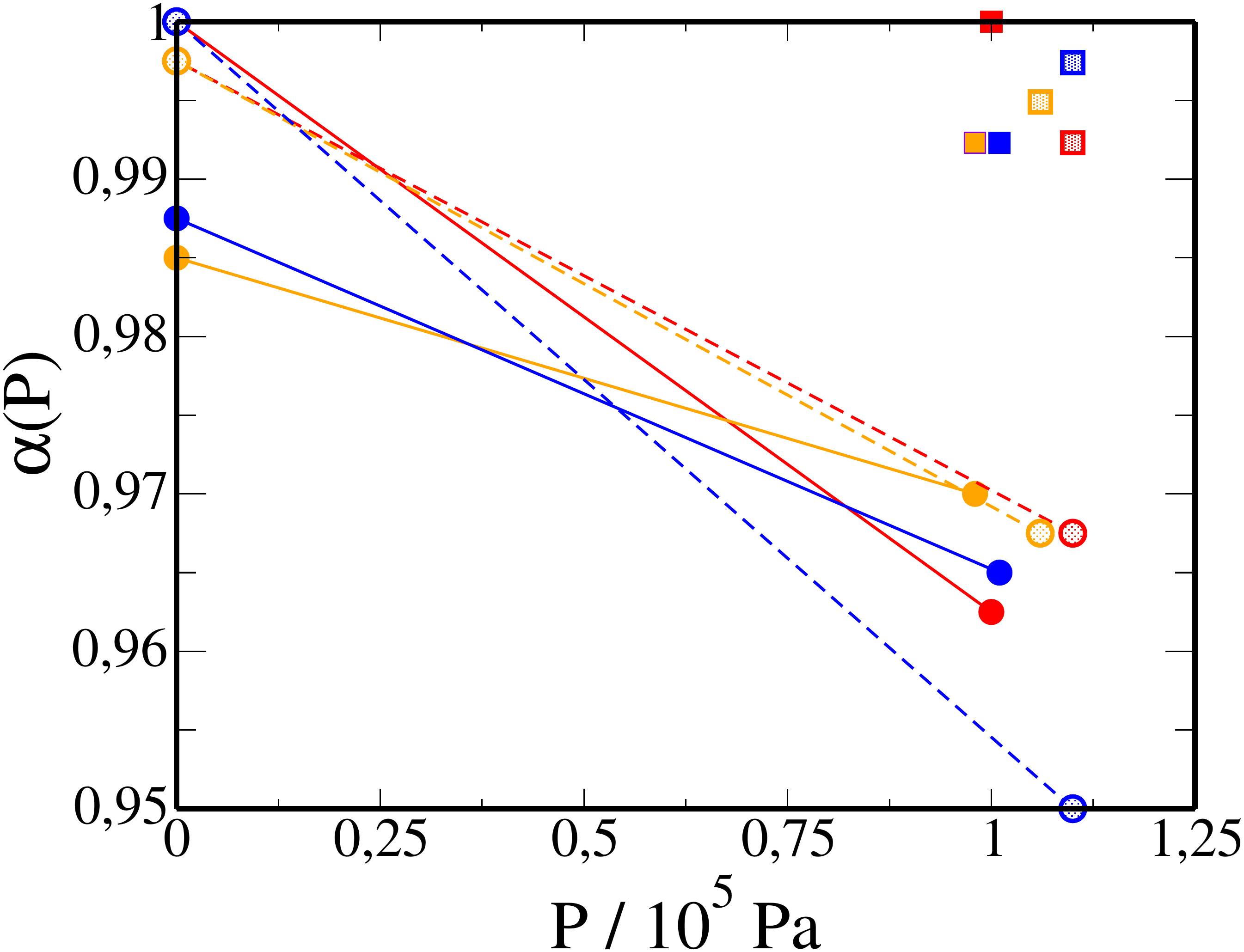}}
   \caption{
	  Attachment coefficients at three different temperatures measured
	  as a function of nitrogen partial pressure for T=230~K (blue), 
          260 (orange) and 270 (red).
          Filled circles correspond to data for the basal plane, and empty
circles for the pI plane. Squares with similar color code show
attachment coefficients calculated without counting trajectories that resulted
in the back-scattering of water molecules with bulk nitrogen gas, as explained in the text.
	}
	 \label{fig:alfa_temp_press}
\end{figure}

An attractive way to disentangle the  impingement rate from gas phase
collision events is to carry out explicit computer
simulations. In our study, we place  water molecules with center of mass and
angular velocities selected from a thermal Boltzmann distribution 
at a distance 2~nm away from an equilibrated ice
surface, and shoot the water molecule against the surface,
as explained in section \ref{sec:attach}.  Results for $\alpha$ collected over
400 trajectories and 20 different thermal ice configurations. 
 are shown in Figure \ref{fig:alfa_temp_press} at zero and one bar
for three different temperatures. Detailed statistical information
on the trajectories may be found in the supplementary material.

At zero nitrogen pressure, the trajectories sampled are essentially ballistic,
because the vapor pressure is extremely small for the TIP4P/ice model (this
limitation is shared by most point charge models) and the probability of
collisions with evaporated water vapor molecules is negligible for such short trajectories.
The results of Figure \ref{fig:alfa_temp_press} show that this direct microscopic
attachment coefficient is almost equal to unity for both the basal and pI
surface planes. In fact, for the pI plane all impinging molecules were
attached to the ice surface except one, at T=260~K. For the basal plane
$\alpha$ is also very close to unity, but five and six water molecules were
reflected back at temperatures of T=230~K and T=260~K, respectively. A look
at the non-sticking trajectories shows that in these cases the reflection
occurs only for gracing trajectories with a small z-component of the velocity.
The molecule then flies at a small distance from the surface and is eventually
reflected back by  electrostatic repulsion at a distance (rather than in
   a hard-sphere like collision between molecules at close contact).

At a pressure of one bar, we find that $\alpha$ remains still very close to
unity, but has decreased nevertheless significantly from the values at zero
pressure.  Out of  $400$ molecules shot against the ice/vapor interface, about
15 per batch were reflected and  did not accommodate within the ice surface
(see table IV of the supplementary material for detailed collision
statistics).  This remains a small number, but is
 about three times as large as in absence of nitrogen gas. In principle,
the reason for the decrease in attachment rates could be  1) nitrogen disrupts 
the ice surface sufficiently that the attachment rates change 2)  coverage of 
nitrogen on the ice surface is large, and water molecules are reflected
by the adsorbed nitrogen 3) nitrogen gas bulk density is sufficient large
that collisions in the bulk decrease the probability that
water vapor molecules will reach the ice/vapor interface.
Clearly, for the  study of previous sections, the first
two hypotheses seem unlikely: we have found that nitrogen barely
changes the structure of the ice surface, and that the coverage
of nitrogen corresponds to about one molecule every 100~nm$^2$.

In order to shed further light into the problem, we reanalyzed the
small set of trajectories which did not result in attachment of
water and found that the overwhelming majority of trajectories resulted
in reversal of the velocity component of the water molecules more
than three molecular diameters away from the outer film/vapor surface
(supplementary material, Table IV). Accordingly, it is concluded that
the decrease of the attachment coefficients in presence of nitrogen
is mainly due to collisions of water molecules with nitrogen gas
from the bulk phase.
To further illustrate the point,  we calculate corrected attachment coefficients,
using only those trajectories which do not involve scattering
from gas phase nitrogen. The new results are shown in
Fig.\ref{fig:alfa_temp_press} as squares, and are now fully consistent 
with the attachment coefficients found in absence of nitrogen. 
From this analysis we confirm that the active role of nitrogen
gas is to reduce the rate at which water molecules can impinge
on the ice surface, and  is not related to a significant reduction
of attachment probabilities of those water molecules that do collide
with the surface.

The results for attachment coefficients at zero pressure are in full agreement with previous
trajectory studies from molecular simulation, which have consistently
observed that the fate of vapor water molecules
directed towards the ice surface is to stick onto the ice
surface,\cite{batista05,neshyba09,pfalzgraf11} irrespective of the
rate of impingement.\cite{neshyba16,hudait16} Our study of trajectories
in the presence of nitrogen gas seem to indicate that a similar fate awaits
water molecules unless an unfavorable collision with free nitrogen gas molecules
occurs. Our results thus clearly favor (microscopic) attachment coefficients
of unity, with little evidence of a significant temperature dependence. This 
is also consistent with some 
recent experiments for water vapor collision rates on ice.\cite{skrotzki13}

Why is there then so much scatter in the literature? Clearly, one reason
is the difficulty to disentangle pure surface processes from gas phase
diffusion processes. However, another large source of discrepancy likely lies 
in the definition of the accommodation coefficient itself. Here, we have
tried to make clear the distinction between the surface attachment kinetics,
which refers to the probability of a colliding molecule to stick onto the
ice surface (which we term here the attachment coefficient, $\alpha$), from the effective 
accommodation coefficient, $\alpha_{\rm eff}$ which accounts
for the net balance between global condensation and evaporation kinetics
as given in terms of the surface density. Further sources of confusion can
arise if $\alpha_{\rm eff}$ is measured in the diffusive regime with a flux $J_{max}$
given in terms  of the density at 'large' distances from the
sample.\cite{kong14} We believe the 
literature does not always clarify which is the actual coefficient that is measured, 
and this likely is one reason for the  great scatter of experimental
results.\cite{iupac09}

Additional support for a (microscopic) attachment coefficient of unity is
given by the study of water gas molecules impinging onto super-cooled water.
At least close to the triple point, where ice is expected to have a premelted
liquid layer of nanometer size,\cite{conde08,benet16} the direct impingement
of $\rm{H_{2}O}$(g) onto liquid water should not differ significantly, while precisely at the
triple point, the effective coefficient should likewise be exactly the
same, since then ice and water must have the same evaporation rates. 
The advantage of experiments performed in water is that the water/vapor
interface here is much simpler in structure, so that one does not have
to account for the complicated premelting mediated crystal growth mechanisms 
(c.f., \cite{haynes92,neshyba16,murata19}), and only the evaporation flux and diffusion
processes needs to be considered. Despite of this somewhat simpler situation,
attachment coefficients on liquid water are also difficult to measure and
experiments exhibit a large scatter. However, by studying the uptake of a flux of
deuterated water it is possible to disentangle the attachment rates from
the evaporative flux, and the results obtained this way do seem to
support  a value of $\alpha=1$,\cite{li01b} in agreement with
other recent experiments and computer simulations
for attachment dynamics on liquid water.\cite{davidovits04,morita04}

\section{Conclusions}

\label{sec:concl}

 In this paper we address the long standing controversy of nitrogen's role
   in determining ice crystal growth rates and premelting film
   thickness.\cite{beckmann82b,beckmann83,kuroda84b,libbrecht17,jayaweera71,kuroda84,gonda70,lamb72,vandenheuvel59,sei89,hallett58,kobayashi67,bluhm02,elbaum91,elbaum91b,elbaum93,lied94,dosch95,mitsui19,bartels-rausch13}
   Indeed, a large number of authors through the years have stressed the
   strong influence of either air or nitrogen in determining both the
   crystal growth rates,\cite{beckmann82b,beckmann83,kuroda84b,libbrecht17} and the premelting layer
   thickness.\cite{elbaum93,lied94,dosch95,bluhm02,mitsui19} As far as crystal
   growth rates are concerned, two
possible explanations have been put forward.
1) Either the nitrogen adsorption is
strong enough to disrupt the ice surface and change the attachment
kinetics,\cite{beckmann82b,beckmann83,kuroda84b,libbrecht17} or
2) nitrogen plays a passive role merely by slowing down crystal growth rates due
to a diffusion limited flow of
water,\cite{kuroda84,gonda70,lamb72,vandenheuvel59,sei89,hallett58,kobayashi67}
but unfortunately no consensus has been achieved from
experiments.\cite{beckmann82b,beckmann83,kuroda84b,libbrecht17,kuroda84,gonda70,lamb72,vandenheuvel59,sei89,hallett58,kobayashi67} As for  the premelting layer thickness,
impurities are suspected to influence the order of magnitude differences
measured experimentally, but whether nitrogen is one such important factor
remains unsolved.\cite{elbaum93,lied94,dosch95,bluhm02,mitsui19}
Here we address these important controversies by performing computer
simulation experiments with 
widely accepted 
point charge models of water and
nitrogen. The direct measurement of nitrogen adsorption and attachment rates, as
well as premelting layer thickness
that can be performed by simulations give us the opportunity to
unambiguously clarify these important problems  of atmospheric science.
To the best of our knowledge this is the first simulation work
aimed at clarifying this long standing problem.


In our study we have carried out a detailed characterization of the ice surface
in a controlled atmosphere of nitrogen gas. Results have been obtained
at conditions found in the International Standard Atmosphere in a range of altitudes 
between 2750 to 9000~m. This  covers  the relevant temperature range 
between 270 to 230~K expected for ice crystal growth in the atmosphere,
and a  pressure range between 0.5 and 1 bar well in the upper limit of relevant 
pressure for cirrus cloud formation.  


From our study, we find i) the adsorbed amount of nitrogen gas on the ice
surface is very small,  corresponding to ca.  one nitrogen molecule  per
100~nm$^2$.  ii) nitrogen adsorbs essentially onto the premelting layer of
ice and does hardly dissolve within. iii) The structure of the
pristine ice/vapor surface is mediated by a thin liquid film, and does
not change in any significant manner by the presence of small amounts of
adsorbed nitrogen. iv) The thickness of the premelting
layer in the temperature range studied varies between 0.4 to 1~nm, and does not
show any significant dependence on nitrogen pressure. v) The adsorption enthalpy of
nitrogen on the ice/air interface is approximately 7 kJ/mol, much lower
than is required for surface poisoning or step free energy changes
to occur.
  vi) the variation of ice/air surface tension due to nitrogen is in the
10$^{-2}$ mN/m range, and accounts for less than 0.05\% of the full surface
tension in pure water vapor. vii) Basal and pI surfaces of pristine ice exhibit significant 
differences.  Particularly, the basal surface appears to show a disordered distribution
of steps on the surface. However, nitrogen does not appear to 
shift this anisotropy in any significant manner.  viii) Direct attachment
rates of water molecules onto the ice surface are very close to unity,
and do not change significantly by the presence of small amounts
of adsorbed nitrogen. In essence, at temperatures relevant to ice formation
and nitrogen pressures well in the upper limit of interest, the ice/vapor
interface appears as a high-energy inert substrate for adsorbed nitrogen gas,
which does not have any significant role on the structural properties of the
ice/air interface, and the direct surface attachment kinetics.
ix) However, our study shows conclusively that the nitrogen densities
typical in the atmosphere are sufficiently large to deflect impinging 
water molecules directed to the ice surface from distances far smaller than the
mean free path of water in pure vapor. This
highlights the role of nitrogen gas in slowing down the flux of water vapor
towards the crystal surface, and indicates that reported differences of crystal
growth rate in a nitrogen atmosphere must be essentially related to diffusion
limited processes.\cite{beckmann82b,beckmann83,kuroda84b,libbrecht17}
As a final comment, notice that the conclusions drawn here stem mainly
from the low adsorption energies of nitrogen on ice. Since our own
results provide an adsorption energy  barely 2.5~kJ/mol smaller
than the upper   bound of 9.6~kJ/mol reported in Temperature Programmed 
Desorption experiments,\cite{fayolle15,minissale16,nguyen18} 
we expect that improvements on the force field employed here will not change
our conclusions qualitatively.

\section*{Acknowledgment}
We gratefully acknowledge funds from the  Spanish
Agencia Estatal de Investigaci\'on 
under Grant No. FIS2017-89361-C3-2-P.


\begin{mcitethebibliography}{104}
\providecommand*{\natexlab}[1]{#1}
\providecommand*{\mciteSetBstSublistMode}[1]{}
\providecommand*{\mciteSetBstMaxWidthForm}[2]{}
\providecommand*{\mciteBstWouldAddEndPuncttrue}
  {\def\EndOfBibitem{\unskip.}}
\providecommand*{\mciteBstWouldAddEndPunctfalse}
  {\let\EndOfBibitem\relax}
\providecommand*{\mciteSetBstMidEndSepPunct}[3]{}
\providecommand*{\mciteSetBstSublistLabelBeginEnd}[3]{}
\providecommand*{\EndOfBibitem}{}
\mciteSetBstSublistMode{f}
\mciteSetBstMaxWidthForm{subitem}
{(\emph{\alph{mcitesubitemcount}})}
\mciteSetBstSublistLabelBeginEnd{\mcitemaxwidthsubitemform\space}
{\relax}{\relax}

\bibitem[Bartels-Rausch(2013)]{bartels-rausch13}
T.~Bartels-Rausch, \emph{Nature}, 2013, \textbf{494}, 27--29\relax
\mciteBstWouldAddEndPuncttrue
\mciteSetBstMidEndSepPunct{\mcitedefaultmidpunct}
{\mcitedefaultendpunct}{\mcitedefaultseppunct}\relax
\EndOfBibitem
\bibitem[Loyola \emph{et~al.}(2018)Loyola, Gimeno~Garc\'{\i}a, Lutz, Argyrouli,
  Romahn, Spurr, Pedergnana, Doicu, Molina~Garc\'{\i}a, and
  Sch\"ussler]{loyola18}
D.~G. Loyola, S.~Gimeno~Garc\'{\i}a, R.~Lutz, A.~Argyrouli, F.~Romahn, R.~J.~D.
  Spurr, M.~Pedergnana, A.~Doicu, V.~Molina~Garc\'{\i}a and O.~Sch\"ussler,
  \emph{Atmospheric Measurement Techniques}, 2018, \textbf{11}, 409--427\relax
\mciteBstWouldAddEndPuncttrue
\mciteSetBstMidEndSepPunct{\mcitedefaultmidpunct}
{\mcitedefaultendpunct}{\mcitedefaultseppunct}\relax
\EndOfBibitem
\bibitem[Baran(2012)]{baran12}
A.~J. Baran, \emph{Atmospheric Research}, 2012, \textbf{112}, 45--69\relax
\mciteBstWouldAddEndPuncttrue
\mciteSetBstMidEndSepPunct{\mcitedefaultmidpunct}
{\mcitedefaultendpunct}{\mcitedefaultseppunct}\relax
\EndOfBibitem
\bibitem[Warren and Brandt(2008)]{warren08}
S.~G. Warren and R.~E. Brandt, \emph{J. Geophys. Research}, 2008, \textbf{113},
  D14220\relax
\mciteBstWouldAddEndPuncttrue
\mciteSetBstMidEndSepPunct{\mcitedefaultmidpunct}
{\mcitedefaultendpunct}{\mcitedefaultseppunct}\relax
\EndOfBibitem
\bibitem[Hesse \emph{et~al.}(2012)Hesse, Macke, Havemann, Baran, Ulanowski, and
  Kaye]{hesse12}
E.~Hesse, A.~Macke, S.~Havemann, A.~Baran, Z.~Ulanowski and P.~Kaye, \emph{J.
  Quantitative Spectroscopy and Radiative Transfer}, 2012, \textbf{113}, 342 --
  347\relax
\mciteBstWouldAddEndPuncttrue
\mciteSetBstMidEndSepPunct{\mcitedefaultmidpunct}
{\mcitedefaultendpunct}{\mcitedefaultseppunct}\relax
\EndOfBibitem
\bibitem[Neshyba \emph{et~al.}(2013)Neshyba, Lowen, Benning, Lawson, and
  Rowe]{neshyba13}
S.~Neshyba, B.~Lowen, M.~Benning, A.~Lawson and P.~M. Rowe, \emph{J. Geophys.
  Res.: Atmos.}, 2013, \textbf{118}, 3309--3318\relax
\mciteBstWouldAddEndPuncttrue
\mciteSetBstMidEndSepPunct{\mcitedefaultmidpunct}
{\mcitedefaultendpunct}{\mcitedefaultseppunct}\relax
\EndOfBibitem
\bibitem[Voigtl{\"a}nder \emph{et~al.}(2018)Voigtl{\"a}nder, Chou, Bieligk,
  Claus, Hartmann, Herenz, Ritter, and Ulanowski]{voigtlander18}
J.~Voigtl{\"a}nder, C.~Chou, H.~Bieligk, T.~Claus, S.~Hartmann, P.~Herenz,
  D.~N.~G. Ritter and F.~S.~Z. Ulanowski, \emph{Atmosph. Chem. Phys.}, 2018,
  \textbf{18}, 13687--13702\relax
\mciteBstWouldAddEndPuncttrue
\mciteSetBstMidEndSepPunct{\mcitedefaultmidpunct}
{\mcitedefaultendpunct}{\mcitedefaultseppunct}\relax
\EndOfBibitem
\bibitem[J\"arvinen \emph{et~al.}(2018)J\"arvinen, Jourdan, Neubauer, Yao, Liu,
  Andreae, Lohmann, Wendisch, McFarquhar, Leisner, and Schnaiter]{jarvinen18}
E.~J\"arvinen, O.~Jourdan, D.~Neubauer, B.~Yao, C.~Liu, M.~O. Andreae,
  U.~Lohmann, M.~Wendisch, G.~M. McFarquhar, T.~Leisner and M.~Schnaiter,
  \emph{Atmosph. Chem. Phys.}, 2018, \textbf{18}, 15767--15781\relax
\mciteBstWouldAddEndPuncttrue
\mciteSetBstMidEndSepPunct{\mcitedefaultmidpunct}
{\mcitedefaultendpunct}{\mcitedefaultseppunct}\relax
\EndOfBibitem
\bibitem[Abbatt(2003)]{abbat03}
J.~P.~D. Abbatt, \emph{Chemical Reviews}, 2003, \textbf{103}, 4783--4800\relax
\mciteBstWouldAddEndPuncttrue
\mciteSetBstMidEndSepPunct{\mcitedefaultmidpunct}
{\mcitedefaultendpunct}{\mcitedefaultseppunct}\relax
\EndOfBibitem
\bibitem[Conde \emph{et~al.}(2008)Conde, Vega, and Patrykiejew]{conde08}
M.~M. Conde, C.~Vega and A.~Patrykiejew, \emph{J. Chem. Phys.}, 2008,
  \textbf{129}, 014702\relax
\mciteBstWouldAddEndPuncttrue
\mciteSetBstMidEndSepPunct{\mcitedefaultmidpunct}
{\mcitedefaultendpunct}{\mcitedefaultseppunct}\relax
\EndOfBibitem
\bibitem[Yang \emph{et~al.}(2013)Yang, Asta, and Laird]{yang13}
A.~J. Yang, M.~Asta and B.~B. Laird, \emph{Phys. Rev. Lett.}, 2013,
  \textbf{110}, 096102\relax
\mciteBstWouldAddEndPuncttrue
\mciteSetBstMidEndSepPunct{\mcitedefaultmidpunct}
{\mcitedefaultendpunct}{\mcitedefaultseppunct}\relax
\EndOfBibitem
\bibitem[Limmer and Chandler(2014)]{limmer14}
D.~T. Limmer and D.~Chandler, \emph{J. Chem. Phys.}, 2014, \textbf{141},
  18C505\relax
\mciteBstWouldAddEndPuncttrue
\mciteSetBstMidEndSepPunct{\mcitedefaultmidpunct}
{\mcitedefaultendpunct}{\mcitedefaultseppunct}\relax
\EndOfBibitem
\bibitem[Neshyba \emph{et~al.}(2016)Neshyba, Adams, Reed, Rowe, and
  Gladich]{neshyba16}
S.~Neshyba, J.~Adams, K.~Reed, P.~M. Rowe and I.~Gladich, \emph{J. Geophys.
  Res.: Atmos.}, 2016, \textbf{121}, 14,035--14,055\relax
\mciteBstWouldAddEndPuncttrue
\mciteSetBstMidEndSepPunct{\mcitedefaultmidpunct}
{\mcitedefaultendpunct}{\mcitedefaultseppunct}\relax
\EndOfBibitem
\bibitem[Hudait and Molinero(2016)]{hudait16}
A.~Hudait and V.~Molinero, \emph{J. Am. Chem. Soc.}, 2016, \textbf{138},
  8958--8967\relax
\mciteBstWouldAddEndPuncttrue
\mciteSetBstMidEndSepPunct{\mcitedefaultmidpunct}
{\mcitedefaultendpunct}{\mcitedefaultseppunct}\relax
\EndOfBibitem
\bibitem[Kling \emph{et~al.}(2018)Kling, Kling, and Donadio]{kling18}
T.~Kling, F.~Kling and D.~Donadio, \emph{The Journal of Physical Chemistry C},
  2018, \textbf{122}, 24780--24787\relax
\mciteBstWouldAddEndPuncttrue
\mciteSetBstMidEndSepPunct{\mcitedefaultmidpunct}
{\mcitedefaultendpunct}{\mcitedefaultseppunct}\relax
\EndOfBibitem
\bibitem[Pickering \emph{et~al.}(2018)Pickering, Paleico, Sirkin, Scherlis, and
  Factorovich]{pickering18}
I.~Pickering, M.~Paleico, Y.~A.~P. Sirkin, D.~A. Scherlis and M.~H.
  Factorovich, \emph{J. Phys. Chem. B}, 2018, \textbf{122}, 4880--4890\relax
\mciteBstWouldAddEndPuncttrue
\mciteSetBstMidEndSepPunct{\mcitedefaultmidpunct}
{\mcitedefaultendpunct}{\mcitedefaultseppunct}\relax
\EndOfBibitem
\bibitem[Mohandesi and Kusalik(2018)]{mohandesi18}
A.~Mohandesi and P.~G. Kusalik, \emph{J. Cryst. Growth}, 2018, \textbf{483},
  156 -- 163\relax
\mciteBstWouldAddEndPuncttrue
\mciteSetBstMidEndSepPunct{\mcitedefaultmidpunct}
{\mcitedefaultendpunct}{\mcitedefaultseppunct}\relax
\EndOfBibitem
\bibitem[Qiu and Molinero(2018)]{qiu18}
Y.~Qiu and V.~Molinero, \emph{J. Phys. Chem. Lett.}, 2018, \textbf{9},
  5179--5182\relax
\mciteBstWouldAddEndPuncttrue
\mciteSetBstMidEndSepPunct{\mcitedefaultmidpunct}
{\mcitedefaultendpunct}{\mcitedefaultseppunct}\relax
\EndOfBibitem
\bibitem[Sandler \emph{et~al.}(1994)Sandler, Jung, Szczesniak, and
  Buch]{sandler94}
P.~Sandler, J.~o. Jung, M.~M. Szczesniak and V.~Buch, \emph{J. Chem. Phys.},
  1994, \textbf{101}, 1378--1391\relax
\mciteBstWouldAddEndPuncttrue
\mciteSetBstMidEndSepPunct{\mcitedefaultmidpunct}
{\mcitedefaultendpunct}{\mcitedefaultseppunct}\relax
\EndOfBibitem
\bibitem[Girardet and Toubin(2001)]{girardet01}
C.~Girardet and C.~Toubin, \emph{Surf. Sci. Rep.}, 2001, \textbf{44}, 159 --
  238\relax
\mciteBstWouldAddEndPuncttrue
\mciteSetBstMidEndSepPunct{\mcitedefaultmidpunct}
{\mcitedefaultendpunct}{\mcitedefaultseppunct}\relax
\EndOfBibitem
\bibitem[Hudait and Molinero(2017)]{hudait17}
A.~Hudait and M.~T. A.~V. Molinero, \emph{J. Am. Chem. Soc.}, 2017,
  \textbf{139}, 10095--10103\relax
\mciteBstWouldAddEndPuncttrue
\mciteSetBstMidEndSepPunct{\mcitedefaultmidpunct}
{\mcitedefaultendpunct}{\mcitedefaultseppunct}\relax
\EndOfBibitem
\bibitem[Waldner \emph{et~al.}(2018)Waldner, Artiglia, Kong, Orlando,
  Huthwelker, Ammann, and Bartels-Rausch]{waldner18}
A.~Waldner, L.~Artiglia, X.~Kong, F.~Orlando, T.~Huthwelker, M.~Ammann and
  T.~Bartels-Rausch, \emph{Phys. Chem. Chem. Phys.}, 2018, \textbf{20},
  24408--24417\relax
\mciteBstWouldAddEndPuncttrue
\mciteSetBstMidEndSepPunct{\mcitedefaultmidpunct}
{\mcitedefaultendpunct}{\mcitedefaultseppunct}\relax
\EndOfBibitem
\bibitem[Sazaki \emph{et~al.}(2012)Sazaki, Zepeda, Nakatsubo, Yokomine, and
  Furukawa]{sazaki12}
G.~Sazaki, S.~Zepeda, S.~Nakatsubo, M.~Yokomine and Y.~Furukawa, \emph{Proc.
  Nat. Acad. Sci.}, 2012, \textbf{109}, 1052--1055\relax
\mciteBstWouldAddEndPuncttrue
\mciteSetBstMidEndSepPunct{\mcitedefaultmidpunct}
{\mcitedefaultendpunct}{\mcitedefaultseppunct}\relax
\EndOfBibitem
\bibitem[Asakawa \emph{et~al.}(2015)Asakawa, Sazaki, Nagashima, Nakatsubo, and
  Furukawa]{asakawa15}
H.~Asakawa, G.~Sazaki, K.~Nagashima, S.~Nakatsubo and Y.~Furukawa,
  \emph{Crystal Growth \& Design}, 2015, \textbf{15}, 3339--3344\relax
\mciteBstWouldAddEndPuncttrue
\mciteSetBstMidEndSepPunct{\mcitedefaultmidpunct}
{\mcitedefaultendpunct}{\mcitedefaultseppunct}\relax
\EndOfBibitem
\bibitem[Asakawa \emph{et~al.}(2016)Asakawa, Sazaki, Nagashima, Nakatsubo, and
  Furukawa]{asakawa16}
H.~Asakawa, G.~Sazaki, K.~Nagashima, S.~Nakatsubo and Y.~Furukawa, \emph{Proc.
  Nat. Acad. Sci.}, 2016, \textbf{113}, 1749--1753\relax
\mciteBstWouldAddEndPuncttrue
\mciteSetBstMidEndSepPunct{\mcitedefaultmidpunct}
{\mcitedefaultendpunct}{\mcitedefaultseppunct}\relax
\EndOfBibitem
\bibitem[Murata \emph{et~al.}(2016)Murata, Asakawa, Nagashima, Furukawa, and
  Sazaki]{murata16}
K.-i. Murata, H.~Asakawa, K.~Nagashima, Y.~Furukawa and G.~Sazaki, \emph{Proc.
  Nat. Acad. Sci.}, 2016, \textbf{113}, E6741--E6748\relax
\mciteBstWouldAddEndPuncttrue
\mciteSetBstMidEndSepPunct{\mcitedefaultmidpunct}
{\mcitedefaultendpunct}{\mcitedefaultseppunct}\relax
\EndOfBibitem
\bibitem[Mitsui and Aoki(2019)]{mitsui19}
T.~Mitsui and K.~Aoki, \emph{Phys. Rev. E}, 2019, \textbf{99}, 010801\relax
\mciteBstWouldAddEndPuncttrue
\mciteSetBstMidEndSepPunct{\mcitedefaultmidpunct}
{\mcitedefaultendpunct}{\mcitedefaultseppunct}\relax
\EndOfBibitem
\bibitem[Sander(2015)]{sander15}
R.~Sander, \emph{Atmosph. Chem. Phys.}, 2015, \textbf{15}, 4399--4981\relax
\mciteBstWouldAddEndPuncttrue
\mciteSetBstMidEndSepPunct{\mcitedefaultmidpunct}
{\mcitedefaultendpunct}{\mcitedefaultseppunct}\relax
\EndOfBibitem
\bibitem[Adams \emph{et~al.}(1967)Adams, Dormant, and Orem]{adamson67}
A.~W. Adams, L.~M. Dormant and M.~Orem, \emph{J. Colloid. Interface Sci.},
  1967, \textbf{25}, 206--217\relax
\mciteBstWouldAddEndPuncttrue
\mciteSetBstMidEndSepPunct{\mcitedefaultmidpunct}
{\mcitedefaultendpunct}{\mcitedefaultseppunct}\relax
\EndOfBibitem
\bibitem[Schmitt \emph{et~al.}(1987)Schmitt, Ocampo, and Klinger]{schmitt87}
B.~Schmitt, J.~Ocampo and J.~Klinger, \emph{J. Phys. Colloques}, 1987,
  \textbf{48}, C1--519--C1--525\relax
\mciteBstWouldAddEndPuncttrue
\mciteSetBstMidEndSepPunct{\mcitedefaultmidpunct}
{\mcitedefaultendpunct}{\mcitedefaultseppunct}\relax
\EndOfBibitem
\bibitem[Hoff \emph{et~al.}(1998)Hoff, Gregor, Mackay, Wania, and Jia]{hoff98}
J.~T. Hoff, D.~Gregor, D.~Mackay, F.~Wania and C.~Q. Jia, \emph{Env. Sci.
  Tech.}, 1998, \textbf{32}, 58--62\relax
\mciteBstWouldAddEndPuncttrue
\mciteSetBstMidEndSepPunct{\mcitedefaultmidpunct}
{\mcitedefaultendpunct}{\mcitedefaultseppunct}\relax
\EndOfBibitem
\bibitem[Hanot and Domin\'e(1999)]{hanot99}
L.~Hanot and F.~Domin\'e, \emph{Env. Sci. Tech.}, 1999, \textbf{33},
  4250--4255\relax
\mciteBstWouldAddEndPuncttrue
\mciteSetBstMidEndSepPunct{\mcitedefaultmidpunct}
{\mcitedefaultendpunct}{\mcitedefaultseppunct}\relax
\EndOfBibitem
\bibitem[Legagneux \emph{et~al.}(2002)Legagneux, Cabanes, and
  Domin\"e]{legagneux02}
L.~Legagneux, A.~Cabanes and F.~Domin\"e, \emph{J. Geophys. Res.: Atmos.},
  2002, \textbf{107}, ACH 5--1--ACH 5--15\relax
\mciteBstWouldAddEndPuncttrue
\mciteSetBstMidEndSepPunct{\mcitedefaultmidpunct}
{\mcitedefaultendpunct}{\mcitedefaultseppunct}\relax
\EndOfBibitem
\bibitem[Fayolle \emph{et~al.}(2015)Fayolle, Balfe, Loomis, Bergner, Graninger,
  and Oberg]{fayolle15}
E.~C. Fayolle, J.~Balfe, R.~Loomis, J.~Bergner, D.~Graninger and M.~R. K.~I.
  Oberg, \emph{Astrophys J. Lett}, 2015, \textbf{816}, L28\relax
\mciteBstWouldAddEndPuncttrue
\mciteSetBstMidEndSepPunct{\mcitedefaultmidpunct}
{\mcitedefaultendpunct}{\mcitedefaultseppunct}\relax
\EndOfBibitem
\bibitem[Ninissale \emph{et~al.}(2016)Ninissale, Congiu, and
  Dulieu]{minissale16}
M.~Ninissale, E.~Congiu and F.~Dulieu, \emph{Astron. Astrophys.}, 2016,
  \textbf{585}, A146\relax
\mciteBstWouldAddEndPuncttrue
\mciteSetBstMidEndSepPunct{\mcitedefaultmidpunct}
{\mcitedefaultendpunct}{\mcitedefaultseppunct}\relax
\EndOfBibitem
\bibitem[Nguyen \emph{et~al.}(2018)Nguyen, Baouche, Congiu, Diana, Pagani, and
  Dulieu]{nguyen18}
T.~Nguyen, S.~Baouche, E.~Congiu, S.~Diana, L.~Pagani and F.~Dulieu,
  \emph{Astron. Astrophys.}, 2018, \textbf{619}, A111\relax
\mciteBstWouldAddEndPuncttrue
\mciteSetBstMidEndSepPunct{\mcitedefaultmidpunct}
{\mcitedefaultendpunct}{\mcitedefaultseppunct}\relax
\EndOfBibitem
\bibitem[Beckmann(1982)]{beckmann82b}
W.~Beckmann, \emph{J. Cryst. Growth}, 1982, \textbf{58}, 443 -- 451\relax
\mciteBstWouldAddEndPuncttrue
\mciteSetBstMidEndSepPunct{\mcitedefaultmidpunct}
{\mcitedefaultendpunct}{\mcitedefaultseppunct}\relax
\EndOfBibitem
\bibitem[Beckmann \emph{et~al.}(1983)Beckmann, Lacmann, and
  Bierfreund]{beckmann83}
W.~Beckmann, R.~Lacmann and A.~Bierfreund, \emph{J. Phys. Chem.}, 1983,
  \textbf{87}, 4142--4146\relax
\mciteBstWouldAddEndPuncttrue
\mciteSetBstMidEndSepPunct{\mcitedefaultmidpunct}
{\mcitedefaultendpunct}{\mcitedefaultseppunct}\relax
\EndOfBibitem
\bibitem[Kuroda and Gonda(1984)]{kuroda84b}
T.~Kuroda and T.~Gonda, \emph{J. Met. Soc. Jap.}, 1984, \textbf{62},
  563--572\relax
\mciteBstWouldAddEndPuncttrue
\mciteSetBstMidEndSepPunct{\mcitedefaultmidpunct}
{\mcitedefaultendpunct}{\mcitedefaultseppunct}\relax
\EndOfBibitem
\bibitem[Elbaum \emph{et~al.}(1993)Elbaum, Lipson, and Dash]{elbaum93}
M.~Elbaum, S.~G. Lipson and J.~G. Dash, \emph{J. Cryst. Growth}, 1993,
  \textbf{129}, 491--505\relax
\mciteBstWouldAddEndPuncttrue
\mciteSetBstMidEndSepPunct{\mcitedefaultmidpunct}
{\mcitedefaultendpunct}{\mcitedefaultseppunct}\relax
\EndOfBibitem
\bibitem[Bluhm \emph{et~al.}(2002)Bluhm, Ogletree, Fadley, Hussain, and
  Salmeron]{bluhm02}
H.~Bluhm, D.~F. Ogletree, C.~S. Fadley, Z.~Hussain and M.~Salmeron, \emph{J.
  Phys.: Condens. Matter}, 2002, \textbf{14}, L227--L233\relax
\mciteBstWouldAddEndPuncttrue
\mciteSetBstMidEndSepPunct{\mcitedefaultmidpunct}
{\mcitedefaultendpunct}{\mcitedefaultseppunct}\relax
\EndOfBibitem
\bibitem[Libbrecht(2017)]{libbrecht17}
K.~G. Libbrecht, \emph{Annu.Rev.Mater.Res}, 2017, \textbf{47}, 271--295\relax
\mciteBstWouldAddEndPuncttrue
\mciteSetBstMidEndSepPunct{\mcitedefaultmidpunct}
{\mcitedefaultendpunct}{\mcitedefaultseppunct}\relax
\EndOfBibitem
\bibitem[Jayaweera(1971)]{jayaweera71}
K.~O. L.~F. Jayaweera, \emph{J. Atmos. Sci.}, 1971, \textbf{28}, 728--736\relax
\mciteBstWouldAddEndPuncttrue
\mciteSetBstMidEndSepPunct{\mcitedefaultmidpunct}
{\mcitedefaultendpunct}{\mcitedefaultseppunct}\relax
\EndOfBibitem
\bibitem[Kuroda and Gonda(1984)]{kuroda84}
T.~Kuroda and T.~Gonda, \emph{J. Met. Soc. Jap.}, 1984, \textbf{62},
  552--561\relax
\mciteBstWouldAddEndPuncttrue
\mciteSetBstMidEndSepPunct{\mcitedefaultmidpunct}
{\mcitedefaultendpunct}{\mcitedefaultseppunct}\relax
\EndOfBibitem
\bibitem[Gonda and Kombayashi(1970)]{gonda70}
T.~Gonda and M.~Kombayashi, \emph{J. Met. Soc. Jap.}, 1970, \textbf{48},
  440--450\relax
\mciteBstWouldAddEndPuncttrue
\mciteSetBstMidEndSepPunct{\mcitedefaultmidpunct}
{\mcitedefaultendpunct}{\mcitedefaultseppunct}\relax
\EndOfBibitem
\bibitem[Lamb and D.Scott(1972)]{lamb72}
D.~Lamb and W.~D.Scott, \emph{J. Cryst. Growth}, 1972, \textbf{12},
  21--31\relax
\mciteBstWouldAddEndPuncttrue
\mciteSetBstMidEndSepPunct{\mcitedefaultmidpunct}
{\mcitedefaultendpunct}{\mcitedefaultseppunct}\relax
\EndOfBibitem
\bibitem[Beckmann and Lacmann(1982)]{beckmann82}
W.~Beckmann and R.~Lacmann, \emph{J. Cryst. Growth}, 1982, \textbf{58},
  433--442\relax
\mciteBstWouldAddEndPuncttrue
\mciteSetBstMidEndSepPunct{\mcitedefaultmidpunct}
{\mcitedefaultendpunct}{\mcitedefaultseppunct}\relax
\EndOfBibitem
\bibitem[Gonda and Sei(1987)]{gonda87}
T.~Gonda and T.~Sei, \emph{J. Phys. Colloques}, 1987, \textbf{48},
  C1--355--C1--359\relax
\mciteBstWouldAddEndPuncttrue
\mciteSetBstMidEndSepPunct{\mcitedefaultmidpunct}
{\mcitedefaultendpunct}{\mcitedefaultseppunct}\relax
\EndOfBibitem
\bibitem[Sei and Gonda(1989)]{sei89}
T.~Sei and T.~Gonda, \emph{J. Cryst. Growth}, 1989, \textbf{94}, 697--707\relax
\mciteBstWouldAddEndPuncttrue
\mciteSetBstMidEndSepPunct{\mcitedefaultmidpunct}
{\mcitedefaultendpunct}{\mcitedefaultseppunct}\relax
\EndOfBibitem
\bibitem[Libbrecht and Rickerby(2013)]{libbrecht12}
K.~G. Libbrecht and M.~E. Rickerby, \emph{J. Cryst. Growth}, 2013,
  \textbf{377}, 1--8\relax
\mciteBstWouldAddEndPuncttrue
\mciteSetBstMidEndSepPunct{\mcitedefaultmidpunct}
{\mcitedefaultendpunct}{\mcitedefaultseppunct}\relax
\EndOfBibitem
\bibitem[Akutsu \emph{et~al.}(2001)Akutsu, Akutsu, and Yamamoto]{akutsu01}
N.~Akutsu, Y.~Akutsu and T.~Yamamoto, \emph{Phys. Rev. B}, 2001, \textbf{64},
  085415\relax
\mciteBstWouldAddEndPuncttrue
\mciteSetBstMidEndSepPunct{\mcitedefaultmidpunct}
{\mcitedefaultendpunct}{\mcitedefaultseppunct}\relax
\EndOfBibitem
\bibitem[van~den Heuvel and Mason(1959)]{vandenheuvel59}
A.~P. van~den Heuvel and B.~J. Mason, \emph{Nature}, 1959, \textbf{184},
  519--520\relax
\mciteBstWouldAddEndPuncttrue
\mciteSetBstMidEndSepPunct{\mcitedefaultmidpunct}
{\mcitedefaultendpunct}{\mcitedefaultseppunct}\relax
\EndOfBibitem
\bibitem[Lamb and Hobbs(1971)]{lamb71}
D.~Lamb and P.~V. Hobbs, \emph{J. Atmos. Sci.}, 1971, \textbf{28},
  1506--1509\relax
\mciteBstWouldAddEndPuncttrue
\mciteSetBstMidEndSepPunct{\mcitedefaultmidpunct}
{\mcitedefaultendpunct}{\mcitedefaultseppunct}\relax
\EndOfBibitem
\bibitem[Hallett and Mason(1958)]{hallett58}
J.~Hallett and B.~J. Mason, \emph{Proc. R. Soc. Lond. A}, 1958, \textbf{247},
  440--453\relax
\mciteBstWouldAddEndPuncttrue
\mciteSetBstMidEndSepPunct{\mcitedefaultmidpunct}
{\mcitedefaultendpunct}{\mcitedefaultseppunct}\relax
\EndOfBibitem
\bibitem[Kobayashi(1967)]{kobayashi67}
T.~Kobayashi, Physics of Snow and Ice: Proceedings, 1967\relax
\mciteBstWouldAddEndPuncttrue
\mciteSetBstMidEndSepPunct{\mcitedefaultmidpunct}
{\mcitedefaultendpunct}{\mcitedefaultseppunct}\relax
\EndOfBibitem
\bibitem[Elbaum(1991)]{elbaum91}
M.~Elbaum, \emph{Phys. Rev. Lett.}, 1991, \textbf{67}, 2982--2985\relax
\mciteBstWouldAddEndPuncttrue
\mciteSetBstMidEndSepPunct{\mcitedefaultmidpunct}
{\mcitedefaultendpunct}{\mcitedefaultseppunct}\relax
\EndOfBibitem
\bibitem[Elbaum and Schick(1991)]{elbaum91b}
M.~Elbaum and M.~Schick, \emph{Phys. Rev. Lett.}, 1991, \textbf{66},
  1713--1716\relax
\mciteBstWouldAddEndPuncttrue
\mciteSetBstMidEndSepPunct{\mcitedefaultmidpunct}
{\mcitedefaultendpunct}{\mcitedefaultseppunct}\relax
\EndOfBibitem
\bibitem[Lied \emph{et~al.}(1994)Lied, Dosch, and Bilgram]{lied94}
A.~Lied, H.~Dosch and J.~H. Bilgram, \emph{Phys. Rev. Lett.}, 1994,
  \textbf{72}, 3554--3557\relax
\mciteBstWouldAddEndPuncttrue
\mciteSetBstMidEndSepPunct{\mcitedefaultmidpunct}
{\mcitedefaultendpunct}{\mcitedefaultseppunct}\relax
\EndOfBibitem
\bibitem[Dosch \emph{et~al.}(1995)Dosch, Lied, and Bilgram]{dosch95}
H.~Dosch, A.~Lied and J.~H. Bilgram, \emph{Surf. Sci.}, 1995, \textbf{327},
  145--164\relax
\mciteBstWouldAddEndPuncttrue
\mciteSetBstMidEndSepPunct{\mcitedefaultmidpunct}
{\mcitedefaultendpunct}{\mcitedefaultseppunct}\relax
\EndOfBibitem
\bibitem[Dash \emph{et~al.}(2006)Dash, Rempel, and Wettlaufer]{dash06}
J.~G. Dash, A.~W. Rempel and J.~S. Wettlaufer, \emph{Rev. Mod. Phys.}, 2006,
  \textbf{78}, 695--741\relax
\mciteBstWouldAddEndPuncttrue
\mciteSetBstMidEndSepPunct{\mcitedefaultmidpunct}
{\mcitedefaultendpunct}{\mcitedefaultseppunct}\relax
\EndOfBibitem
\bibitem[Li and Somorjai(2007)]{li07b}
Y.~Li and G.~A. Somorjai, \emph{J. Phys. Chem. C}, 2007, \textbf{111},
  9631--9637\relax
\mciteBstWouldAddEndPuncttrue
\mciteSetBstMidEndSepPunct{\mcitedefaultmidpunct}
{\mcitedefaultendpunct}{\mcitedefaultseppunct}\relax
\EndOfBibitem
\bibitem[Michaelides and Slater(2017)]{michaelides17}
A.~Michaelides and B.~Slater, \emph{Proc. Nat. Acad. Sci.}, 2017, \textbf{114},
  195--197\relax
\mciteBstWouldAddEndPuncttrue
\mciteSetBstMidEndSepPunct{\mcitedefaultmidpunct}
{\mcitedefaultendpunct}{\mcitedefaultseppunct}\relax
\EndOfBibitem
\bibitem[Wettlaufer(1999)]{wettlaufer99}
J.~Wettlaufer, \emph{Phys. Rev. Lett.}, 1999, \textbf{82}, 2516--2519\relax
\mciteBstWouldAddEndPuncttrue
\mciteSetBstMidEndSepPunct{\mcitedefaultmidpunct}
{\mcitedefaultendpunct}{\mcitedefaultseppunct}\relax
\EndOfBibitem
\bibitem[Berendsen \emph{et~al.}(1995)Berendsen, van~der Spoel, and van
  Drunen]{gromacs3}
H.~Berendsen, D.~van~der Spoel and R.~van Drunen, \emph{Comp. Phys. Comm.},
  1995, \textbf{91}, 43 -- 56\relax
\mciteBstWouldAddEndPuncttrue
\mciteSetBstMidEndSepPunct{\mcitedefaultmidpunct}
{\mcitedefaultendpunct}{\mcitedefaultseppunct}\relax
\EndOfBibitem
\bibitem[Hess \emph{et~al.}(2008)Hess, Kutzner, van~der Spoel, and
  Lindahl]{gromacs4}
B.~Hess, C.~Kutzner, D.~van~der Spoel and E.~Lindahl, \emph{J. Chem. Theo.
  Comp.}, 2008, \textbf{4}, 435--447\relax
\mciteBstWouldAddEndPuncttrue
\mciteSetBstMidEndSepPunct{\mcitedefaultmidpunct}
{\mcitedefaultendpunct}{\mcitedefaultseppunct}\relax
\EndOfBibitem
\bibitem[Bussi \emph{et~al.}(2007)Bussi, Donadio, and Parrinello]{bussi07}
G.~Bussi, D.~Donadio and M.~Parrinello, \emph{J. Chem. Phys.}, 2007,
  \textbf{126}, 014101\relax
\mciteBstWouldAddEndPuncttrue
\mciteSetBstMidEndSepPunct{\mcitedefaultmidpunct}
{\mcitedefaultendpunct}{\mcitedefaultseppunct}\relax
\EndOfBibitem
\bibitem[Allen and Tildesley(2017)]{allen17}
M.~Allen and D.~Tildesley, \emph{Computer Simulation of Liquids}, Clarendon
  Press, Oxford, 2nd edn., 2017\relax
\mciteBstWouldAddEndPuncttrue
\mciteSetBstMidEndSepPunct{\mcitedefaultmidpunct}
{\mcitedefaultendpunct}{\mcitedefaultseppunct}\relax
\EndOfBibitem
\bibitem[Frenkel and Smit(2002)]{frenkel02}
D.~Frenkel and B.~Smit, \emph{Understanding Molecular Simulation}, Academic
  Press, San Diego, 2nd edn., 2002\relax
\mciteBstWouldAddEndPuncttrue
\mciteSetBstMidEndSepPunct{\mcitedefaultmidpunct}
{\mcitedefaultendpunct}{\mcitedefaultseppunct}\relax
\EndOfBibitem
\bibitem[Abascal \emph{et~al.}(2005)Abascal, Sanz, Fernandez, and
  Vega]{abascal05}
J.~L.~F. Abascal, E.~Sanz, R.~G. Fernandez and C.~Vega, \emph{J. Chem. Phys.},
  2005, \textbf{122}, 234511\relax
\mciteBstWouldAddEndPuncttrue
\mciteSetBstMidEndSepPunct{\mcitedefaultmidpunct}
{\mcitedefaultendpunct}{\mcitedefaultseppunct}\relax
\EndOfBibitem
\bibitem[Potoff and Siepmann(2001)]{siepmann01}
J.~J. Potoff and J.~I. Siepmann, \emph{AIChE Journal}, 2001, \textbf{47},
  1676--1682\relax
\mciteBstWouldAddEndPuncttrue
\mciteSetBstMidEndSepPunct{\mcitedefaultmidpunct}
{\mcitedefaultendpunct}{\mcitedefaultseppunct}\relax
\EndOfBibitem
\bibitem[Sadlej \emph{et~al.}(1995)Sadlej, Rowland, Devlin, and Buch]{sadlej95}
J.~Sadlej, B.~Rowland, J.~P. Devlin and V.~Buch, \emph{J. Chem. Phys.}, 1995,
  \textbf{102}, 4804--4818\relax
\mciteBstWouldAddEndPuncttrue
\mciteSetBstMidEndSepPunct{\mcitedefaultmidpunct}
{\mcitedefaultendpunct}{\mcitedefaultseppunct}\relax
\EndOfBibitem
\bibitem[Tulegenov \emph{et~al.}(2007)Tulegenov, Wheatley, Hodges, and
  Harvey]{tulegenov07}
A.~S. Tulegenov, R.~J. Wheatley, M.~P. Hodges and A.~H. Harvey, \emph{J. Chem.
  Phys.}, 2007, \textbf{126}, 094305\relax
\mciteBstWouldAddEndPuncttrue
\mciteSetBstMidEndSepPunct{\mcitedefaultmidpunct}
{\mcitedefaultendpunct}{\mcitedefaultseppunct}\relax
\EndOfBibitem
\bibitem[Devlin and Buch(1995)]{devlin95}
J.~P. Devlin and V.~Buch, \emph{J. Phys. Chem.}, 1995, \textbf{99},
  16534--16548\relax
\mciteBstWouldAddEndPuncttrue
\mciteSetBstMidEndSepPunct{\mcitedefaultmidpunct}
{\mcitedefaultendpunct}{\mcitedefaultseppunct}\relax
\EndOfBibitem
\bibitem[Manca and Allouche(2001)]{manca01}
C.~Manca and A.~Allouche, \emph{The Journal of Chemical Physics}, 2001,
  \textbf{114}, 4226--4234\relax
\mciteBstWouldAddEndPuncttrue
\mciteSetBstMidEndSepPunct{\mcitedefaultmidpunct}
{\mcitedefaultendpunct}{\mcitedefaultseppunct}\relax
\EndOfBibitem
\bibitem[Estrada-Torres \emph{et~al.}(2007)Estrada-Torres, Iglesias-Silva,
  Ramos-Estrada, and Hall]{estrada07}
R.~Estrada-Torres, G.~A. Iglesias-Silva, M.~Ramos-Estrada and K.~R. Hall,
  \emph{Fluid Phase Equilibria}, 2007, \textbf{258}, 148 -- 154\relax
\mciteBstWouldAddEndPuncttrue
\mciteSetBstMidEndSepPunct{\mcitedefaultmidpunct}
{\mcitedefaultendpunct}{\mcitedefaultseppunct}\relax
\EndOfBibitem
\bibitem[Rowlinson and Widom(1982)]{rowlinson82b}
J.~Rowlinson and B.~Widom, \emph{Molecular Theory of Capillarity}, Clarendon,
  Oxford, 1982\relax
\mciteBstWouldAddEndPuncttrue
\mciteSetBstMidEndSepPunct{\mcitedefaultmidpunct}
{\mcitedefaultendpunct}{\mcitedefaultseppunct}\relax
\EndOfBibitem
\bibitem[Benet \emph{et~al.}(2016)Benet, Llombart, Sanz, and
  MacDowell]{benet16}
J.~Benet, P.~Llombart, E.~Sanz and L.~G. MacDowell, \emph{Phys. Rev. Lett.},
  2016, \textbf{117}, 096101\relax
\mciteBstWouldAddEndPuncttrue
\mciteSetBstMidEndSepPunct{\mcitedefaultmidpunct}
{\mcitedefaultendpunct}{\mcitedefaultseppunct}\relax
\EndOfBibitem
\bibitem[Benet \emph{et~al.}(2019)Benet, Llombart, Sanz, and
  MacDowell]{benet19}
J.~Benet, P.~Llombart, E.~Sanz and L.~G. MacDowell, \emph{Mol. Phys.}, 2019,
  1--19\relax
\mciteBstWouldAddEndPuncttrue
\mciteSetBstMidEndSepPunct{\mcitedefaultmidpunct}
{\mcitedefaultendpunct}{\mcitedefaultseppunct}\relax
\EndOfBibitem
\bibitem[Lechner and Dellago(2008)]{lechner08}
W.~Lechner and C.~Dellago, \emph{J. Chem. Phys.}, 2008, \textbf{129},
  114707\relax
\mciteBstWouldAddEndPuncttrue
\mciteSetBstMidEndSepPunct{\mcitedefaultmidpunct}
{\mcitedefaultendpunct}{\mcitedefaultseppunct}\relax
\EndOfBibitem
\bibitem[Steinhardt \emph{et~al.}(1983)Steinhardt, Nelson, and
  Ronchetti]{steinhardt83}
P.~J. Steinhardt, D.~R. Nelson and M.~Ronchetti, \emph{Phys. Rev. B}, 1983,
  \textbf{28}, 784--805\relax
\mciteBstWouldAddEndPuncttrue
\mciteSetBstMidEndSepPunct{\mcitedefaultmidpunct}
{\mcitedefaultendpunct}{\mcitedefaultseppunct}\relax
\EndOfBibitem
\bibitem[Benet \emph{et~al.}(2014)Benet, MacDowell, and Sanz]{benet14}
J.~Benet, L.~G. MacDowell and E.~Sanz, \emph{J. Chem. Phys.}, 2014,
  \textbf{141}, 034701\relax
\mciteBstWouldAddEndPuncttrue
\mciteSetBstMidEndSepPunct{\mcitedefaultmidpunct}
{\mcitedefaultendpunct}{\mcitedefaultseppunct}\relax
\EndOfBibitem
\bibitem[Benet \emph{et~al.}(2014)Benet, MacDowell, and Sanz]{benet14c}
J.~Benet, L.~G. MacDowell and E.~Sanz, \emph{Phys. Chem. Chem. Phys.}, 2014,
  \textbf{16}, 22159--22166\relax
\mciteBstWouldAddEndPuncttrue
\mciteSetBstMidEndSepPunct{\mcitedefaultmidpunct}
{\mcitedefaultendpunct}{\mcitedefaultseppunct}\relax
\EndOfBibitem
\bibitem[Espinosa \emph{et~al.}(2016)Espinosa, Vega, and Sanz]{espinosa16}
J.~R. Espinosa, C.~Vega and E.~Sanz, \emph{J. Phys. Chem. C}, 2016,
  \textbf{120}, 8068--8075\relax
\mciteBstWouldAddEndPuncttrue
\mciteSetBstMidEndSepPunct{\mcitedefaultmidpunct}
{\mcitedefaultendpunct}{\mcitedefaultseppunct}\relax
\EndOfBibitem
\bibitem[Nguyen and Molinero(2014)]{nguyen15}
A.~H. Nguyen and V.~Molinero, \emph{J. Phys. Chem. B}, 2014, \textbf{119}, 9369
  -- 9376\relax
\mciteBstWouldAddEndPuncttrue
\mciteSetBstMidEndSepPunct{\mcitedefaultmidpunct}
{\mcitedefaultendpunct}{\mcitedefaultseppunct}\relax
\EndOfBibitem
\bibitem[S{\'a}nchez \emph{et~al.}(2017)S{\'a}nchez, Kling, Ishiyama, van
  Zadel, Bisson, Mezger, Jochum, Cyran, Smit, Bakker, Shultz, Morita, Donadio,
  Nagata, Bonn, and Backus]{sanchez17}
M.~A. S{\'a}nchez, T.~Kling, T.~Ishiyama, M.-J. van Zadel, P.~J. Bisson,
  M.~Mezger, M.~N. Jochum, J.~D. Cyran, W.~J. Smit, H.~J. Bakker, M.~J. Shultz,
  A.~Morita, D.~Donadio, Y.~Nagata, M.~Bonn and E.~H.~G. Backus, \emph{Proc.
  Nat. Acad. Sci.}, 2017, \textbf{114}, 227--232\relax
\mciteBstWouldAddEndPuncttrue
\mciteSetBstMidEndSepPunct{\mcitedefaultmidpunct}
{\mcitedefaultendpunct}{\mcitedefaultseppunct}\relax
\EndOfBibitem
\bibitem[Jorge \emph{et~al.}(2010)Jorge, Jedlovszky, and Cordeiro]{jorge10}
M.~Jorge, P.~Jedlovszky and M.~N. D.~S. Cordeiro, \emph{J. Phys. Chem. C},
  2010, \textbf{114}, 11169--11179\relax
\mciteBstWouldAddEndPuncttrue
\mciteSetBstMidEndSepPunct{\mcitedefaultmidpunct}
{\mcitedefaultendpunct}{\mcitedefaultseppunct}\relax
\EndOfBibitem
\bibitem[Sega \emph{et~al.}(2015)Sega, Fabian, and Jedlovszky]{sega15}
M.~Sega, B.~Fabian and P.~Jedlovszky, \emph{J. Chem. Phys.}, 2015,
  \textbf{143}, 114709\relax
\mciteBstWouldAddEndPuncttrue
\mciteSetBstMidEndSepPunct{\mcitedefaultmidpunct}
{\mcitedefaultendpunct}{\mcitedefaultseppunct}\relax
\EndOfBibitem
\bibitem[Shepherd \emph{et~al.}(2012)Shepherd, Koc, and Molinero]{shepherd12}
T.~D. Shepherd, M.~A. Koc and V.~Molinero, \emph{J. Phys. Chem. C}, 2012,
  \textbf{116}, 12172--12180\relax
\mciteBstWouldAddEndPuncttrue
\mciteSetBstMidEndSepPunct{\mcitedefaultmidpunct}
{\mcitedefaultendpunct}{\mcitedefaultseppunct}\relax
\EndOfBibitem
\bibitem[Battino \emph{et~al.}(1984)Battino, Rettich, and Tominaga]{battino84}
R.~Battino, T.~R. Rettich and T.~Tominaga, \emph{J. Phys. Chem. Ref. Data},
  1984, \textbf{13}, 563--600\relax
\mciteBstWouldAddEndPuncttrue
\mciteSetBstMidEndSepPunct{\mcitedefaultmidpunct}
{\mcitedefaultendpunct}{\mcitedefaultseppunct}\relax
\EndOfBibitem
\bibitem[Fletcher(1970)]{fletcher70}
N.~H. Fletcher, \emph{The Chemical Physics of Ice}, Cambridge University Press,
  1970\relax
\mciteBstWouldAddEndPuncttrue
\mciteSetBstMidEndSepPunct{\mcitedefaultmidpunct}
{\mcitedefaultendpunct}{\mcitedefaultseppunct}\relax
\EndOfBibitem
\bibitem[Pruppacher and Klett(2010)]{pruppacher10}
H.~R. Pruppacher and J.~D. Klett, \emph{Microphysics of Clouds and
  Precipitation}, Springer, Heidelberg, 2010\relax
\mciteBstWouldAddEndPuncttrue
\mciteSetBstMidEndSepPunct{\mcitedefaultmidpunct}
{\mcitedefaultendpunct}{\mcitedefaultseppunct}\relax
\EndOfBibitem
\bibitem[Watkins \emph{et~al.}(2011)Watkins, Pand, Wang, Michaelides,
  VandeVondele, and Slater]{watkins11}
M.~Watkins, D.~Pand, E.~G. Wang, A.~Michaelides, J.~VandeVondele and B.~Slater,
  \emph{Nature Materials}, 2011, \textbf{10}, 794--798\relax
\mciteBstWouldAddEndPuncttrue
\mciteSetBstMidEndSepPunct{\mcitedefaultmidpunct}
{\mcitedefaultendpunct}{\mcitedefaultseppunct}\relax
\EndOfBibitem
\bibitem[Haynes \emph{et~al.}(1992)Haynes, Tro, and George]{haynes92}
D.~R. Haynes, N.~J. Tro and S.~M. George, \emph{J. Phys. Chem.}, 1992,
  \textbf{96}, 8502--8509\relax
\mciteBstWouldAddEndPuncttrue
\mciteSetBstMidEndSepPunct{\mcitedefaultmidpunct}
{\mcitedefaultendpunct}{\mcitedefaultseppunct}\relax
\EndOfBibitem
\bibitem[Neshyba \emph{et~al.}(2009)Neshyba, Nugent, Roeselova, and
  Jungwirth]{neshyba09}
S.~Neshyba, E.~Nugent, M.~Roeselova and P.~Jungwirth, \emph{J. Phys. Chem. C},
  2009, \textbf{113}, 4597--4604\relax
\mciteBstWouldAddEndPuncttrue
\mciteSetBstMidEndSepPunct{\mcitedefaultmidpunct}
{\mcitedefaultendpunct}{\mcitedefaultseppunct}\relax
\EndOfBibitem
\bibitem[Delval and Rossi(2004)]{delval04}
C.~Delval and M.~J. Rossi, \emph{Phys. Chem. Chem. Phys}, 2004, \textbf{6},
  4665--4676\relax
\mciteBstWouldAddEndPuncttrue
\mciteSetBstMidEndSepPunct{\mcitedefaultmidpunct}
{\mcitedefaultendpunct}{\mcitedefaultseppunct}\relax
\EndOfBibitem
\bibitem[Skrotzki \emph{et~al.}(2013)Skrotzki, Connolly, Schnaiter, Saathoff,
  M\"ohler, Wagner, Niemand, Ebert, and Leisner]{skrotzki13}
J.~Skrotzki, P.~Connolly, M.~Schnaiter, H.~Saathoff, O.~M\"ohler, R.~Wagner,
  M.~Niemand, V.~Ebert and T.~Leisner, \emph{Atmosph. Chem. Phys.}, 2013,
  \textbf{13}, 4451--4466\relax
\mciteBstWouldAddEndPuncttrue
\mciteSetBstMidEndSepPunct{\mcitedefaultmidpunct}
{\mcitedefaultendpunct}{\mcitedefaultseppunct}\relax
\EndOfBibitem
\bibitem[Kong \emph{et~al.}(2014)Kong, Papagiannakopoulos, Thomson, Markovi\"c,
  and Pettersson]{kong14}
X.~Kong, P.~Papagiannakopoulos, E.~S. Thomson, N.~Markovi\"c and J.~B.~C.
  Pettersson, \emph{J. Phys. Chem. A}, 2014, \textbf{118}, 3973--3979\relax
\mciteBstWouldAddEndPuncttrue
\mciteSetBstMidEndSepPunct{\mcitedefaultmidpunct}
{\mcitedefaultendpunct}{\mcitedefaultseppunct}\relax
\EndOfBibitem
\bibitem[IUPAC(2009)]{iupac09}
IUPAC, \emph{Task Group on Atmospheric Chemical Kinetic DataEvaluation,
  http://iupac.pole-ether.fr}, Iupac technical report, 2009\relax
\mciteBstWouldAddEndPuncttrue
\mciteSetBstMidEndSepPunct{\mcitedefaultmidpunct}
{\mcitedefaultendpunct}{\mcitedefaultseppunct}\relax
\EndOfBibitem
\bibitem[Batista \emph{et~al.}(2005)Batista, Ayotte,
  Bili\ifmmode~\acute{c}\else \'{c}\fi{}, Kay, and J\'onsson]{batista05}
E.~R. Batista, P.~Ayotte, A.~Bili\ifmmode~\acute{c}\else \'{c}\fi{}, B.~D. Kay
  and H.~J\'onsson, \emph{Phys. Rev. Lett.}, 2005, \textbf{95}, 223201\relax
\mciteBstWouldAddEndPuncttrue
\mciteSetBstMidEndSepPunct{\mcitedefaultmidpunct}
{\mcitedefaultendpunct}{\mcitedefaultseppunct}\relax
\EndOfBibitem
\bibitem[Pfalzgraff \emph{et~al.}(2011)Pfalzgraff, Neshyba, and
  Roeselova]{pfalzgraf11}
W.~Pfalzgraff, S.~Neshyba and M.~Roeselova, \emph{J. Phys. Chem. A}, 2011,
  \textbf{115}, 6184--6193\relax
\mciteBstWouldAddEndPuncttrue
\mciteSetBstMidEndSepPunct{\mcitedefaultmidpunct}
{\mcitedefaultendpunct}{\mcitedefaultseppunct}\relax
\EndOfBibitem
\bibitem[Murata \emph{et~al.}(2019)Murata, Nagashima, and Sazaki]{murata19}
K.-i. Murata, K.~Nagashima and G.~Sazaki, \emph{Phys. Rev. Lett.}, 2019,
  \textbf{122}, 026102\relax
\mciteBstWouldAddEndPuncttrue
\mciteSetBstMidEndSepPunct{\mcitedefaultmidpunct}
{\mcitedefaultendpunct}{\mcitedefaultseppunct}\relax
\EndOfBibitem
\bibitem[Li \emph{et~al.}(2001)Li, Davidovits, Kolb, and Worsnop]{li01b}
Y.~Q. Li, P.~Davidovits, C.~E. Kolb and D.~R. Worsnop, \emph{J. Phys. Chem. A},
  2001, \textbf{105}, 10627--10634\relax
\mciteBstWouldAddEndPuncttrue
\mciteSetBstMidEndSepPunct{\mcitedefaultmidpunct}
{\mcitedefaultendpunct}{\mcitedefaultseppunct}\relax
\EndOfBibitem
\bibitem[Davidovits \emph{et~al.}(2004)Davidovits, Worsnop, Jayne, Kolb,
  Winkler, Vrtala, Wagner, Kulmala, Lehtinen, Vesala, and
  Mozurkewich]{davidovits04}
P.~Davidovits, D.~R. Worsnop, J.~T. Jayne, C.~E. Kolb, P.~Winkler, A.~Vrtala,
  P.~E. Wagner, M.~Kulmala, K.~E.~J. Lehtinen, T.~Vesala and M.~Mozurkewich,
  \emph{Geo. Phys. Res. Lett.}, 2004, \textbf{31}, \relax
\mciteBstWouldAddEndPuncttrue
\mciteSetBstMidEndSepPunct{\mcitedefaultmidpunct}
{\mcitedefaultendpunct}{\mcitedefaultseppunct}\relax
\EndOfBibitem
\bibitem[Morita \emph{et~al.}(2004)Morita, Sugiyama, Kameda, Koda, and
  Hanson]{morita04}
A.~Morita, M.~Sugiyama, H.~Kameda, S.~Koda and D.~R. Hanson, \emph{J. Phys.
  Chem. B}, 2004, \textbf{108}, 9111--9120\relax
\mciteBstWouldAddEndPuncttrue
\mciteSetBstMidEndSepPunct{\mcitedefaultmidpunct}
{\mcitedefaultendpunct}{\mcitedefaultseppunct}\relax
\EndOfBibitem
\end{mcitethebibliography}


\providecommand*{\mcitethebibliography}{\thebibliography}
\csname @ifundefined\endcsname{endmcitethebibliography}
{\let\endmcitethebibliography\endthebibliography}{}




\setcounter{page}{1}
\setcounter{table}{0}
\setcounter{figure}{0}
\pagenumbering{arabic}

{\centering
{\large Supporting Information for}

{\Large Structure and water attachment rates of ice in the atmosphere: role of
   nitrogen
\\ by \\}
{\large P. Llombart, R. Bergua, E. G. Noya and  L. G. MacDowell}

{\normalsize
Instituto de Qu\'{\i}mica F\'{\i}sica Rocasolano
, CSIC, Calle Serrano 119, 28006 Madrid, Spain \\
and \\ Departamento de Qu\'{\i}mica F\'{\i}sica, Facultad de Ciencias
Qu\'{\i}micas, Universidad Complutense, Madrid, 28040, Spain.}

}

\vspace*{1cm}




\begin{table}[hb]
\footnotesize
\begin{tabular}{|ccccccccccccccccccccc|}
\hline
\hline
 $\rho$  & T & P  & $\rho$ & T & P & $\rho$ & T & P & $\rho$ & T & P & $\rho$ & T & P & $\rho$ & T & P & $\rho$ & T & P  \\ \hline
 \multicolumn{1}{|l|}{\multirow{9}{*}{0.56}}& 230        & 0.39   & \multicolumn{1}{|l|}{\multirow{9}{*}{0.79}} & 230        & 0.55  & \multicolumn{1}{|l|}{\multirow{9}{*}{1.03 }}& 230        & 0.71 &\multicolumn{1}{|l|}{\multirow{9}{*}{1.26}} & 230        & 0.87  & \multicolumn{1}{|l|}{\multirow{9}{*}{1.50}} & 230        & 1.01   & \multicolumn{1}{|l|}{\multirow{9}{*}{1.73}} & 230        & 1.14  &\multicolumn{1}{|l|}{\multirow{9}{*}{1.96}} & 230      & 1.25           \\
 \multicolumn{1}{|l|}{}                    & 235        & 0.40   &   \multicolumn{1}{|l|}{}        & 235        & 0.55  &  \multicolumn{1}{|l|}{}                        & 235        & 0.70 &  \multicolumn{1}{|l|}{}                       & 235        & 0.88   &    \multicolumn{1}{|l|}{}                      & 235        & 0.99 &    \multicolumn{1}{|l|}{}                     & 235        & 1.17  &      \multicolumn{1}{|l|}{}                    & 235        & 1.35           \\
 \multicolumn{1}{|l|}{}                    & 240        & 0.39   &   \multicolumn{1}{|l|}{}        & 240        & 0.57  &  \multicolumn{1}{|l|}{}                        & 240        & 0.72 &  \multicolumn{1}{|l|}{}                       & 240        & 0.88   &    \multicolumn{1}{|l|}{}                      & 240        & 1.08 &    \multicolumn{1}{|l|}{}                     & 240        & 1.24  &      \multicolumn{1}{|l|}{}                    & 240        & 1.37           \\  
 \multicolumn{1}{|l|}{}                    & 245        & 0.42   &   \multicolumn{1}{|l|}{}        & 245        & 0.59  &  \multicolumn{1}{|l|}{}                        & 245        & 0.74 &  \multicolumn{1}{|l|}{}                       & 245        & 0.91   &    \multicolumn{1}{|l|}{}                      & 245        & 1.08 &    \multicolumn{1}{|l|}{}                     & 245        & 1.25  &      \multicolumn{1}{|l|}{}                    & 245        & 1.42           \\
 \multicolumn{1}{|l|}{}                    & 250        & 0.42   &   \multicolumn{1}{|l|}{}        & 250        & 0.58  &  \multicolumn{1}{|l|}{}                        & 250        & 0.75 &  \multicolumn{1}{|l|}{}                       & 250        & 0.94   &    \multicolumn{1}{|l|}{}                      & 250        & 1.11 &    \multicolumn{1}{|l|}{}                     & 250        & 1.29  &      \multicolumn{1}{|l|}{}                    & 250        & 1.42           \\
 \multicolumn{1}{|l|}{}                    & 255        & 0.42   &   \multicolumn{1}{|l|}{}        & 255        & 0.59  &  \multicolumn{1}{|l|}{}                        & 255        & 0.78 &  \multicolumn{1}{|l|}{}                       & 255        & 0.94   &    \multicolumn{1}{|l|}{}                      & 255        & 1.09 &    \multicolumn{1}{|l|}{}                     & 255        & 1.29  &      \multicolumn{1}{|l|}{}                    & 255        & 1.45           \\
 \multicolumn{1}{|l|}{}                    & 260        & 0.43   &   \multicolumn{1}{|l|}{}        & 260        & 0.60  &  \multicolumn{1}{|l|}{}                        & 260        & 0.80 &  \multicolumn{1}{|l|}{}                       & 260        & 0.95   &    \multicolumn{1}{|l|}{}                      & 260        & 1.16 &    \multicolumn{1}{|l|}{}                     & 260        & 1.34  &      \multicolumn{1}{|l|}{}                    & 260        & 1.48           \\ 
 \multicolumn{1}{|l|}{}                    & 265        & 0.44   &   \multicolumn{1}{|l|}{}        & 265        & 0.62  &  \multicolumn{1}{|l|}{}                        & 265        & 0.80 &  \multicolumn{1}{|l|}{}                       & 265        & 1.00   &    \multicolumn{1}{|l|}{}                      & 265        & 1.11 &    \multicolumn{1}{|l|}{}                     & 265        & 1.33  &      \multicolumn{1}{|l|}{}                    & 265        & 1.52           \\
  \multicolumn{1}{|l|}{}                   & 270        & 0.46   &    \multicolumn{1}{|l|}{}       & 270        & 0.64  &   \multicolumn{1}{|l|}{}                       & 270        & 0.82 &   \multicolumn{1}{|l|}{}                      & 270        & 1.02   &     \multicolumn{1}{|l|}{}                     & 270        & 1.19 &     \multicolumn{1}{|l|}{}                    & 270        & 1.39  &       \multicolumn{1}{|l|}{}                   & 270        & 1.55      \\  \hline
\hline
\end{tabular}
\caption{
         Equation of State for nitrogen. Pressure (P) is expressed in units of $10^{5}$ Pa, 
	   temperature (T) in K and density ($\rho$) in kg m$^{-3}$.  
         }
\label{tab:eos}
\end{table}

\begin{table}[hb]
  \centering
  \begin{tabular}{cc}
     \hline
     \hline
        T  &  $B_{2}(T)$ \\ 
       $K$ &  $10^{3}$~$kg^{-1}m^{3}$ \\
     \hline
                               270               &                      -3.2         \\          
                              265               &                      -11.9        \\          
                              260               &                      -5.9         \\          
                              255               &                      -13.8        \\          
                              250               &                      -4.5$^{\ast}$         \\          
                              245               &                      -3.4$^{\ast}$         \\          
                              240               &                      -4.7$^{\ast}$       \\          
                              235               &                      -13.0        \\          
                              230               &               	   -20.9        \\
     \hline
  \end{tabular}
  \caption{ Second virial coefficient $B_2$ as a function of temperature. 
           \textsuperscript{$\ast$} Three out layers were not included in the fit 
             for $B_2(T)$.
          }
\label{tab:systems2}
\end{table}

\begin{table*}[ht]
\footnotesize
  \begin{tabular}{cccccp{4cm}p{5cm}}
     \hline
     \hline
      $Face$      &  $T$      &  $Lx$         &   $Ly$         & $Lz$                  &     Number of $\rm{N_{2}}$ & $\rho_{\rm{N_{2}}}^{bulk}$ \\
                  &  $K$      &  $nm$         &   $nm$         & $nm$                  &                            & $kg m^{-3}$ \\
     \hline
      \multirow{10}{*}{Basal}    & \multirow{2}{*}{270} &   \multirow{2}{*}{7.26698} & \multirow{2}{*}{6.29362} & 15.00000 &      0             &       0          \\
                                 &                      &                            &                          & 36.85765 & 46, 44, 42, 37, 32 & 0.95, 1.10, 1.24, 1.31, 1.37\\         
                                 & \multirow{2}{*}{260} &   \multirow{2}{*}{7.26350} & \multirow{2}{*}{6.29060} & 15.00000 &      0             &       0                \\ 
                                 &                      &                            &                          & 36.85765 & 49, 46, 43, 36, 29 & 0.86, 1.06, 1.27, 1.36, 1.46\\        
                                 & \multirow{2}{*}{250} &   \multirow{2}{*}{7.25949} & \multirow{2}{*}{6.28713} & 15.00000 &      0             &       0             \\
                                 &                      &                            &                          & 36.85765 & 53, 50, 47, 30, 25 & 0.74, 0.89, 1.40, 1.48, 1.56\\        
                                 & \multirow{2}{*}{240} &   \multirow{2}{*}{7.25609} & \multirow{2}{*}{6.28412} & 15.00000 &      0             &       0             \\
                                 &                      &                            &                          & 36.88880 & 48, 45, 37, 21     & 0.62, 1.09, 1.33, 1.41\\        
                                 & \multirow{2}{*}{230} &   \multirow{2}{*}{7.25229} & \multirow{2}{*}{6.28104} & 15.00000 &      0             &       0             \\
                                 &                      &                            &                          & 36.58888 & 50, 33, 17         & 0.50, 0.98 1.48\\        
     \hline
      \multirow{10}{*}{pI}       & \multirow{2}{*}{270} &   \multirow{2}{*}{7.26707} & \multirow{2}{*}{5.91452} & 15.00000 &      0             & 0 \\
                                 &                      &                            &                          & 36.85765 & 46, 44, 42, 37, 32 & 1.02, 1.18, 1.34, 1.40, 1.46\\        
                                 & \multirow{2}{*}{260} &   \multirow{2}{*}{7.26329} & \multirow{2}{*}{5.91143} & 15.00000 &      0             &       0 \\
                                 &                      &                            &                          & 36.85765 & 49, 46, 43, 36, 29 & 0.92, 1.15, 1.37, 1.46, 1.55\\        
                                 & \multirow{2}{*}{250} &   \multirow{2}{*}{7.25957} & \multirow{2}{*}{5.90841} & 15.00000 &      0             &       0 \\
                                 &                      &                            &                          & 36.85765 & 53, 50, 47, 30, 25 & 0.79, 0.95, 1.42, 1.52, 1.58\\        
                                 & \multirow{2}{*}{240} &   \multirow{2}{*}{7.25604} & \multirow{2}{*}{5.90554} & 15.00000 &      0             &       0 \\
                                 &                      &                            &                          & 35.58888 & 48, 45, 37, 21     & 0.69, 1.22, 1.48, 1.58 \\        
                                 & \multirow{2}{*}{230} &   \multirow{2}{*}{7.25229} & \multirow{2}{*}{5.90249} & 15.00000 &      0             &       0 \\
                                 &                      &                            &                          & 35.85765 & 50, 33, 17         & 0.55, 1.07, 1.62 \\        
     \hline
     \hline
  \end{tabular}
  \caption{Summary of thermodynamic conditions and system sizes for the
     simulations of the ice interface in presence of nitrogen.
           }
\label{tab:systems1}
\end{table*}

\begin{table}
\footnotesize
\begin{tabular}{cccccc}
\hline
\hline
$ Facet $     & $T$ / K    & N$_{N_{2}}$  &  N$_{\rm ns}$  & N$_{ev}$  &  $\alpha $  \\
\hline
Basal        & 270        &  42             &      14 (14)        &            2                &     0.9650   \\
Basal        & 270        &  0              &      0              &            0                &     1.0000   \\
Basal        & 260        &  43             &      12 (9)         &            2                &     0.9700   \\
Basal        & 260        &  0              &      6              &            1                &     0.9850   \\
Basal        & 230        &  50             &      14 (11)        &            0                &     0.9650   \\
Basal        & 230        &  0              &      5              &            0                &     0.9875   \\
pI           & 270        &  42             &      13 (10)        &            2                &     0.9675   \\
pI           & 270        &  0              &      1              &            1                &     0.9975   \\
pI           & 260        &  43             &      13 (11)        &            2                &     0.9675   \\
pI           & 260        &  0              &      1              &            0                &     0.9975   \\
pI           & 230        &  50             &      20 (19)        &            0                &     0.9500   \\
pI           & 230        &  0              &      0              &            0                &     1.0000   \\
\hline
\hline
\end{tabular}
\caption{Table with detailed information of the collision statistics.
$N_{N_2}$ is the number
of nitrogen molecules present in the simulation box.  N$_{ns}$ is the number of
water molecules shot  a distance of 2 nm away from the surface and not
sticking into the surface. Shown in parenthesis is the number
of molecules which were reflected back to the gas phase by collisions with nitrogen 
gas molecules.  N$_{ev}$ provides the number of evaporation events observed during the
simulations.
}
\label{tab:alpha_temp_press}
\end{table}

\clearpage

\begin{figure}
\centering
\resizebox*{16cm}{!}{\includegraphics{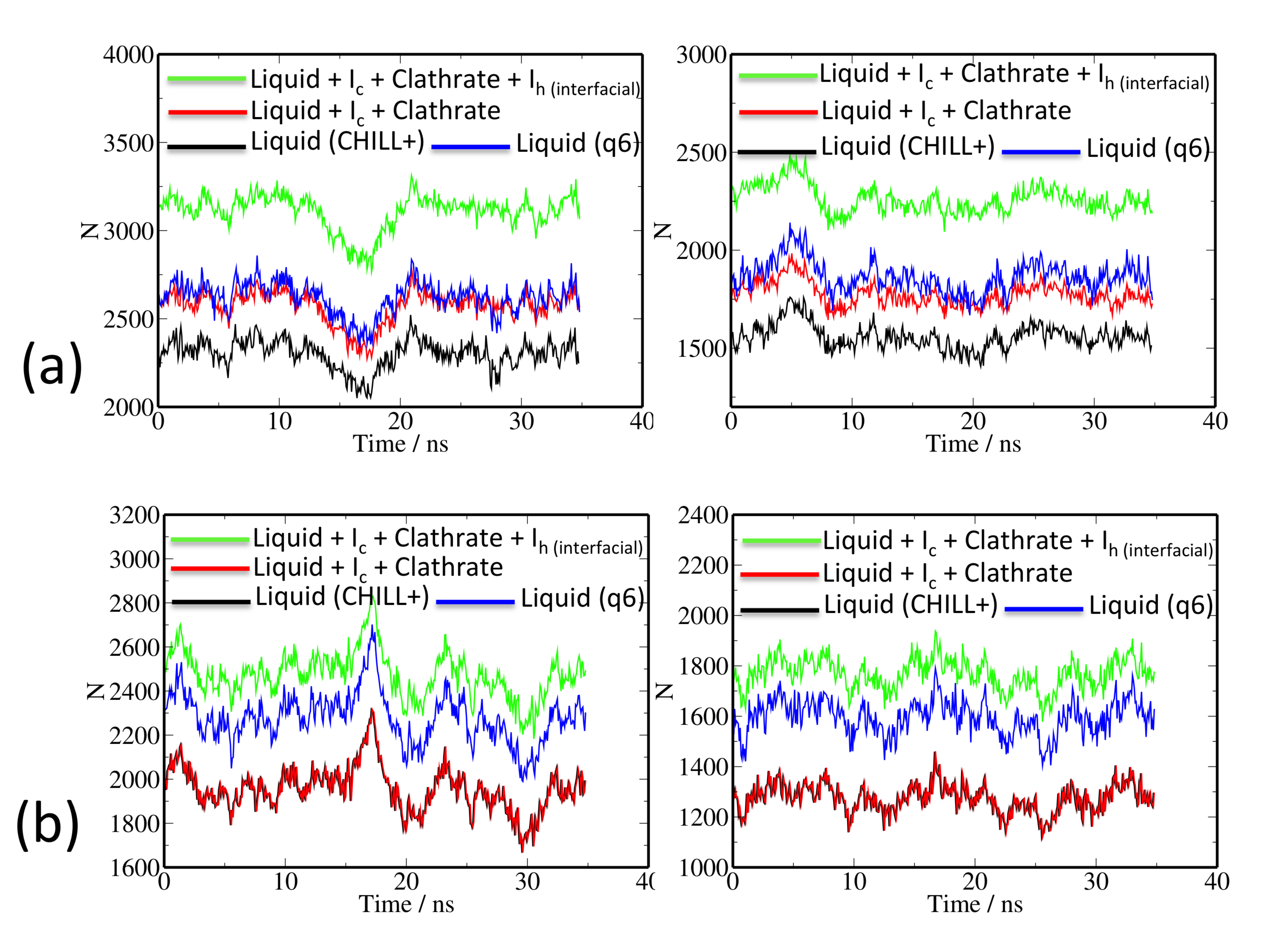}}
\caption{
   Comparison of the number of liquid-like molecules as determined from
   the $\bar q_{6}$ parameter and the CHILL+ algorithm.$^{84}$.
a) Results for the Basal plane. b) Results for the prismatic plane. 
Plots are shown from left to right at 270, 260, 250, 240 and 230~K, respectively.
Blue: Runing number of liquid molecules during a simulation as obtained from
the $\bar q_{6}$ parameter used in this work. 
Black: Runing number of liquid molecules as a function of time as  extracted
with the  CHILL+ algorithm. Notice that the blue and black lines run almost
parallel to each other, with a constant offset of about 12\%. 
Other possible choices to determine the thickness of the premelting layer
remain also largely correlated.
Red: Runing number of molecules in liquid, cubic and clathrate like
environments as obtained from the  CHILL+ algorithm. 
Green: Runing number of molecules in liquid, cubic, clathrate, and interfacial
hexagonal environments as obtained from the CHILL+ algorithm.
}
 \label{fig:chill_q6}
\end{figure}

\begin{figure}
\centering
\resizebox*{6cm}{!}{\includegraphics{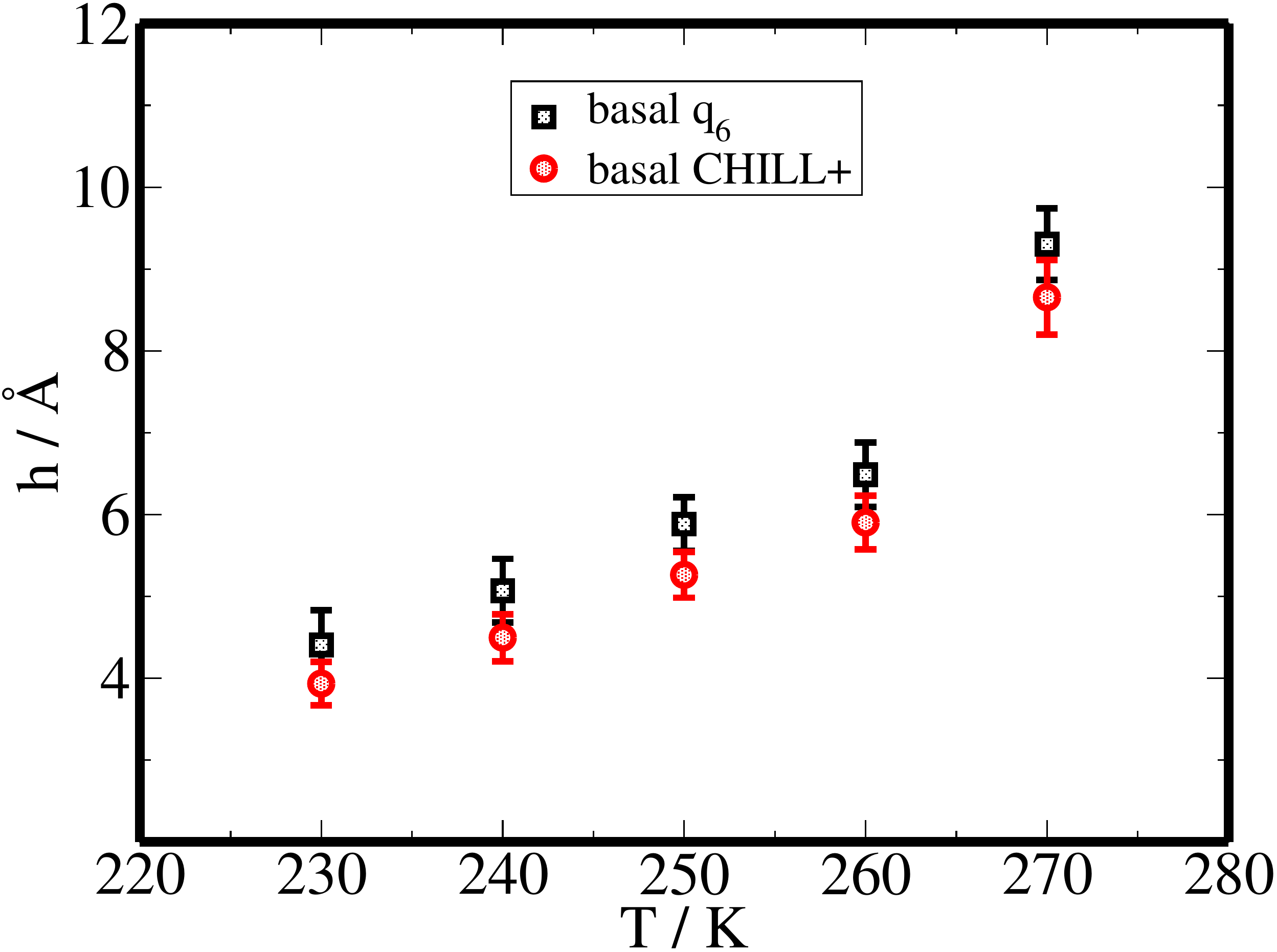}}
\resizebox*{6cm}{!}{\includegraphics{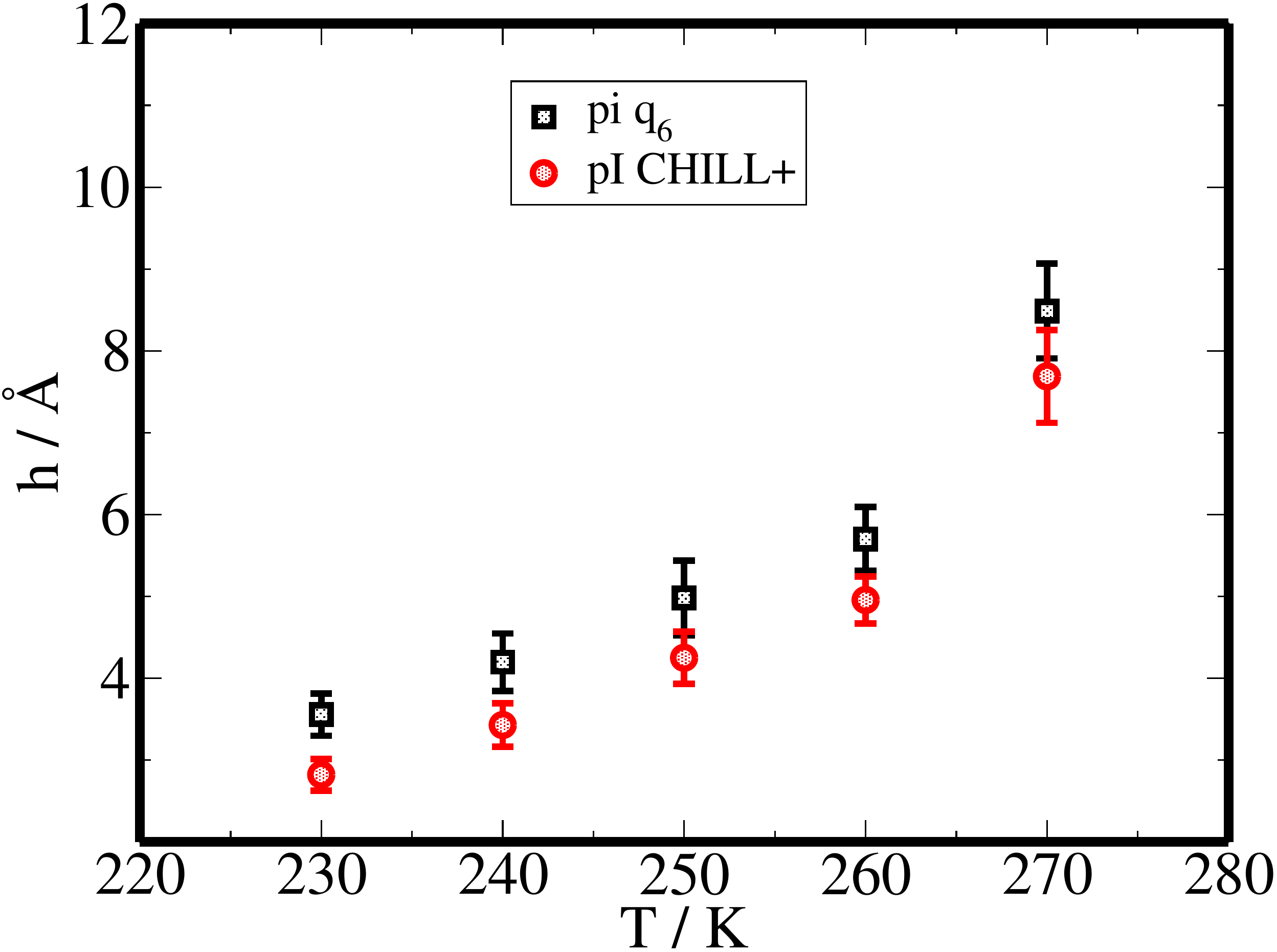}}
\caption{Left: Basal. Right: pI. Premelting thickness at zero nitrogen pressure as
   calculated using the number of liquid like molecules from $\bar q_{6}$
   (squares)
   used in this work and the CHILL+ algorithm (circles). Notice that
   the calculation of film heights differs by an almost constant offset.
}
 \label{fig:chill_q6_2}
\end{figure}

\end{document}